\documentclass[aps,prb,longbibliography,twocolumn,floatfix,superscriptaddress,amsmath,amssymb]{revtex4-2}

\usepackage{graphicx} % Required for inserting images
\usepackage[usenames,dvipsnames]{xcolor}
\usepackage{amssymb}
\usepackage{graphicx}
\usepackage{amsmath}
\usepackage{bbm}
\usepackage[colorlinks,bookmarks=true,citecolor=blue,linkcolor=red,urlcolor=blue]{hyperref}
\usepackage{braket}
\usepackage[english]{babel}
\usepackage{blindtext}
\usepackage{physics}
\usepackage{mathrsfs}
\usepackage{footmisc}
\usepackage{booktabs}
\usepackage{tikz}
\usetikzlibrary{arrows.meta,positioning}

\newcommand{\e}{\mathrm{e}}

\usepackage{mathtools}
\usepackage[caption=false]{subfig}

\usepackage{dsfont}

\definecolor{mypurp}{rgb}{0.35, 0, 0.7}

\usepackage{amsthm}
\usepackage{txfonts}

\usepackage{soul}
\usepackage[export]{adjustbox}

\usepackage{comment}

\usetikzlibrary{positioning}

\newcommand{\ii}{\mathrm{i}}

\newcommand{\MPO}{\ensuremath{\mathrm{MPO}}}
\newcommand{\iiMPS}{\ensuremath{\mathrm{iMPS}}}
\newcommand{\iiMPO}{\ensuremath{\mathrm{iMPO}}}

\begin{document}
\title{Lattice Symmetry Remnants in Spatial and Temporal Entanglement of \\ Quantum Critical Spin Chains}

\newcommand{\gscaep}[0]{Graduate School of China Academy of Engineering Physics, Beijing 100193, China}

\author{Xiao-Han Yang}
\affiliation{Hefei National Laboratory, University of Science and Technology of China, Hefei 230088, China}
\affiliation{Hefei National Research Center for Physical Sciences at the Microscale and School of Physical Sciences, University of Science and Technology of China, Hefei 230026, China}
\affiliation{\gscaep}

\author{Rui-Zhen Huang}
\email{huangrzh@icloud.com}
\affiliation{\gscaep}

\begin{abstract}
Quantum critical lattice models retain universal information about their lattice symmetries beyond the bulk data of their infrared conformal field theory (CFT). We probe these lattice symmetry remnants by comparing the spatial and temporal entanglement of two Euclidean-evolved states from the same critical model. At large length scales, short-range initial states define conformal boundaries and boundary CFT (BCFT) organizes the universal low-lying entanglement spectrum. In the temporal construction, the translation invariance along Euclidean time requires a uniform state. Moreover, the time slice $\delta \tau \rightarrow 0$ forces a split-gate basis, resulting in a new regular uniform symmetry-preserving transfer representation. For the Ising and Potts chains, onsite symmetries result in the same BCFT structure in the spatial and temporal constructions. The spin-$1/2$ XXZ chain instead realizes the Lieb--Schultz--Mattis (LSM) constraint differently in the two Euclidean quantization. A $U(1)$-preserving spatial state has a doubled unit cell and inequivalent entanglement cuts, whereas the temporal state remains uniform and alternates between translation relation related Dirichlet sectors with integer and half-integer compact-boson zero modes. Breaking $U(1)$ removes the zero mode and the alternation. Thus microscopic symmetry and anomaly data remain visible in universal entanglement by constraining the boundary realization of the infrared CFT.
\end{abstract}

\maketitle
%\tableofcontents

\section{Introduction}

At a quantum critical point, the long-distance fixed-point physics of a lattice model is usually described by a conformal field theory~\cite{difrancesco_1997_CFT_book, blumenhagen_2009_CFT_book}. A lattice realization nevertheless contains additional exact microscopic structure. It specifies a tensor-product Hilbert space, the representation carried by each microscopic unit cell, and the action of crystalline symmetries such as translation. These data determine how symmetries are implemented and constrain the symmetry-preserving perturbations available near the fixed point. They can therefore leave universal remnants in the infrared even when the bulk scaling behavior is governed by the same CFT.

The Lieb--Schultz--Mattis (LSM) constraint provides a particularly sharp example~\cite{Lieb_1961_LSM,oshikawa_2000_commensurability,tasaki_2022_LSM_topo,cheng_2023_lsm,else_thorngren_2020_lsm}. For a spin-$1/2$ chain at zero magnetization, one-site translation and the internal spin-rotation symmetry have an anomalous low-energy realization. Accordingly, a one-site injective, short-range MPS cannot preserve both symmetries, and more generally no unique gapped ground state can preserve both with a one-site unit cell~\cite{Lieb_1961_LSM,oshikawa_2000_commensurability,tasaki_2022_LSM_topo,cheng_2023_lsm,else_thorngren_2020_lsm,sanz_2009_mps_lsm,cirac_2021_mps_review}. At the critical point, the exact ground state is gapless and may remain translation invariant, but this anomalous symmetry realization must still be matched by the long-distance theory. The microscopic symmetry representation carried by the unit cell therefore remains part of the physical specification of the critical lattice system. This motivates a basic question of how such lattice information can be exposed in universal observables at criticality.

Entanglement along Euclidean evolution provides a natural setting for addressing this question. Starting from a short-range state and evolving with a critical Hamiltonian generates a family of states that continuously connects the microscopic preparation to the long-distance critical regime. The Euclidean evolution length supplies a controllable scale before the asymptotic ground-state limit is reached. Entanglement probes this evolution at complementary levels. Its leading logarithmic growth measures bulk criticality, while the detailed universal low-lying entanglement spectrum can retain information about the microscopic preparation. By varying the initial state and its symmetry, one can therefore track how exact lattice constraints remain visible at large length scales in universal entanglement data.

We study this idea using two Euclidean-evolution states associated with the same two-dimensional tensor network. The \emph{spatial state}
\begin{equation}
    \ket{\Psi_{\mathrm{sp}}(\tau)}
    =\mathcal N_{\mathrm{sp}}(\tau)\,
      \e^{-\tau H_x}\ket{\Psi_{\mathrm{sp},0}}
    \label{eq:spatial_state_def}
\end{equation}
lives on the physical spin chain and is generated by ordinary imaginary-time evolution. Quantizing the same Euclidean network in the orthogonal direction gives the \emph{temporal state}
\begin{equation}
    \ket{\Psi_{\mathrm{tm}}(L)}
    =\mathcal N_{\mathrm{tm}}(L)\,
      \mathbb T_x^L\ket{\Psi_{\mathrm{tm},0}},
    \label{eq:temporal_state_def}
\end{equation}
which lives on a transverse Hilbert space assembled from Euclidean-time slices and evolves under the spatial quantum transfer operator $\mathbb T_x$. These two constructions are Euclidean-evolution trajectories rather than generic real-time quenches. The temporal state is an auxiliary transfer state, so its bipartite entanglement across Euclidean-time segments is best viewed as a diagnostic of the Euclidean path integral rather than as entanglement between two physical times. Related transverse-contraction, temporal-MPS, and space--time-dual approaches have been studied in both Euclidean and real-time settings~\cite{banuls_2009_transverse,mullerhermes_2012_folding,hastings_2015_time_space,tirrito_2018_temporal_mps,sonner_2021_temporal_entanglement,lerose_2021_temporal_scaling,giudice_2022_temporal_entanglement,lerose_2023_spacetime_duality,foligno_2023_temporal_entanglement,carignano_2024_temporal_entropy,carignano_tagliacozzo_2025_temporal_cft}.

At a Lorentz-invariant continuum fixed point, a $90^\circ$ Euclidean rotation relates the two quantization directions. The lattice regulator, however, resolves the two slices into different microscopic tensor-product Hilbert spaces. Onsite internal symmetries and crystalline translation can consequently be represented differently in the two lattice quantizations. For example, the translation invariance along the temporal direction can not be spontaneously broken which enforces the temporal state to be uniform.  Comparing the spatial and temporal entanglement therefore gives a direct way to test how microscopic symmetry information is carried into the infrared. This distinction becomes especially consequential when translation participates in an LSM constraint.

Boundary conformal field theory (BCFT) enters naturally once the Euclidean-evolved states are written as path integrals~\cite{cardy_1984_surface,cardy_2008_bcft,calabrese_cardy_2006_quench,calabrese_cardy_2007_quenches,calabrese_cardy_2016_quench_review}. The short-range initial state terminates the Euclidean path integral and therefore defines a microscopic boundary condition. At length scales large compared with those of the microscopic preparation, this boundary is described by a conformal boundary $B_O$. The lattice bipartition supplies a second microscopic boundary, fixed by the tensor-product factorization and the chosen Schmidt cut. Its long-distance description is the entanglement boundary $B_E$. Because the cut is held fixed throughout our boundary-engineering protocol, varying the initial state changes $B_O$ while leaving $B_E$ unchanged. After mapping the reduced-density-matrix geometry to an annulus, the universal low-lying entanglement spectrum is organized by the open-channel BCFT between $B_O$ and $B_E$. Exact microscopic symmetries restrict the allowed boundary perturbations and hence the conformal boundaries accessible from a given initial state. The entanglement spectrum thus provides a spectroscopic probe of symmetry-compatible conformal boundaries.

In the scaling window
\begin{equation}
    1\ll \ell\ll\xi,
    \label{eq:scaling_window}
\end{equation}
where $\ell=\tau/L$ for the spatial/temporal state and $\xi$ is the correlation length of the evolved state, the universal low-lying entanglement spectrum realizes the open-channel Hilbert space $\mathcal H_{B_OB_E}$ up to an overall scale, an additive shift, and irrelevant lattice corrections. The entanglement entropy has the corresponding logarithmic coefficient fixed by the bulk central charge $c$. Changing the microscopic initial state can therefore select different conformal boundary sectors in this long-distance description. In the tensor network framework, $\xi$ is upper-bounded by the finite bond dimension $D$~\cite{frank_2009_ee_scaling,tagliacozzo_2008_fes}. At $\ell\gtrsim\xi$ the initial boundary is screened and the spectrum crosses over to a boundary selected by the finite-entanglement deformation~\cite{tagliacozzo_2008_fes,frank_2009_ee_scaling,pirvu_2012_finite_size_entanglement,stojevic_2015_fes,cho_2017_universal_ES,huang_2024_emergent_bc}.

With this framework, we ask how the same microscopic symmetry constraint is represented in the two lattice quantizations. We first construct the transfer operator for the temporal state. Interestingly, the temporal translation invariance and the time slice $\delta \tau \rightarrow 0$ together forces a split-gate basis, resulting in a regular, uniform-preserving transfer representation.  The Ising and three-state Potts chains provide onsite-symmetry control cases, for which the spatial and temporal constructions approach the same BCFT structure. The XXZ chain provides the central LSM example. Bosonization fixes the lattice translation action on compact-boson primaries as $V_{e,w}\mapsto(-1)^{e+w}V_{e,w}$ and places a $U(1)_z$-preserving elementary boundary in a translation-related Dirichlet pair. The spatial preparation realizes this constraint through a doubled short-range unit cell and inequivalent entanglement cuts. The temporal state remains uniform along Euclidean time, while successive spatial transfer steps move its entanglement spectrum between the two translation-related boundary sectors. These are complementary lattice manifestations of the same infrared symmetry constraint.

The paper is organized as follows. Section~\ref{sec:bcft_framework} reviews the necessary BCFT framework and describes spatial and temporal Euclidean-evolution entanglement through a common annulus construction with conformal boundaries. Section~\ref{sec:tensor_network} gives the lattice transfer construction. Section~\ref{sec:onsite_models} presents Ising and Potts benchmarks. Section~\ref{sec:xxz} develops the compact-boson description, the LSM constraint on its conformal boundaries, and the two lattice realizations of that constraint. Section~\ref{sec:crossover} discusses finite-entanglement crossover, followed by conclusions in Sec.~\ref{sec:conclusions}.

\section{BCFT framework for spatial and temporal entanglement}
\label{sec:bcft_framework}

The finite time or equivalently length in the Euclidean evolution provides a controllable scale over which one can follow how information encoded in a short-range microscopic preparation survives into the universal critical regime. In the scaling window, the preparation boundary and the fixed microscopic lattice cut are described at long distances by conformal boundaries $B_O$ and $B_E$, respectively. After a conformal map to an annulus, the universal low-lying reduced-density-matrix spectrum is generated by the open-channel BCFT between them. We first review the boundary-state and Cardy-consistency ingredients needed for this construction, and then formulate the entanglement problem in a way that applies equally to spatial and temporal Euclidean evolution.

\subsection{Conformal boundaries and Cardy consistency}

A boundary conformal field theory is a CFT defined on a manifold with a boundary~\cite{cardy_1984_surface,cardy_1986_boundary_operator,cardy_2008_bcft}. On the upper half-plane, conformal invariance requires the energy flux through the real axis to vanish. In complex coordinates this condition is
\begin{equation}
    T(x)=\bar T(x),
    \qquad x\in\mathbb R.
    \label{eq:bcft_stress_gluing}
\end{equation}
It identifies the left- and right-moving stress tensors at the boundary, leaving a single diagonal Virasoro algebra acting on boundary fields. In the closed-channel quantization, the same condition is imposed on a conformal boundary state $\ket{B_a}$ as
\begin{equation}
    \left(L_n-\bar L_{-n}\right)\ket{B_a}=0,
    \qquad n\in\mathbb Z.
    \label{eq:boundary_state_gluing}
\end{equation}
Additional chiral currents may be preserved up to an automorphism, but Eqs.~\eqref{eq:bcft_stress_gluing} and \eqref{eq:boundary_state_gluing} are the minimal Virasoro conditions meant by a conformal boundary below.

The bulk Hilbert space is specified by a modular invariant pairing of left- and right-moving representations,
\begin{equation}
    \mathcal H_{\mathrm{bulk}}
    =\bigoplus_{i,j}M_{ij}\,
      \mathcal V_i\otimes\overline{\mathcal V}_j,
    \qquad M_{ij}\in\mathbb Z_{\geq0}.
    \label{eq:bulk_modular_invariant}
\end{equation}
For every sector compatible with the gluing condition, one can construct an Ishibashi state~\cite{ishibashi_1989_boundary_crosscap}. In the diagonal, multiplicity-free case it has the schematic form
\begin{equation}
    \left|i\right\rangle\!\rangle
    =\sum_{N,\alpha}
      \ket{i,N,\alpha}\otimes
      U\,\overline{\ket{i,N,\alpha}},
    \label{eq:ishibashi_state}
\end{equation}
where $N$ labels the descendant level, $\alpha$ resolves its degeneracy, and the antiunitary map $U$ implements the gluing. Ishibashi states solve Eq.~\eqref{eq:boundary_state_gluing} representation by representation, but they are not themselves normalizable physical states. A conformal boundary state is a consistent linear combination
\begin{equation}
    \ket{B_a}
    =\sum_{i,\mu}B_a^{i,\mu}
      \left|i,\mu\right\rangle\!\rangle,
    \label{eq:general_boundary_state}
\end{equation}
where $\mu$ accounts for possible multiplicities in the bulk modular invariant.

The consistency condition on the coefficients $B_a^{i,\mu}$ follows by evaluating an annulus in two directions. %Let the annulus have width $W$ and circumference $\beta$. 
The open- and closed-channel Hamiltonians are
\begin{align}
    H_{ab}^{\mathrm{op}}(W)
    &=\frac{\pi}{W}\left(L_0-\frac{c}{24}\right),
    \nonumber\\
    H^{\mathrm{cl}}(\beta)
    &=\frac{2\pi}{\beta}
      \left(L_0+\bar L_0-\frac{c}{12}\right)
    \label{eq:annulus_hamiltonians}
\end{align}
where $W$ denotes the width and $\beta$ is the circumference of the annulus. Quantizing around the circumference gives the open-channel trace, whereas quantizing across the width gives a closed-channel amplitude between boundary states:
\begin{align}
    Z_{ab}(\beta,W)
    &=\Tr_{\mathcal H_{ab}}
      \e^{-\beta H_{ab}^{\mathrm{op}}(W)}
      \nonumber\\
    &=\bra{B_a}\e^{-W H^{\mathrm{cl}}(\beta)}
      \ket{B_b}.
    \label{eq:annulus_open_closed}
\end{align}
The open-channel Hilbert space decomposes into representations of the preserved chiral algebra,
\begin{equation}
    \mathcal H_{ab}
    =\bigoplus_i n_{ab}^{i}\,\mathcal V_i,
    \qquad n_{ab}^{i}\in\mathbb Z_{\geq0},
    \label{eq:open_channel_hilbert_space}
\end{equation}
and therefore, with $q=\exp(-\pi\beta/W)$,
\begin{equation}
    Z_{ab}(q)
    =\Tr_{\mathcal H_{ab}}q^{L_0-c/24}
    =\sum_i n_{ab}^{i}\,\chi_i(q).
    \label{eq:bcft_open_partition}
\end{equation}
Boundary fields $\psi_i^{ab}$ are in one-to-one correspondence with the states in $\mathcal H_{ab}$. When $a\neq b$, they change the boundary condition from $a$ to $b$ and are called boundary-condition-changing operators. Their conformal weights and descendant multiplicities are precisely the universal data encoded in Eq.~\eqref{eq:bcft_open_partition}.

Equality of the two lines of Eq.~\eqref{eq:annulus_open_closed} after a modular transformation is the Cardy consistency condition~\cite{cardy_1989_bc_fusion,cardy_2008_bcft}. For a diagonal rational CFT, the canonical Cardy solution is
\begin{equation}
    \ket{B_a}
    =\sum_i\frac{S_{ai}}{\sqrt{S_{0i}}}
      \left|i\right\rangle\!\rangle,
    \qquad
    n_{ab}^{i}=\mathcal N_{ia}^{\ \ b},
    \label{eq:cardy_boundary_state}
\end{equation}
up to the standard orientation and conjugation convention for the fusion coefficient. The simple Cardy formula is specific to this diagonal rational setting. For a non-diagonal modular invariant, for a boundary preserving only a subalgebra, or for a non-rational theory, the allowed boundary states and multiplicities must instead be obtained from the corresponding boundary classification~\cite{behrend_2000_rcft_boundaries}. This distinction is relevant below. The three-state Potts model has the non-diagonal $D_4$ invariant, while the compact boson has continuous boundary parameters and infinitely many zero-mode sectors. An annulus determines the boundary operator content for the chosen pair $(a,b)$. It does not by itself reconstruct the full bulk CFT.

Finally, a microscopic boundary need not initially be conformal. Near a boundary fixed point, its Euclidean action takes the form
\begin{equation}
    S_{\partial}
    =S_{\partial,*}
     +\sum_\alpha\lambda_\alpha
      \int d\tau\,\psi_\alpha(\tau),
    \qquad [\lambda_\alpha]=1-h_\alpha.
    \label{eq:boundary_perturbations}
\end{equation}
A boundary operator with $h_\alpha<1$ is relevant and can drive a boundary renormalization-group (RG) flow between conformal boundaries. This provides the effective-theory interpretation of the long-distance boundary description: distinct short-range microscopic preparations can be described by the same conformal boundary $B_a$. The universal constant associated with that boundary is characterized by the Affleck--Ludwig factor $g_a=\langle 0 \vert B_a \rangle $, which enters the entanglement entropy~\cite{cardy_1984_surface,affleck_ludwig_1991_boundary_entropy,cardy_2008_bcft}.

\subsection{Entanglement of Euclidean-evolved states as a BCFT}

We now review how the entanglement of a Euclidean-evolved state is described within BCFT~\cite{calabrese_cardy_2006_quench,calabrese_cardy_2007_quenches,calabrese_cardy_2016_quench_review}, in a form that applies equally to the two Euclidean orientations considered here. Consider
\begin{equation}
    \ket{\Psi(\ell)}
    =\mathcal N(\ell)\,
      \e^{-\ell H}\ket{\Psi_0},
    \label{eq:unified_euclidean_state}
\end{equation}
where $\ell$ is the Euclidean evolution length and $H$ is the generator in the chosen quantization direction. For the spatial state, $\ell=\tau$ and $H=H_x$. For the temporal state, $\ell=L$ and $H=H_\tau$ generates translations in the spatial direction. On the lattice this evolution is represented by the quantum transfer operator $\mathbb T_x$~\cite{suzuki_1976_transfer_matrix_1,betsuyaku_1985_transfer_matrix_2,wang_1997_tmrg}. Increasing $\ell$ probes progressively larger length scales. Within the finite-length scaling regime, the universal entanglement structure is described by a BCFT whose preparation boundary retains information about the microscopic initial state.

For a bipartition of the one-dimensional slice into $A\cup\bar A$, the normalized reduced density matrix is~\cite{calabrese_cardy_2004_entanglement,calabrese_lefevre_2008_entanglement_spectrum}
\begin{equation}
    \rho_A
    =\tr_{\bar A}\!\left(
      \ket{\Psi(\ell)}\bra{\Psi(\ell)}
      \right).
    \label{eq:reduced_density_matrix_def}
\end{equation}
We define the modular, or entanglement Hamiltonian $\mathcal{H}_E$ by
\begin{equation}
    K_A\equiv-\log\rho_A,
    \qquad
    \mathcal H_E\equiv\frac{1}{2\pi}K_A.
\label{eq:entanglement_hamiltonian_def}
\end{equation}
Eigenvalues of $\mathcal{H}_E$ are called entanglement spectrum.
%The two operators contain the same spectral information. %the factor $2\pi$ is only a convention. 
Below we show $\mathcal{H}_E$ is the open channel Hamiltonian of a BCFT and  entanglement spectra are the corresponding energy levels.
%Below we quote BCFT levels using the rescaled Hamiltonian $\mathcal H_E$ and use $-\log\lambda_j$ for the unrescaled entanglement levels.

The origin of the BCFT is most transparent in the ket--bra path integral for Eq.~\eqref{eq:reduced_density_matrix_def}. The microscopic initial state supplies the boundary condition on the two outer edges of a strip of width $2\ell$. A generic short-range state is not literally a conformal boundary state. At length scales large compared with the microscopic preparation scales, its boundary condition is described by a conformal boundary $B_O$. Within long-distance matrix elements, one may therefore replace the normalizable microscopic state by a regularized conformal boundary state~\cite{calabrese_cardy_2006_quench,calabrese_cardy_2007_quenches,calabrese_cardy_2016_quench_review},
\begin{align}
    \ket{\Psi_0}
    &\overset{\mathrm{IR}}{\sim}
      \mathcal N_0
      \e^{-\ell_0 H}
      \ket{B_O},
    \nonumber\\
    \ket{\Psi(\ell)}
    &\overset{\mathrm{IR}}{\sim}
      \widetilde{\mathcal N}
      \e^{-(\ell+\ell_0)H}
      \ket{B_O},
    \qquad \ell>0.
    \label{eq:regularized_boundary_state}
\end{align}
The extrapolation length $\ell_0$ gives a nonuniversal shift of the strip width, while irrelevant boundary operators generate subleading corrections. The entanglement cut provides the second microscopic boundary. In the lattice problem, the tensor-product factorization fixes this boundary directly at the lattice scale, hence no independent entanglement regulator needs to be chosen. At long distances, the microscopic cut is also described by a conformal boundary $B_E$. Thus $B_O$ and $B_E$ encode the state-preparation boundary and the fixed lattice bipartition, respectively.

To see explicitly how the reduced density matrix becomes a BCFT Hamiltonian, we use the complex coordinate $z=u+\ii v$, where $u$ parametrizes the one-dimensional slice supporting the state and $v\in[-\ell,\ell]$ runs across the ket--bra strip. For the temporal orientation, $(u,v)=(\tau,x)$ and $\ell=L$; for the spatial orientation, $(u,v)=(x,\tau)$ and $\ell=\tau$. The two outer edges $v=\pm\ell$ carry the infrared boundary $B_O$. Tracing the half-line $u<0$ glues the two sides of the path integral there and leaves a slit along $u>0$ at $v=0$. For the continuum conformal map, we represent the lattice-scale cut by a small disk of radius $\epsilon\sim a$ around the entangling point, which is the short-distance regulator for the already fixed microscopic cut and carries $B_E$ in the infrared~\cite{lauchli_2013_opContent,cho_2017_universal_ES,yu_2022_conformal_bc,huang_2024_emergent_bc}.

\begin{figure}[t]
    \centering
    \includegraphics[width=0.96\linewidth]{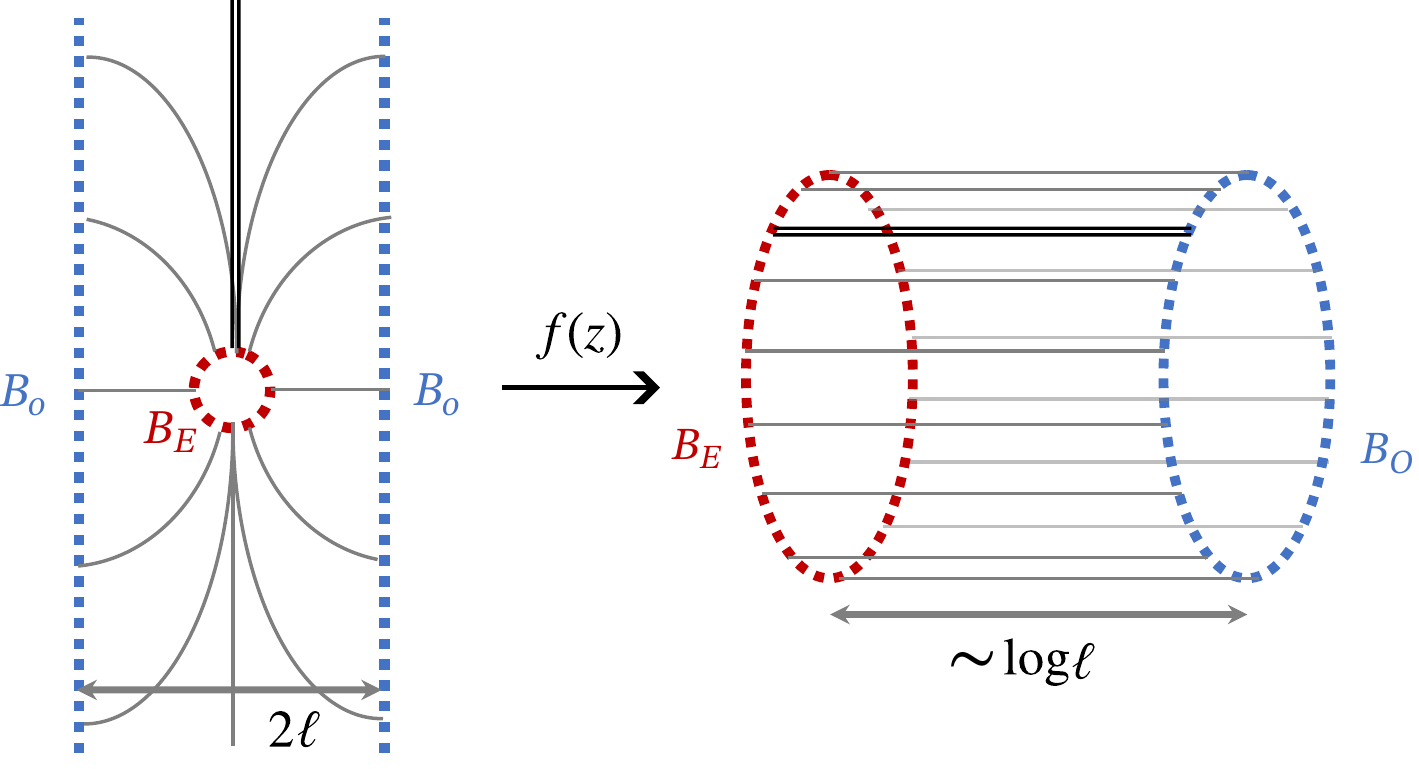}
    \caption{Conformal map Eq.~\ref{eq:annulus_map} of the reduced density matrix in either Euclidean orientation. The slit strip (left), with initial-state determined boundary $B_O$ and the entanglement cut represented by a short-distance regulator, is mapped to the finite cylinder.}
    \label{fig:conformal_mapping}
\end{figure}

The conformal map in Fig.~\ref{fig:conformal_mapping}
\begin{equation}
    w=f(z)=\ii\log\!\left[
       \frac{\sinh(\pi z/4\ell)}{\cosh(\pi z/4\ell)}
    \right]
    \label{eq:annulus_map}
\end{equation}
sends the slit strip to an annulus, or equivalently a finite cylinder~\cite{Cardy_2016_EntanglementHamiltonian}. With the standard logarithm branch, its circumference is $2\pi$ and its axial width is
\begin{equation}
    W(\ell,\epsilon)
    =\log\!\left(\frac{4\ell}{\pi\epsilon}\right).
    \label{eq:annulus_width}
\end{equation}

The path integral from one bank of the slit to the other becomes translation once around the cylinder. Quantizing in this circumferential direction therefore gives the normalized reduced density matrix as an open-channel BCFT evolution operator,
\begin{equation}
    \rho_A
    =\frac{1}{Z_{OE}}
      \exp\!\left[-\frac{2\pi^2}{W}
      \left(L_0-\frac{c}{24}\right)\right],
    \label{eq:rho_bcft}
\end{equation}
acting on the open-channel Hilbert space
\begin{equation}
    \mathcal H_{B_OB_E}
    =\bigoplus_i n_{OE}^{i}\,\mathcal V_i.
    \label{eq:orientation_open_channel}
\end{equation}
Equation~\eqref{eq:rho_bcft} is the precise sense in which the entanglement Hamiltonian is a BCFT Hamiltonian after the conformal map. The initial state and the entanglement cut define the two ends of an annulus, and the universal low-lying reduced-density-matrix spectrum is generated by the open-channel BCFT between them. In the universal low-energy sector,
\begin{equation}
    \begin{aligned}
    K_A
    &=\frac{2\pi^2}{W}
      \left(L_0-\frac{c}{24}\right)
      +\log Z_{OE},\\
    \mathcal H_E
    &=\frac{\pi}{W}
      \left(L_0-\frac{c}{24}\right)
      +\mathrm{const.}
    \label{eq:entanglement_hamiltonian_bcft}
\end{aligned}
\end{equation}
up to corrections from irrelevant operators. Consequently, the rescaled entanglement energy levels $\varepsilon$ obey
\begin{equation}
    \varepsilon_{h,N} \sim \frac{1}{W} \left(h + N\right), \qquad 
    \Delta_{h,N}
    = h+N
    \label{eq:entanglement_gaps}
\end{equation}
where $N\in\mathbb Z_{\geq0}$ is a descendant level, $h$ runs over the conformal families in $\mathcal H_{B_OB_E}$ and $\Delta_{h,N}$ is by rescaling $\varepsilon_{h,N}$ with the lowest energy gap. The same annulus gives the von Neumann entropy
\begin{equation}
    S_A
    =\frac{c}{6}\log\!\left(\frac{2\ell}{\pi\epsilon}\right)
     +\log g_O+\log g_E+s_{\mathrm{uv}}+\cdots,
    \label{eq:entropy_scaling}
\end{equation}
where $g_O$ and $g_E$ are Affleck--Ludwig boundary degeneracies and $s_{\mathrm{uv}}$ is a regulator-dependent constant. Equations~\eqref{eq:entanglement_gaps} and \eqref{eq:entropy_scaling} apply to the universal low-lying sector in the window~\eqref{eq:scaling_window}, not to the entire lattice entanglement spectrum.

\subsection{Microscopic symmetry constraints and the two Euclidean quantizations}

Microscopic lattice information enters the BCFT description through the boundary conditions associated with the initial preparation and the entanglement cut. The bulk CFT supplies the candidate conformal boundaries and their operator contents, while exact microscopic symmetries restrict the allowed boundary perturbations. Let $G$ be a symmetry preserved jointly by the bulk evolution and $\ket{\Psi_0}$. The effective boundary description must respect $G$, so its boundary couplings are restricted to the $G$-invariant subspace. In the notation of Eq.~\eqref{eq:boundary_perturbations},
\begin{equation}
    g\!\left(\sum_\alpha
      \lambda_\alpha\psi_\alpha\right)g^{-1}
    =\sum_\alpha\lambda_\alpha\psi_\alpha,
    \qquad g\in G.
    \label{eq:symmetry_restricted_boundary_rg}
\end{equation}
Thus symmetry removes all relevant directions outside the invariant subspace and can, when a unique stable compatible boundary exists, select the infrared boundary $B_O$. Conversely, a symmetry-breaking initial state permits additional relevant boundary couplings and can be described by a different conformal boundary at large scales. This is a constraint rather than a universal one-to-one classification: several conformal boundaries may preserve the same symmetry, and a symmetry may permute a set of boundaries rather than fix one member. Non-diagonal, non-rational, and anomalous theories therefore require the corresponding boundary classification in addition to symmetry. In the numerical sections below, the low-lying conformal towers provide the additional identification of $B_O$.

In the continuum description, the entanglement boundary $B_E$ requires a short-distance regulator. In the lattice problem, it is already selected separately by the microscopic cut and the symmetry it preserves. Once the lattice factorization and entanglement prescription are fixed, this microscopic boundary is fixed as well. Our boundary-engineering protocol therefore varies $B_O$ through the initial state or microscopic boundary term while leaving $B_E$ unchanged. This separation makes changes of the conformal tower content directly traceable to the long-distance boundary condition associated with the initial state.

With the microscopic boundary restriction in hand, we can compare the two Euclidean quantizations. At the Lorentz-invariant fixed points considered here, after setting the velocity to unity, a $90^\circ$ Euclidean rotation exchanges the spatial and imaginary-time directions. The spatial and temporal states are therefore two quantizations of the same continuum path integral along orthogonal directions, and the continuum CFT contains no intrinsic spatial--temporal distinction in the universal entanglement structure. In the spatial protocol the reduced density matrix is that of a physical spatial subsystem, while in the temporal protocol it belongs to an auxiliary transfer state and is a diagnostic of the Euclidean path integral. If the two microscopic constructions are described at long distances by the same pair of conformal boundaries $(B_O,B_E)$, they have the same entropy coefficient $c/6$, open-channel operator content, descendant multiplicities, and universal entanglement scaling, up to nonuniversal length and energy normalizations.

A lattice, however, specifies more than the bulk CFT --- it fixes the tensor-product Hilbert space and the microscopic action of internal and crystalline symmetries. Orthogonal quantization can change how these symmetries are represented, particularly when translation participates in an anomaly. Through Eq.~\eqref{eq:symmetry_restricted_boundary_rg}, this can change which infrared conformal boundaries are accessible even though the bulk CFT is the same. Any spatial--temporal inequivalence in our examples therefore traces back to the microscopic symmetry realization of the lattice regulator and serves as a diagnostic of the underlying lattice constraint. We make this lattice input explicit in Sec.~\ref{sec:tensor_network}.

\section{Lattice framework for spatial and temporal Euclidean evolution}
\label{sec:tensor_network}

To determine these microscopic restrictions, we now retain the lattice regulator and work directly with the Trotterized two-dimensional tensor network. It is useful to regard this network as a \emph{lattice Euclidean spacetime}: slicing it along the two orthogonal directions defines two one-dimensional tensor-product Hilbert spaces and their corresponding evolution operators. Tensor-network methods then provide efficient representations of the resulting states.

What matters for the infrared boundary problem is the symmetry action inherited by each slice. Onsite internal symmetries descend to the orthogonal transfer problem in a controlled composite representation, whereas crystalline translation remains tied to the original spatial lattice and is represented differently after transverse quantization. These symmetry actions constrain the accessible conformal boundaries and provide the microscopic input tested by the spatial--temporal comparison.

\subsection{Two lattice quantizations of the Euclidean path integral}

Consider a nearest-neighbor Hamiltonian
\begin{equation}
    H_x=\sum_i h_{i,i+1},
\end{equation}
and split it into commuting even- and odd-bond pieces. With $\delta\tau=\beta/N_\tau$, a first-order Trotter decomposition gives
\begin{align}
    Z
    &=\lim_{N_\tau\to\infty}
      \Tr\!\left\{
      \left[
      \e^{-\delta\tau H_{\mathrm{odd}}}
      \e^{-\delta\tau H_{\mathrm{even}}}
      +O(\delta\tau^2)
      \right]^{N_\tau}
      \right\}.
    \label{eq:trotter_network}
\end{align}
The local building block is the two-site gate
\begin{equation}
    G_{\sigma_i'\sigma_{i+1}';\sigma_i\sigma_{i+1}}
    =\bra{\sigma_i'\sigma_{i+1}'}
      \e^{-\delta\tau h_{i,i+1}}
      \ket{\sigma_i\sigma_{i+1}}.
    \label{eq:local_gate}
\end{equation}
The checkerboard contraction of these rank-four tensors is a discrete representation of the same Euclidean path integral used in Sec.~\ref{sec:bcft_framework}. At finite $\delta\tau$ it already admits two natural quantizations~\cite{suzuki_1976_transfer_matrix_1,betsuyaku_1985_transfer_matrix_2,wang_1997_tmrg,haegeman_2017_transfer_mpo}.

A horizontal cut at fixed Euclidean time intersects the physical lattice sites. Contracting the network vertically therefore gives the usual imaginary-time evolution of the spatial state,
\begin{equation}
    \ket{\Psi_{\mathrm{sp}}(\tau+\delta\tau)}
    \propto
    \mathbb T_\tau(\delta\tau)
    \ket{\Psi_{\mathrm{sp}}(\tau)},
    \qquad
    \mathbb T_\tau
    =\e^{-\delta\tau H_{\mathrm{odd}}}
     \e^{-\delta\tau H_{\mathrm{even}}}
    \label{eq:spatial_lattice_transfer}
\end{equation}
which can be efficiently simulated using tensor network states~\cite{vidal_2007_tebd,schollwock_2011_dmrg_mps,haegeman_2017_transfer_mpo}. A vertical cut at fixed $x$, by contrast, intersects the Trotter legs of the network. Those legs form a different one-dimensional Hilbert space, the Hilbert space of the temporal transfer state. Contracting one spatial column defines the quantum transfer operator $\mathbb T_x$, and
\begin{equation}
    \ket{\Psi_{\mathrm{tm}}(L+1)}
    \propto
    \mathbb T_x\ket{\Psi_{\mathrm{tm}}(L)}.
    \label{eq:temporal_lattice_transfer}
\end{equation}
Thus Eqs.~\eqref{eq:spatial_lattice_transfer} and \eqref{eq:temporal_lattice_transfer} are the lattice counterparts of choosing the two Euclidean quantization directions in the CFT. They come from the same two-dimensional network, but they act on different microscopic tensor-product spaces.

\begin{figure}[t]
    \centering
    \includegraphics[width=0.75\linewidth]{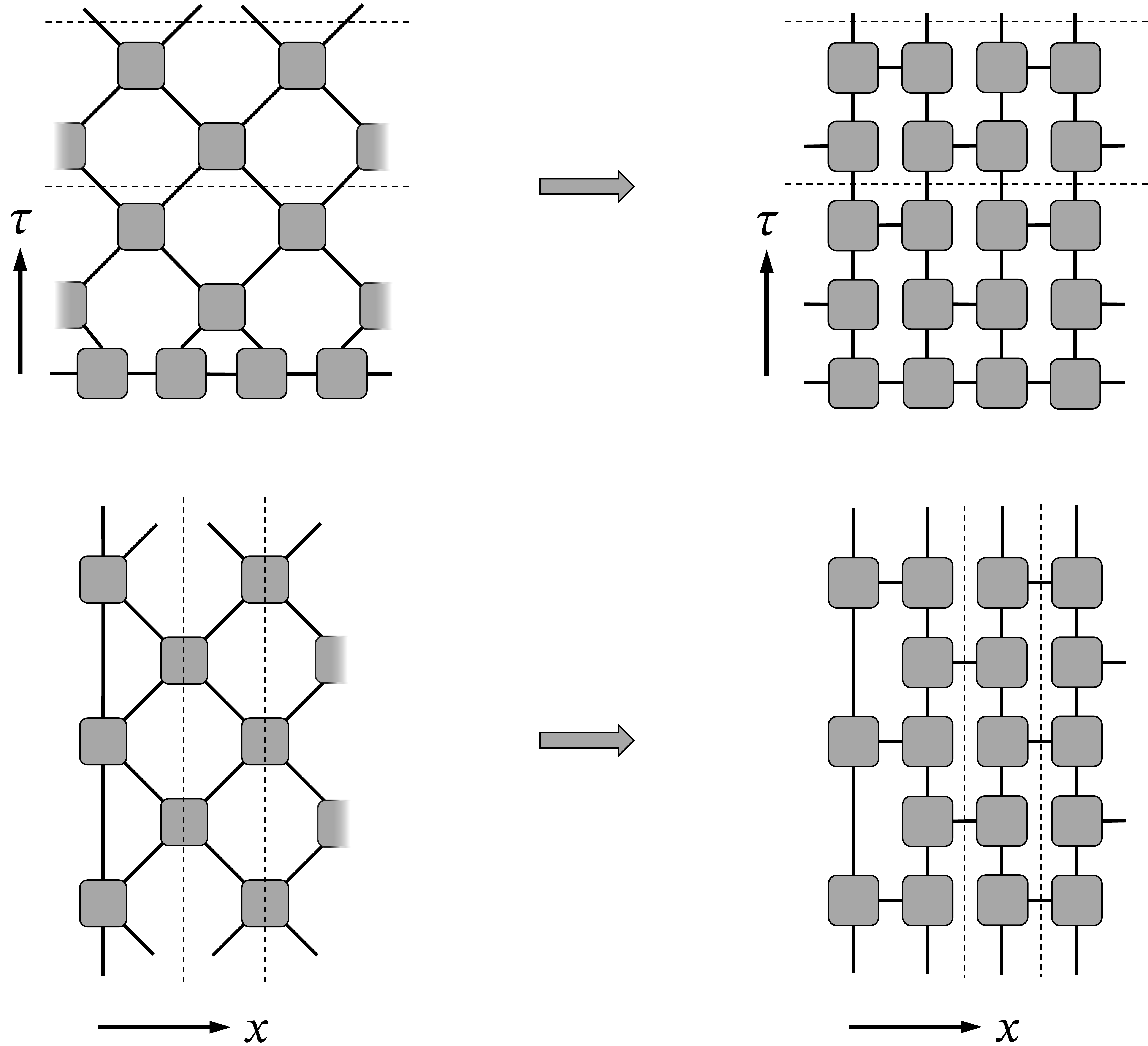}
    \caption{Lattice Euclidean-evolution framework for the spatial (upper) and temporal (lower) state. The checkerboard network of Trotter gates is factorized and regrouped into a transverse \iiMPO{}. One application of the resulting spatial or transfer operator advances the state by one step.}
    \label{fig:evolution_rdm}
\end{figure}

Figure~\ref{fig:evolution_rdm} summarizes this lattice picture. Importantly, no finite-bond-dimension approximation is involved in exchanging the contraction direction. Before compression, the two transfer problems are simply two exact reorganizations of the Trotterized partition-function network. The lattice constant $a$, which we set to unity below, also supplies the ultraviolet regulator at the entanglement cut. Hence the entanglement boundary $B_E$ of Sec.~\ref{sec:bcft_framework} is generated microscopically by the chosen lattice cut rather than by introducing an independent continuum cutoff by hand. The same cut is used when we vary the initial state, so $B_E$ remains fixed across the different $B_O$ realized below.

\subsection{Regular transverse transfer operator in the \texorpdfstring{$\delta\tau\to0$}{delta tau to 0} limit}

\begin{figure}[t]
    \centering
    \resizebox{0.9\linewidth}{!}{\begin{tikzpicture}[
  x=1.35cm,
  y=1.25cm,
  line width=0.75pt,
  line cap=round,
  line join=round,
  connection/.style={dash pattern=on 1.8pt off 1.8pt},
]

% Panel (a)
\node[anchor=west, font=\Large] at (-2.65,5.35) {(a)};

% Central wavy line: three left-facing and three right-facing bulges.
\draw
  (0,5.15)
  .. controls (-0.28,5.02) and (-0.42,4.88) .. (-0.42,4.75)
  .. controls (-0.42,4.62) and (-0.28,4.47) .. (0,4.34)
  .. controls ( 0.28,4.21) and ( 0.42,4.07) .. ( 0.42,3.94)
  .. controls ( 0.42,3.81) and ( 0.28,3.66) .. (0,3.53)
  .. controls (-0.28,3.40) and (-0.42,3.26) .. (-0.42,3.13)
  .. controls (-0.42,3.00) and (-0.28,2.85) .. (0,2.72)
  .. controls ( 0.28,2.59) and ( 0.42,2.45) .. ( 0.42,2.32)
  .. controls ( 0.42,2.19) and ( 0.28,2.04) .. (0,1.91)
  .. controls (-0.28,1.78) and (-0.42,1.64) .. (-0.42,1.51)
  .. controls (-0.42,1.38) and (-0.28,1.23) .. (0,1.10)
  .. controls ( 0.28,0.97) and ( 0.42,0.83) .. ( 0.42,0.70)
  .. controls ( 0.42,0.57) and ( 0.28,0.42) .. (0,0.29);

% Three small left-facing arcs.
\foreach \ymid in {4.75,3.13,1.51}{
  \draw
    (-1.40,{\ymid+0.40})
    .. controls (-0.94,{\ymid+0.30}) and (-0.75,{\ymid+0.13})
    .. (-0.75,\ymid)
    .. controls (-0.75,{\ymid-0.13}) and (-0.94,{\ymid-0.30})
    .. (-1.40,{\ymid-0.40});
  \draw[connection] (-0.75,\ymid) -- (-0.42,\ymid);
}

% Three small right-facing arcs.
\foreach \ymid in {3.94,2.32,0.70}{
  \draw
    (1.40,{\ymid+0.40})
    .. controls (0.94,{\ymid+0.30}) and (0.75,{\ymid+0.13})
    .. (0.75,\ymid)
    .. controls (0.75,{\ymid-0.13}) and (0.94,{\ymid-0.30})
    .. (1.40,{\ymid-0.40});
  \draw[connection] (0.42,\ymid) -- (0.75,\ymid);
}

% Dotted blue rectangle around the central wave.
\draw[blue!75!black, densely dotted, line width=0.7pt]
  (-0.52,0.15) rectangle (0.52,5.30);

% Panel (b)
\node[anchor=west, font=\Large] at (2.05,5.35) {(b)};

\begin{scope}[shift={(0.9,0)}]
% Opposing arcs; the named midpoint coordinates are their closest tips.
\draw
  (2.555,4.50)
  .. controls (2.984,4.39) and (3.14,4.28) .. (3.14,4.16)
  .. controls (3.14,4.04) and (2.984,3.93) .. (2.555,3.82);
\draw
  (4.245,4.50)
  .. controls (3.816,4.39) and (3.66,4.28) .. (3.66,4.16)
  .. controls (3.66,4.04) and (3.816,3.93) .. (4.245,3.82);
\draw[connection] (3.14,4.16) -- (3.66,4.16);
\end{scope}

% Panel (c)
\node[anchor=west, font=\Large] at (2.05,2.85) {(c)};

\begin{scope}[shift={(0.9,0)}]
% Left S-like object.  Coordinates L1 and L2 mark the two tips.
\draw
  (2.62,2.25)
  .. controls (2.236,2.13) and (2.236,1.90) .. coordinate (L1) (2.62,1.72)
  .. controls (3.004,1.54) and (3.004,1.32) .. (2.62,1.14);

% Right S-like object, mirrored horizontally.
\draw
  (4.18,2.25)
  .. controls (4.564,2.13) and (4.564,1.90) .. coordinate (R1) (4.18,1.72)
  .. controls (3.796,1.54) and (3.796,1.32) .. (4.18,1.14);

% The visual extrema occur near these points on the Bezier curves.
\draw[connection] (1.90,2.02) -- (2.332,2.02);
\draw[connection] (2.908,1.37) -- (3.34,1.37);
\draw[connection] (3.46,1.37) -- (3.892,1.37);
\draw[connection] (4.468,2.02) -- (4.90,2.02);
\end{scope}

\end{tikzpicture}}
    
    \caption{Why the transverse representation requires a nontrivial local factorization. (a) The full checkerboard Euclidean network. (b) As $\delta\tau\to0$, a physical two-site gate approaches the identity, but after transverse reshuffling its leading contribution is the rank-one maximally entangled projector in Eq.~\eqref{eq:transverse_projector_main}. (c) Splitting each physical gate first and regrouping neighboring half-gates produces the two checkerboard local tensors of the transverse \MPO{}. After the appropriate local basis/gauge transformation they define a regular uniform transfer operator.}
    \label{fig:gate_mpo}
\end{figure}
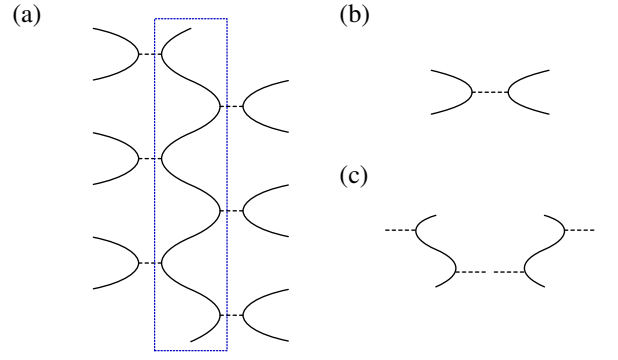

The transverse quantization contains an additional lattice subtlety that is absent in the ordinary imaginary-time evolution~\cite{ref_appendix}. The continuous translation invariance along the temporal direction can not be broken, enforcing the temporal state to be uniform for arbitrary small time slice $\delta \tau$ when constructing transfer gates in Fig.~\ref{fig:gate_mpo}. It turns out to define the regular transfer evolution representation, we need to choose the split-gate basis~\cite{ref_appendix}. For small Trotter step,
\begin{equation}
    G=I_i\otimes I_{i+1}
      -\delta\tau\,h_{i,i+1}
      +O(\delta\tau^2).
    \label{eq:gate_small_dt_main}
\end{equation}
In the physical orientation the leading term is the identity operator. After a $90^\circ$ index reshuffling, the same tensor becomes a rank-one map. If the transverse composite indices are $(a',a)$ and $(b',b)$, then
\begin{align}
    \widetilde G_{(b',b);(a',a)}
    &=\delta_{b'b}\delta_{a'a}+O(\delta\tau)
    \nonumber\\
    &\propto
      \lvert I\rangle \langle I\rvert
      +O(\delta\tau),
    \qquad
    \lvert I\rangle=\sum_s\ket{s\,s}.
    \label{eq:transverse_projector_main}
\end{align}
This is the projector shown schematically in Fig.~\ref{fig:gate_mpo}(b). If the reshuffled gate is treated directly as an ordinary local two-site evolution operator, its transverse representation becomes rank deficient as $\delta\tau\to0$: the rank-one identity contribution dominates the local Schmidt structure and obscures the nontrivial transfer dynamics. For finite length temporal state, it is usually fine even for a small $\tau$ and the two-dimensional partition function itself remains regular. However, when the length increases, it becomes highly unstable and the transfer operator behaves like a rank-1 projector. 

%the difficulty lies in the local factorization used to define the one-dimensional transverse Hilbert space.

Our construction avoids applying this rank-one object as a local evolution gate. We first factorize the physical gate by an SVD,
\begin{equation}
    G_{a'b';ab}
    =\sum_\mu
      L_{a'a;\mu}\,R_{\mu;b'b},
    \label{eq:gate_svd_main}
\end{equation}
with the singular values absorbed symmetrically into $L$ and $R$. We then contract $R$ from one plaquette with $L$ from the neighboring plaquette along the Trotter direction. The resulting rank-four tensors form the local tensors of the spatial transfer operator $\mathbb T_x$, as illustrated in Fig.~\ref{fig:evolution_rdm}(a) and Fig.~\ref{fig:gate_mpo}(c). This operation is an exact regrouping of the original two-dimensional tensor network before any MPS truncation~\cite{pirvu_2010_mpo,haegeman_2017_transfer_mpo}. The maximally entangled factors associated with the reshuffled identity are thereby absorbed into local basis and normalization factors, while the $O(\delta\tau)$ pieces accumulate into the nontrivial transfer dynamics~\cite{ref_appendix}. 

Because of the checkerboard Trotterization, the two graphical local tensors in Fig.~\ref{fig:gate_mpo}(c) appear different. They are related by the local change of basis associated with the two SVD factorizations. After this identification the $\beta\to\infty$ transfer problem can be represented by a translation-invariant infinite \MPO{} acting on a uniform temporal \iiMPS{}. The resulting representation remains well conditioned and translationally invariant even for arbitrarily small Trotter steps.

\subsection{Internal symmetry in the temporal quantization}

An onsite internal symmetry is inherited exactly by the quantum transfer matrix under transverse quantization~\cite{suzuki_1976_transfer_matrix_1,betsuyaku_1985_transfer_matrix_2,wang_1997_tmrg,haegeman_2017_transfer_mpo,cirac_2021_mps_review}. We summarize this inheritance because it supplies the microscopic symmetry action used below to engineer Euclidean-evolved states and constrain their BCFT boundaries. Consider a unitary onsite internal symmetry $G$ acting on each physical spin by $u_g$, $g\in G$, and suppose
\begin{equation}
    [h_{i,i+1},u_g\otimes u_g]=0.
\end{equation}
Every Trotter gate is also symmetric.

The important point for the temporal quantization is that its local Hilbert space is not the original single-spin Hilbert space. Before the regular factorization described above, a transverse local degree of freedom is naturally a ket--bra pair,
\begin{equation}
    \mathcal H_{\tau}^{\mathrm{raw}}
    \simeq \mathcal H\otimes\mathcal H^*,
\end{equation}
and therefore carries the operator-space representation
\begin{equation}
    u_g^{\mathrm{raw}}
    =u_g\otimes\bar u_g,
    \label{eq:transverse_symmetry_action}
\end{equation}
equivalently the adjoint action $X\mapsto u_gXu_g^\dagger$. The SVD and regrouping in Eq.~\eqref{eq:gate_svd_main} then introduce a local change of basis, which we denote schematically by $\mathcal B$. Thus the onsite action carried by the final temporal degree of freedom is the representation equivalent to
\begin{equation}
    \widetilde u_g
    \simeq
    \mathcal B\,
    (u_g\otimes\bar u_g)\,
    \mathcal B^{-1}.
    \label{eq:temporal_symmetry_basis}
\end{equation}
The symbol $\simeq$ emphasizes equivalence of representations: the final temporal transfer operator is not expressed in the original lattice-spin basis. Nevertheless, the symmetry is exact. With $\widetilde U_g$ denoting the corresponding global action on the temporal chain,
\begin{equation}
    [\mathbb T_x,\widetilde U_g]=0,
    \qquad g\in G.
    \label{eq:temporal_transfer_symmetry}
\end{equation}

For $G=U(1)$ this general statement reduces to a simple charge relation. If a nonzero physical gate element obeys
\begin{equation}
    q_1+q_2=q_3+q_4,
\end{equation}
then the same constraint, viewed transversely, is
\begin{equation}
    q_1-q_3=-q_2+q_4.
    \label{eq:u1_transverse_charge}
\end{equation}
The transverse charge is therefore a ket--bra charge difference. %Equation~\eqref{eq:u1_transverse_charge} is only the Abelian example of the general representation-theoretic statement in Eqs.~\eqref{eq:transverse_symmetry_action}--\eqref{eq:temporal_symmetry_basis}.

\subsection{Symmetry-selected initial states and conformal boundaries}

Equation~\eqref{eq:temporal_transfer_symmetry} gives direct control over the symmetry sector of the temporal Euclidean-evolution trajectory. We may choose an initial temporal state that preserves the full group $G$, only a subgroup $H\subseteq G$, or no nontrivial internal symmetry. If
\begin{equation}
    \widetilde U_h
    \ket{\Psi_{\mathrm{tm},0}^{(H)}}
    =\ket{\Psi_{\mathrm{tm},0}^{(H)}},
    \qquad h\in H,
    \label{eq:symmetric_temporal_initial}
\end{equation}
then every state along the lattice Euclidean evolution preserves the same subgroup,
\begin{equation}
    \widetilde U_h
    \ket{\Psi_{\mathrm{tm}}^{(H)}(L)}
    =\ket{\Psi_{\mathrm{tm}}^{(H)}(L)},
    \qquad
    \ket{\Psi_{\mathrm{tm}}^{(H)}(L)}
    \propto
    \mathbb T_x^L
    \ket{\Psi_{\mathrm{tm},0}^{(H)}}.
    \label{eq:symmetric_temporal_evolution}
\end{equation}
This realizes the symmetry restriction discussed in Sec.~\ref{sec:bcft_framework} directly on the lattice. The effective boundary couplings associated with the initial state must be $H$ invariant, so its long-distance description must be an $H$-compatible conformal boundary $B_O^{(H)}$. Symmetry does not by itself identify a unique boundary, but it excludes symmetry-forbidden relevant perturbations. The low-lying entanglement spectrum then identifies the boundary realized in the finite-length scaling regime through the open-channel content of $\mathcal H_{B_O^{(H)}B_E}$.

%Figure~\ref{fig:temporal_symmetry_bcft} summarizes the distinction between the standard transfer-matrix symmetry and its role in our problem. 

The physically useful operation is to vary the symmetry of the microscopic initial state while keeping the critical bulk evolution fixed. A fully symmetric initial state, a state preserving only a subgroup, and a symmetry-breaking state generally admit different sets of relevant boundary operators and may therefore realize different long-distance conformal boundaries. In this sense the lattice symmetry provides a practical control knob for engineering BCFT operator content in the entanglement spectrum of the evolved state.

The spatial and temporal constructions test this mechanism in two different microscopic factorizations. For onsite internal symmetries, both orientations inherit the same group and, when their initial boundaries preserve the corresponding symmetry, should approach compatible BCFT data. Lattice translation is different. In the temporal quantization, one application of $\mathbb T_x$ advances the system by one site of the original spatial lattice, so translation can relate successive evolved states or permute boundary sectors rather than act onsite on every temporal degree of freedom. When translation participates in an LSM anomaly with the internal symmetry, this distinction affects how the symmetry-compatible boundary sectors are realized. The Ising and Potts examples below illustrate the onsite case, while the XXZ chain shows how the LSM constraint leaves different spatial and temporal lattice remnants.

\subsection{Tensor-network representation and scaling regimes}

For numerical calculations we represent both Euclidean-evolved states by \iiMPS{} of finite bond dimension $D$~\cite{schollwock_2011_dmrg_mps,cirac_2021_mps_review,ref_appendix}. The spatial evolution in Eq.~\eqref{eq:spatial_lattice_transfer} is implemented with iTEBD-type updates~\cite{vidal_2007_tebd,orus_2008_itebd}. In the temporal quantization, the $\beta\to\infty$ limit makes the transverse chain infinite, so the temporal state is represented by a uniform \iiMPS{} and $\mathbb T_x$ by the uniform \iiMPO{} constructed above. Each spatial transfer step is followed by a variational compression and restoration of canonical form~\cite{haegeman_2016_tdvp,vanderstraeten_2019_tangnet,ref_appendix}. For either orientation, a finite bond dimension induces a finite-entanglement correlation length $\xi$, and the long-evolution fixed point at fixed $D$ can be accessed directly with variational MPS methods such as VUMPS~\cite{zauner_2018_VUMPS,vanderstraeten_2019_tangnet}.

\begin{figure}[t]
    \centering
    \begin{tikzpicture}[
        node distance=5.5mm,
        box/.style={draw, rounded corners=2pt, align=center,
                    inner sep=3.5pt, text width=0.86\linewidth},
        regime/.style={draw, rounded corners=2pt, align=center,
                       inner sep=3pt, text width=0.40\linewidth},
        flow/.style={-Latex, thick},
        every node/.style={font=\scriptsize}]
      \node[box] (state) {\textbf{Finite-$D$ Euclidean-evolved state}\\[1mm]
        $\ket{\Psi_D(\ell)}$};

      \node[regime, below=9mm of state, xshift=-0.23\linewidth] (finiteL) {
        \textbf{Finite-length scaling}\\[1mm]
        $1\ll\ell\ll\xi$\\
        infrared scale: $\ell$\\
        $\mathcal H_E\leftrightarrow
        \mathcal H_{B_OB_E}$\\[-0.6mm]
        \emph{main regime}};

      \node[regime, below=9mm of state, xshift=0.23\linewidth] (finiteD) {
        \textbf{Finite-entanglement scaling}\\[1mm]
        $\ell\gg\xi$\\
        infrared scale: $\xi$\\
        $\mathcal H_E\leftrightarrow
        \mathcal H_{B_DB_E}$};

      \draw[flow] (state.south) -- node[left, pos=0.62, yshift=0.8mm] {$\ell\ll\xi$} (finiteL.north);
      \draw[flow] (state.south) -- node[right, pos=0.62, yshift=0.8mm] {$\ell\gg\xi$} (finiteD.north);
    \end{tikzpicture}
    \caption{Finite-length and finite-entanglement scaling of a finite-$D$ Euclidean-evolved state. The same schematic applies to both protocols, with $\ell=\tau$ for the spatial state and $\ell=L$ for the temporal state. For $1\ll\ell\ll\xi$, the evolution length $\ell$ sets the infrared scale and the low-lying entanglement spectrum probes the BCFT selected by the microscopic initial boundary, $\mathcal H_{B_OB_E}$. For $\ell\gg\xi$, the infrared scale is instead set by $\xi$ and the finite-entanglement deformation can select a different effective boundary $B_D$; the crossover occurs around $\ell\sim\xi$.}
    \label{fig:finite_length_vs_finite_entanglement}
\end{figure}
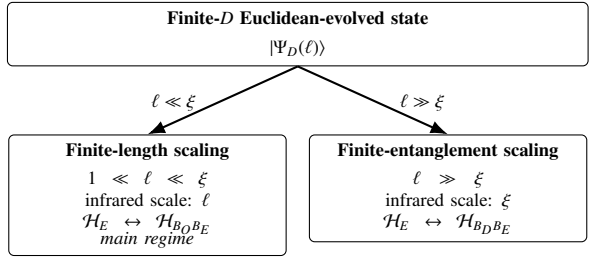

The finite-$D$ evolution can be written schematically as
\begin{align}
    \ket{\Psi(\ell+\delta\ell)}
    &\propto
      \mathcal T_{\delta\ell}\ket{\Psi(\ell)},
    \nonumber\\
    \ket{\Psi_D(\ell+\delta\ell)}
    &\propto
      \mathcal P_D\!\left[
      \mathcal T_{\delta\ell}\ket{\Psi_D(\ell)}
      \right],
    \label{eq:truncated_evolution}
\end{align}
where $\mathcal P_D$ denotes the nonlinear variational compression and $\mathcal T_{\delta\ell}$ one Euclidean-evolution step in the chosen quantization. The competition between the evolution length $\ell$ and the finite-entanglement correlation length $\xi$ produces the two scaling regimes summarized in Fig.~\ref{fig:finite_length_vs_finite_entanglement}.

The left branch of Fig.~\ref{fig:finite_length_vs_finite_entanglement} is the main numerical regime studied in this work. There the finite-entanglement effect is kept subleading by choosing sufficiently large $D$, the microscopic initial boundary remains visible, and the low-lying entanglement spectrum is governed by $\mathcal H_{B_OB_E}$. The right branch describes the finite-entanglement regime: once $\ell$ becomes much larger than $\xi$, the finite-$D$ variational deformation controls the infrared and can drive the spectrum toward a different effective boundary $B_D$~\cite{tagliacozzo_2008_fes,frank_2009_ee_scaling,pirvu_2012_finite_size_entanglement,stojevic_2015_fes}. The boundary data $B_O$ and $B_D$ can depend on the microscopic symmetry realization and on the symmetry constraints imposed on the variational manifold.

Tensor networks therefore supply the computational representation of the lattice framework. The physical ingredients carried forward are the two lattice quantizations, their inherited symmetry actions, and the resulting constraints on the long-distance conformal boundaries.

\section{Onsite lattice symmetry and conformal-boundary selection}
\label{sec:onsite_models}

We now turn from the general framework to boundary selection in concrete critical chains. For each model, we begin with the microscopic lattice symmetry and the symmetry of the initial state, determine the compatible long-distance conformal boundaries, and infer the corresponding open-channel operator content. The lattice entanglement spectrum then provides a direct test of these BCFT predictions.

For a diagonal rational CFT, an elementary Cardy boundary is labeled by a primary $a$, and the annulus with boundaries $a$ and $b$ has
\begin{align}
    Z_{ab}(q)
    &=\sum_c N_{ab}^{\ \ c}\,\chi_c(q),
    \nonumber\\
    \mathcal H_{ab}
    &=\bigoplus_c N_{ab}^{\ \ c}\,\mathcal V_c,
    \label{eq:cardy_boundary_fusion_content}
\end{align}
up to the standard orientation/conjugation convention already discussed in Sec.~\ref{sec:bcft_framework}. Thus, once the physical boundary $B_O$ selected by the initial state and the entanglement boundary $B_E$ are known, the allowed conformal towers are fixed by boundary fusion. In a non-diagonal theory the same logic survives, but the multiplicities are determined by the corresponding boundary classification or NIMrep rather than by the diagonal Cardy formula itself.

The Ising and three-state Potts chains provide clean examples of this principle because their relevant microscopic symmetries are onsite. The same internal group is inherited by the two lattice quantizations, so an initial state in a given symmetry class restricts the allowed boundary perturbations in the same way in either representation. Their spatial--temporal agreement is therefore a consequence of the same lattice symmetry constraint acting on the same infrared boundary theory. These models also show that symmetry constrains rather than uniquely identifies the conformal boundary, most clearly in the Potts theory where distinct conformal boundaries can preserve the same subgroup. They provide the control cases before we turn to the XXZ chain, where one-site translation participates in the microscopic constraint.

\subsection{Internal-symmetry case studies: Ising and three-state Potts}
\label{sec:internal_symmetry_benchmarks}

\noindent {\emph{Ising CFT realized in the critical Ising chain---}}  The critical transverse-field Ising chain is
\begin{equation}
    H_{\mathrm{Ising}}
    =-\sum_i\sigma_i^z\sigma_{i+1}^z
     -g\sum_i\sigma_i^x,
    \qquad g=1,
    \label{eq:ising_hamiltonian}
\end{equation}
with onsite symmetry $\prod_i\sigma_i^x$. Its scaling limit is the diagonal Ising CFT with $c=1/2$~\cite{difrancesco_1997_CFT_book,cardy_1986_boundary_operator,cardy_1989_bc_fusion}. The three primaries are
\begin{equation}
    I,\quad \sigma,\quad\epsilon,
    \qquad
    h_I=0,\quad h_\sigma=\frac{1}{16},\quad h_\epsilon=\frac12,
    \label{eq:ising_weights}
\end{equation}
and the fusion algebra is
\begin{equation}
    \sigma\times\sigma=I+\epsilon,
    \qquad
    \epsilon\times\epsilon=I,
    \qquad
    \epsilon\times\sigma=\sigma.
    \label{eq:ising_fusion}
\end{equation}
Because the modular invariant is diagonal, the elementary conformal boundaries are labeled by the same three primaries. The Cardy boundary $\sigma$ is the free Ising boundary, while $I$ and $\epsilon$ are the two fixed boundaries. The lattice $\mathbb Z_2$ symmetry leaves the free boundary invariant and exchanges the two fixed boundaries,
\begin{equation}
    \mathbb Z_2:\qquad
    I\longleftrightarrow\epsilon,
    \qquad
    \sigma\longmapsto\sigma .
    \label{eq:ising_z2_boundary_action}
\end{equation}
This immediately constrains the conformal boundary associated with the microscopic initial state. A $\mathbb Z_2$-symmetric short-range initial state cannot select either fixed boundary individually. Its preparation boundary is instead described at long distances by the symmetry-preserving free boundary,
\begin{equation}
    \ket{\Psi_0^{\rm sym}}
    \xrightarrow{\text{long distances}}
    B_O=\sigma .
    \label{eq:ising_symmetric_boundary_flow}
\end{equation}
A symmetry-breaking initial state instead selects one member of the fixed pair,
\begin{equation}
    \ket{\Psi_0^{\rm br}}
    \xrightarrow{\text{long distances}}
    B_O=I\ \text{or}\ \epsilon .
    \label{eq:ising_broken_boundary_flow}
\end{equation}
In a conventional boundary-Hamiltonian realization, a boundary magnetic field generates a free-to-fixed boundary flow~\cite{cardy_1986_boundary_operator,affleck_ludwig_1991_boundary_entropy}. Here we realize the two boundary classes directly by choosing symmetric or symmetry-breaking initial states in both Euclidean quantizations.

For the entanglement prescription used here, the entanglement cut realizes the free Ising boundary $B_E=\sigma$~\cite{lauchli_2013_opContent}. The BCFT operator content is therefore fixed before examining the numerical spectrum. For a symmetric initial state,
\begin{align}
    \sigma\times\sigma&=I+\epsilon,
    \nonumber\\
    \mathcal H_{\sigma\sigma}
    &=\mathcal V_I\oplus\mathcal V_\epsilon,
    \label{eq:ising_free_free}
\end{align}
whereas a symmetry-breaking initial state gives
\begin{align}
    I\times\sigma&=\sigma,
    &
    \epsilon\times\sigma&=\sigma,
    \nonumber\\
    \mathcal H_{I\sigma}
    &=\mathcal H_{\epsilon\sigma}
     =\mathcal V_\sigma .
    \label{eq:ising_fixed_free}
\end{align}
The two symmetry classes therefore have sharply different boundary spectra even though the bulk CFT is unchanged. The symmetric case contains the identity and energy towers; in particular the identity module has no level-one state because $L_{-1}\ket0=0$. The symmetry-breaking case contains a single $\sigma$ tower beginning at $h_\sigma=1/16$. The logarithmic entropy follows the common formula Eq.~\eqref{eq:entropy_scaling}; changing the initial-state symmetry changes the boundary contribution and the operator content, not the universal bulk coefficient.

\begin{figure}[t]
    \centering
    \includegraphics[width=\linewidth]{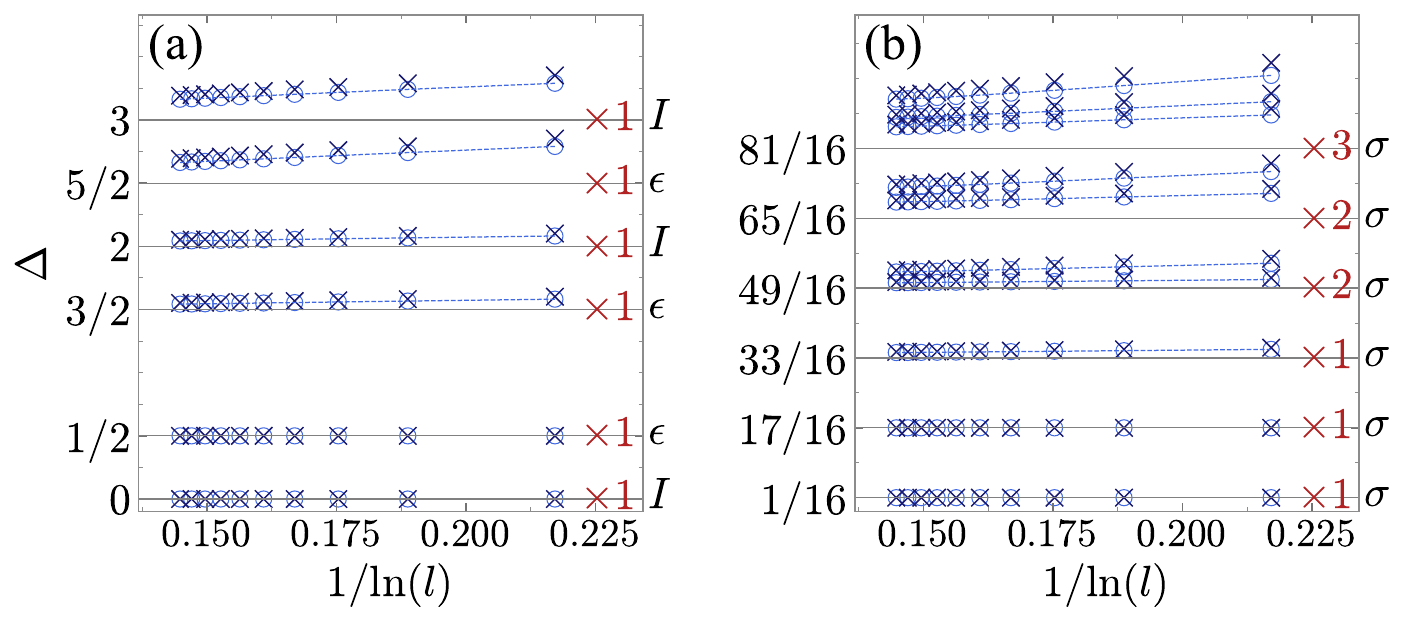}
    \caption{Spatial (circle) and temporal (cross) entanglement in the Euclidean-time evolved states for the critical Ising chain. The two columns correspond to a $\mathbb Z_2$-symmetric and a symmetry-breaking initial state. At large length scales within the finite-length scaling regime, the two preparations are described, respectively, by the free Cardy boundary $B_O=\sigma$ and fixed boundaries $B_O=I/\epsilon$. With the free entanglement boundary $B_E=\sigma$, BCFT fusion predicts the $[I]\oplus[\epsilon]$ and $[\sigma]$ operator contents displayed by the low-lying spectra. }
    \label{fig:Ising_ES_EE}
\end{figure}

The spatial and temporal calculations then provide a direct check of these predictions, as shown in Fig.~\ref{fig:Ising_ES_EE}. In both orientations the symmetric initial state produces the $[I]\oplus[\epsilon]$ spectrum, while the symmetry-breaking initial state produces the $[\sigma]$ spectrum. The agreement of the two lattice quantizations is therefore a consequence of the same symmetry-selected Cardy boundary and the same fusion rule, rather than an independent numerical coincidence.

\emph{Three-state Potts CFT realized in the critical three-state Potts chain---} At its self-dual critical point, the three-state Potts chain is
\begin{equation}
    H_{\mathrm{Potts}}
    =-\sum_i\left(
       Z_i^\dagger Z_{i+1}
       +Z_{i+1}^\dagger Z_i
       +X_i+X_i^\dagger
      \right),
    \label{eq:potts_hamiltonian}
\end{equation}
with
\begin{equation}
    \begin{gathered}
      Z=\begin{pmatrix}
         1&0&0\\0&\omega&0\\0&0&\omega^2
        \end{pmatrix},
      \qquad
      X=\begin{pmatrix}
         0&1&0\\0&0&1\\1&0&0
        \end{pmatrix},
      \qquad \omega=\e^{2\pi\ii/3}.
    \end{gathered}
    \label{eq:potts_operators}
\end{equation}
This critical lattice model flows in the infrared to the $c=4/5$ three-state Potts CFT, equivalently the non-diagonal $D_4$ modular invariant of the $M(6,5)$ minimal model with extended $W_3$ symmetry~\cite{fateev_zamolodchikov_1987_z3,difrancesco_1997_CFT_book,behrend_2000_rcft_boundaries}. It is therefore a useful example in which the microscopic symmetry is elementary but the conformal-boundary structure is substantially richer than in the diagonal Ising theory.

We first make the microscopic internal symmetry explicit. In the clock basis we identify $\ket{A}=\ket0$, $\ket{B}=\ket1$, and $\ket{C}=\ket2$. The onsite symmetry is
\begin{equation}
    S_3=\mathbb Z_3\rtimes\mathbb Z_2,
\end{equation}
where $X$ generates the cyclic permutation of the three colors. A convenient local generator of the reflection, or charge-conjugation, part is
\begin{equation}
    K=\begin{pmatrix}
       1&0&0\\
       0&0&1\\
       0&1&0
      \end{pmatrix},
    \label{eq:potts_local_reflection}
\end{equation}
which leaves $\ket A$ invariant and exchanges $\ket B$ and $\ket C$. The global onsite generators may thus be chosen as
\begin{equation}
    \mathcal R=\prod_i X_i,
    \qquad
    \mathcal K=\prod_i K_i,
\end{equation}
satisfying
\begin{equation}
    \mathcal R^3=\mathcal K^2=1,
    \qquad
    \mathcal K\mathcal R\mathcal K=\mathcal R^{-1}.
    \label{eq:potts_lattice_s3_generators}
\end{equation}
Their action on the clock operators is
\begin{align}
    \mathcal R:\quad &Z_i\mapsto\omega Z_i,
    &X_i&\mapsto X_i,
    \nonumber\\
    \mathcal K:\quad &Z_i\mapsto Z_i^\dagger,
    &X_i&\mapsto X_i^\dagger,
    \label{eq:potts_lattice_s3_action}
\end{align}
which makes the $S_3$ invariance of Eq.~\eqref{eq:potts_hamiltonian} manifest.

We now turn to the conformal boundaries of the infrared theory. The complete Potts BCFT contains eight elementary conformal boundary conditions~\cite{affleck_1998_bc,fuchs_1998_potts_boundaries},
\begin{equation}
    A,\ B,\ C,\qquad
    AB,\ BC,\ CA,\qquad
    \mathrm{free},\qquad
    \mathrm{new}.
    \label{eq:potts_eight_boundaries}
\end{equation}
The first three are fixed boundaries. The next three are mixed boundaries, in which one Potts color is excluded, while free and new are two additional elementary boundary conditions. Under the microscopic $S_3$ action these eight boundaries form four orbits,
\begin{equation}
    \{A,B,C\},\qquad
    \{AB,BC,CA\},\qquad
    \{\mathrm{free}\},\qquad
    \{\mathrm{new}\}.
    \label{eq:potts_four_orbits}
\end{equation}
This orbit structure provides the natural starting point for identifying the microscopic subgroup preserved by each boundary representative.

For the representative fixed and mixed boundaries $A$ and $BC$, the lattice realization is particularly transparent. Consider the local boundary constraints
\begin{equation}
    P_A=\ket A\!\bra A,
    \qquad
    P_{BC}=\ket B\!\bra B+\ket C\!\bra C .
    \label{eq:potts_boundary_projectors}
\end{equation}
Both are invariant under $K$ but not under the three-cycle $X$. Consequently, although $A$ and $BC$ are inequivalent conformal boundaries, they preserve the same microscopic reflection subgroup,
\begin{equation}
    G_A=G_{BC}
    =\langle\mathcal K\rangle
    \simeq\mathbb Z_2 .
    \label{eq:potts_fixed_mixed_stabilizer}
\end{equation}
The other two fixed--mixed pairs preserve the conjugate reflections obtained by acting with powers of $\mathcal R$. The free and new boundaries provide the complementary example. The local state
\begin{equation}
    \ket{+}=\frac{\ket A+\ket B+\ket C}{\sqrt3}
    \label{eq:potts_symmetric_local_state}
\end{equation}
is invariant under both $\mathcal R$ and $\mathcal K$, and a product of such states realizes the free boundary at long distances. The standard lattice realization of the new boundary uses a different $S_3$-invariant boundary coupling~\cite{ref_appendix}. In its strong-coupling description the boundary Hilbert space is the complementary subspace
\begin{equation}
    P_\perp=1-\ket+\!\bra+,
\end{equation}
which is likewise invariant under the full $S_3$. Hence
\begin{equation}
    G_{\mathrm{free}}=G_{\mathrm{new}}=S_3,
    \label{eq:potts_free_new_symmetry}
\end{equation}
even though free and new are distinct BCFT fixed points. Potts therefore gives two explicit examples in which the preserved microscopic symmetry does not determine the conformal boundary:
\begin{equation}
    G_A=G_{BC}=\mathbb Z_2,
    \qquad
    G_{\mathrm{free}}=G_{\mathrm{new}}=S_3,
\end{equation}
but
\begin{equation}
    A\neq BC,
    \quad
    \mathrm{free}\neq\mathrm{new}.
    \label{eq:potts_same_symmetry_distinct_boundaries}
\end{equation}

Three distinct structures organize the relations among these four boundary classes. Ordinary $S_3$ symmetry only permutes boundaries within the fixed triplet or within the mixed triplet. Choosing the convention
\begin{equation}
    \begin{aligned}
        &A\leftrightarrow I,\quad
        B\leftrightarrow\psi,\quad
        C\leftrightarrow\psi^\dagger,\\
        &BC\leftrightarrow\epsilon,\quad
        CA\leftrightarrow\sigma,\quad
        AB\leftrightarrow\sigma^\dagger,
    \end{aligned}
    \label{eq:potts_boundary_primary_correspondence}
\end{equation}
the relation between the fixed and mixed triplets is instead generated by fusion with the neutral primary $\epsilon$,
\begin{equation}
    \epsilon\times I=\epsilon,
    \qquad
    \epsilon\times\psi=\sigma,
    \qquad
    \epsilon\times\psi^\dagger=\sigma^\dagger.
    \label{eq:potts_epsilon_maps_fixed_mixed}
\end{equation}
For example,
\begin{equation}
    A\xrightarrow{\ \epsilon\text{-fusion}\ }BC.
    \label{eq:potts_A_to_BC}
\end{equation}
This is neither an $S_3$ rotation nor an invertible Fourier transformation. The non-invertibility follows from
\begin{equation}
    \epsilon\times\epsilon=I+\epsilon,
    \label{eq:potts_epsilon_noninvertible}
\end{equation}
so a second fusion does not return a single fixed boundary.

The free/new pair carries an analogous non-invertible two-object structure, but now in the boundary algebra appropriate to the $W_3$-violating sector. Fuchs and Schweigert showed that free and new are the two Dirichlet-type Potts boundaries and are governed by a two-dimensional algebra isomorphic to the Lee--Yang fusion algebra~\cite{fuchs_1998_potts_boundaries}. Denoting its nontrivial generator by $\epsilon_{\partial}$,
\begin{equation}
    \epsilon_{\partial}\star\epsilon_{\partial}
    =1+\epsilon_{\partial}.
    \label{eq:potts_dirichlet_boundary_algebra}
\end{equation}
Thus free and new are also related by a non-invertible boundary-algebra structure rather than by a microscopic $S_3$ transformation.

Kramers--Wannier duality acts in the orthogonal direction~\cite{affleck_1998_bc,frohlich_2004_kw_defects}. At the level of boundary universality classes,
\begin{equation}
    \{A,B,C\}
    \xleftrightarrow{\ \mathrm{KW}\ }
    \mathrm{free},
    \qquad
    \{AB,BC,CA\}
    \xleftrightarrow{\ \mathrm{KW}\ }
    \mathrm{new}.
    \label{eq:potts_kw_boundary_pairs}
\end{equation}
The discrete Fourier transform of the three-state clock variables is one ingredient in constructing the lattice dual variables, but the full Potts Kramers--Wannier map is a nonlocal site--bond duality. It should therefore not be identified with the fixed--mixed relation generated by $\epsilon$ fusion.

The same classification also determines the available boundary perturbations and the stability of each conformal boundary. The diagonal annulus spectra for one representative of each class are~\cite{affleck_1998_bc}
\begin{align}
    Z_{A,A}
    &=\chi_I,
    \nonumber\\
    Z_{BC,BC}
    &=\chi_I+\chi_\epsilon,
    \nonumber\\
    Z_{\mathrm{free,free}}
    &=\chi_I+\chi_\psi+\chi_{\psi^\dagger},
    \nonumber\\
    Z_{\mathrm{new,new}}
    &=\chi_I+\chi_\epsilon
      +\chi_\sigma+\chi_{\sigma^\dagger}
      +\chi_\psi+\chi_{\psi^\dagger}.
    \label{eq:potts_diagonal_boundary_spectra}
\end{align}
The fixed boundary has no nontrivial relevant boundary field. The mixed boundary contains one relevant field with $h_\epsilon=2/5$. This perturbation becomes available once its residual $\mathbb Z_2$ is relaxed and selects one of the two allowed colors. The free boundary contains two charged relevant fields $\psi,\psi^\dagger$ with $h=2/3$, which are forbidden when the full $S_3$ symmetry is imposed. The new boundary is qualitatively different even though it preserves the same $S_3$. In addition to the charged fields it contains the neutral $S_3$-invariant field $\epsilon$ with $h=2/5$. It is therefore unstable even within the fully symmetric boundary-coupling space and can flow to the free boundary. The pairs $(A,BC)$ and $(\mathrm{free},\mathrm{new})$ make the central lesson explicit. Symmetry restricts the allowed boundary perturbations, but the operator content is also needed to identify the conformal boundary and determine its stability.

We now apply this boundary classification to the entanglement problem. For the regulated cut used here, the entanglement boundary is the free Potts boundary, $B_E=\mathrm{free}$. Holding $B_E$ fixed turns the low-lying entanglement spectrum into a spectroscopic probe of the physical boundary $B_O$. The four $S_3$-inequivalent classes then give four distinct open-channel partition functions,
\begin{equation}
    \begin{aligned}
        &Z_{\mathrm{free,free}}
        =\chi_I+\chi_\psi+\chi_{\psi^\dagger}
        =\chi_{11}+\chi_{41}+2\chi_{43},\\
        &Z_{\mathrm{fixed,free}}
        =\chi_{44}+\chi_{42},\\
        &Z_{\mathrm{mixed,free}}
        =\chi_{22}+\chi_{32},\\
        &Z_{\mathrm{new,free}}
        =\chi_\epsilon+\chi_\sigma+\chi_{\sigma^\dagger}
        =\chi_{21}+\chi_{31}+2\chi_{33}.
    \label{eq:potts_bcft}
    \end{aligned}
\end{equation}
Their characteristic primary weights are correspondingly distinct: the free--free spectrum contains the identity tower together with the $\psi,\psi^\dagger$ pair at $h=2/3$; the fixed--free spectrum begins at $h=1/8$ with the second primary at $13/8$; the mixed--free spectrum begins at $h=1/40$ with the $21/40$ family; and the new--free spectrum contains the degenerate $\sigma,\sigma^\dagger$ pair at $h=1/15$ together with the neutral $\epsilon$ family at $h=2/5$. Thus the BCFT spectrum distinguishes the same-symmetry pairs that the microscopic subgroup alone cannot separate.

The microscopic initial state determines which of these conformal boundaries describes the preparation at long distances. The fully symmetric product state
\begin{equation}
    \ket{\Psi_0^{\rm free}}
    =\bigotimes_i
      \frac{\ket0_i+\ket1_i+\ket2_i}{\sqrt3}
    \label{eq:potts_symmetric_initial}
\end{equation}
realizes the free boundary in our calculation, whereas the color-polarized state
\begin{equation}
    \ket{\Psi_0^{A}}
    =\bigotimes_i\ket0_i
    \label{eq:potts_fixed_initial}
\end{equation}
selects one member of the fixed triplet. We also engineer short-range initial states that realize a representative mixed boundary and the new boundary in the finite-length scaling regime~\cite{ref_appendix}. Interestingly, the minimal lattice realization of the new boundary initial state is not a product state --- there is minimum entanglement in the state. It can only be realized as a $S_3$ symmetric MPS with $D>1$~\cite{obrien_2020_potts_new_bdy, ref_appendix}. The mixed initial state preserves the same residual $\mathbb Z_2$ as its paired fixed boundary, whereas the new initial state preserves the full $S_3$ and is tuned to realize the unstable symmetric conformal boundary. Their microscopic details are nonuniversal and their infrared identification is made from the tower contents in Eq.~\eqref{eq:potts_bcft}.

\begin{figure*}[t]
    \centering
    \includegraphics[width=\textwidth]{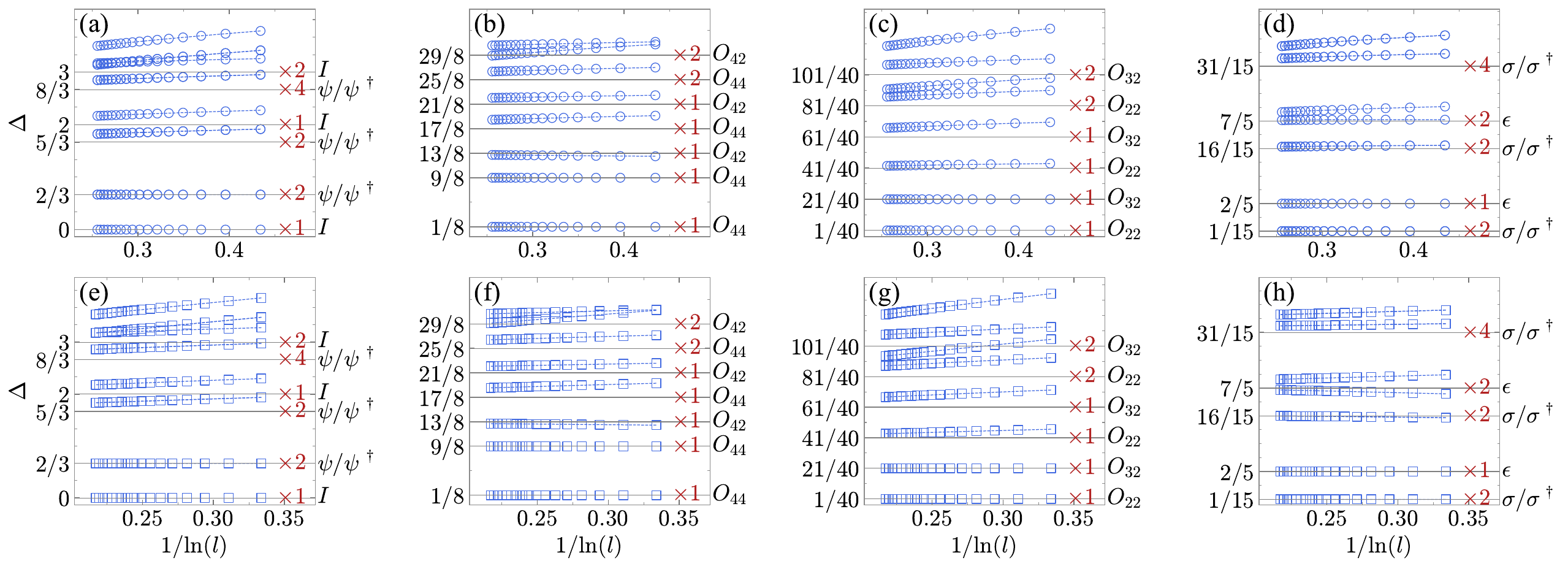}
    \caption{Spatial (upper) and temporal (lower) state entanglement for the critical three-state Potts chain for all four symmetry-inequivalent conformal boundary classes. The four columns correspond to free, fixed, mixed and new physical boundaries respectively, with the entanglement boundary kept free. The rescaled low-lying spectra reproduce the operator contents in Eq.~\eqref{eq:potts_bcft}: the characteristic lowest primary weights are $0$ and $2/3$ (free), $1/8$ and $13/8$ (fixed), $1/40$ and $21/40$ (mixed), and the $h=1/15$ doublet together with $h=2/5$ (new). The comparison makes explicit that fixed and mixed are spectrally distinct despite preserving the same residual $\mathbb Z_2$, while free and new are distinct despite preserving the full $S_3$. Solid straight lines denote the BCFT predictions.}
    \label{fig:potts_spatial}
\end{figure*}

The numerical results then test these BCFT predictions rather than define them. Because $S_3$ is onsite, the spatial and temporal quantizations inherit the same microscopic generators $\mathcal R$ and $\mathcal K$. Figure~\ref{fig:potts_spatial} gives the complete four-class spatial and temporal spectroscopy. After removing the nonuniversal affine normalization of the entanglement Hamiltonian, the four spectra approach precisely the free--free, fixed--free, mixed--free, and new--free towers in Eq.~\eqref{eq:potts_bcft}, including their characteristic primary weights and descendant multiplicities. In particular, the numerical spectra distinguish $A$ from $BC$ despite their common residual $\mathbb Z_2$, and distinguish free from new despite their common full $S_3$ symmetry.

%The entanglement entropies provide the complementary bulk check. For all four boundary classes the logarithmic coefficient is consistent with the same Potts central charge $c=4/5$, as required by Eq.~\eqref{eq:entropy_scaling}. Changing the initial boundary changes the boundary contribution and operator content, but not the bulk conformal anomaly.

The Potts chain therefore provides a more discriminating test of boundary selection than the Ising chain. The lattice $S_3$ symmetry restricts the accessible conformal boundaries, while the BCFT operator content distinguishes boundaries that share the same microscopic symmetry: fixed from mixed at residual $\mathbb Z_2$, and free from new at full $S_3$. Because $S_3$ is onsite, the same boundary classification is recovered in the spatial and temporal quantizations. Potts thus serves as a non-diagonal control case before the LSM constraint is introduced.

\section{LSM constraint and lattice--CFT symmetry matching in the XXZ chain}
\label{sec:xxz}

The XXZ chain provides the central example of how an LSM constraint restricts the infrared boundary theory. The interplay between one-site translation and the internal spin symmetry gives the compact-boson CFT an anomalous symmetry action inherited from the lattice. We begin with the consequences for the spatial initial state and review the compact-boson current algebra and conformal boundaries. Bosonization then determines how the lattice symmetries act on the primary fields and, through them, on the allowed boundaries. The resulting BCFT sectors appear at two inequivalent cuts of a single spatial state and at a fixed cut of successive temporal states. This correspondence connects spatial dimerization to temporal sector pumping.

\subsection{XXZ chain and the spatial LSM benchmark}

We consider the antiferromagnetic spin-$1/2$ XXZ chain
\begin{equation}
    H_{\mathrm{XXZ}}
    =\sum_i\left(
       S_i^x S_{i+1}^x
       +S_i^yS_{i+1}^y
       +g\,S_i^zS_{i+1}^z
      \right),
    \label{eq:xxz_hamiltonian}
\end{equation}
where $S^\alpha$, with $\alpha = x,y,z$, are spin-$1/2$ operators. The Hamiltonian is invariant under $U_z(1)$ rotations generated by
\begin{equation}
    Q_z = \sum_i S_i^z,
    \qquad
    \e^{\ii\alpha Q_z} \, S_j^\pm \, \e^{-\ii\alpha Q_z}
    =\e^{\pm\ii\alpha} S_j^\pm,
    \label{eq:xxz_u1}
\end{equation}
with $S^\pm = S^x \pm i\,S^y$. The Hamiltonian also preserves one-site translation $T_1$ and the spin flip $F=\prod_i 2 S_i^x$. At zero magnetization the spin chain is at half filling,
\begin{equation}
    \nu=\left\langle S_j^z+\frac12\right\rangle=\frac12,
\end{equation}
and the LSM theorem forbids a unique gapped state preserving both $U(1)_z$ and $T_1$~\cite{Lieb_1961_LSM,oshikawa_2000_commensurability,tasaki_2022_LSM_topo,cheng_2023_lsm,else_thorngren_2020_lsm,furuya_oshikawa_2017_global_anomaly}. In tensor-network language, no injective short-range MPS at this filling can preserve both $U(1)_z$ and one-site translation~\cite{sanz_2009_mps_lsm,cirac_2021_mps_review}.

For $-1<g<1$, the infrared theory is a Luttinger liquid~\cite{giamarchi_2003_q1d,lukyanov_1998_xxz}. At the $\mathrm{SU}(2)_1$ Heisenberg endpoint $g=1$, a marginally irrelevant current--current interaction generates logarithmic corrections~\cite{affleck_gepner_1989_spin_chains,lukyanov_1998_xxz}. The Gaussian description below applies to $-1<g<1$; we discuss the approach to $g=1$ only qualitatively.

The LSM constraint thus enters the spatial protocol through the choice of short-range initial state. At half filling, an injective $U(1)_z$-symmetric MPS requires an even number of sites per unit cell, with two sites the minimal choice. For the preparation in Eq.~\eqref{eq:spatial_state_def}, we use a product of nearest-neighbor singlets. Its two-site unit cell gives two inequivalent entanglement cuts, $A$--$B$ and $B$--$A$, whose spectra can be compared within the same state at fixed $\tau$. This microscopic dimerization persists while $U(1)_z$ is preserved and $\ell$ remains finite in the scaling region. Under exact Euclidean-time evolution, it vanishes as $\ell\to\infty$, when the state approaches the translation-invariant critical ground state.

\subsection{Compact-boson CFT and conformal boundaries}
\label{sec:compact_boson}

With the velocity rescaled to unity, the compact-boson action and field identifications are
\begin{equation}
    \begin{aligned}
        S&=\frac{1}{8\pi}\int d\tau\,dx\,
          \left[
           (\partial_\tau\varphi)^2
           +(\partial_x\varphi)^2
          \right], \\[-1mm]
        &\hspace{2mm}\varphi\sim\varphi+2\pi R,
        \qquad
        \vartheta\sim\vartheta+\frac{4\pi}{R},
        \label{eq:compact_action}
    \end{aligned}
\end{equation}
where $\vartheta$ is the dual compact field. The Bethe-ansatz radius is~\cite{giamarchi_2003_q1d,lukyanov_1998_xxz}
\begin{equation}
    R^2=\frac{2}{\pi}\arccos(-g)=\frac{1}{K},
    \label{eq:radius_xxz}
\end{equation}
with $K$ the Luttinger parameter in this convention.

\noindent\emph{Extended $U(1)$ current algebra.---}
The compact boson has two continuous $U(1)$ symmetries: the shift symmetry of $\varphi$ and the winding, or topological, symmetry. In the T-dual description, the latter is the shift symmetry of $\vartheta$. In terms of the left- and right-moving fields,
\begin{equation}
    \varphi=\varphi_L+\varphi_R,
    \qquad
    \vartheta=\varphi_L-\varphi_R,
\end{equation}
these symmetries are described by the affine current algebra $\widehat{U(1)}_L\times\widehat{U(1)}_R$. This algebra extends the Virasoro symmetry, with the stress tensor obtained from the currents through the Sugawara construction. The current operators therefore provide a common description of the bulk spectrum and conformal boundaries~\cite{difrancesco_1997_CFT_book,blumenhagen_2009_CFT_book}.

On the cylinder $w=\tau+\ii x$, the holomorphic scalar field has the mode expansion
\begin{equation}
    \varphi_L(w)
    =q_L-\ii j_0 w
     +\ii\sum_{n\neq0}\frac{j_n}{n}\e^{-nw},
    \label{eq:compact_chiral_field}
\end{equation}
where the coefficients $j_n$ are the modes of the holomorphic current,
\begin{equation}
    j(w)=\ii\,\partial\varphi_L(w)
    =\sum_{n\in\mathbb Z}j_n\e^{-nw},
    \qquad
    [j_m,j_n]=m\,\delta_{m+n,0}.
    \label{eq:compact_current_algebra}
\end{equation}
The antiholomorphic sector has the same form under $(w,j_n,\varphi_L)\to(\bar w,\bar j_n,\varphi_R)$. The Sugawara construction gives the stress tensor and its zero mode,
\begin{equation}
    T=\frac12:j^2:,
    \qquad
    L_0=\frac12j_0^2+\sum_{n>0}j_{-n}j_n,
    \label{eq:compact_sugawara_L0}
\end{equation}
with analogous expressions in the antiholomorphic sector. Each $\widehat{U(1)}$ representation is specified by the zero-mode charge of its primary state, and the negative current modes generate its oscillator descendants.

Compactness quantizes the zero modes. The primaries of the extended current algebra are the vertex operators
\begin{equation}
    V_{e,w}(z,\bar z)
    =:\exp\!\left[
       \ii\frac{e}{R}\varphi
       +\ii\frac{wR}{2}\vartheta
      \right]:,
    \qquad e,w\in\mathbb Z,
    \label{eq:vertex_operator}
\end{equation}
whose chiral charges are
\begin{equation}
    p_L=\frac{e}{R}+\frac{wR}{2},
    \qquad
    p_R=\frac{e}{R}-\frac{wR}{2}.
    \label{eq:compact_chiral_momenta}
\end{equation}
Under the state--operator correspondence, the associated primary states satisfy
\begin{equation}
    j_0\ket{e,w}=p_L\ket{e,w},
    \qquad
    \bar j_0\ket{e,w}=p_R\ket{e,w},
    \label{eq:compact_chiral_charges}
\end{equation}
and the Sugawara construction fixes their conformal weights to be
\begin{equation}
    h_{e,w}=\frac{p_L^2}{2},
    \qquad
    \bar h_{e,w}=\frac{p_R^2}{2}.
    \label{eq:compact_bulk_spectrum}
\end{equation}
The two global $U(1)$ charges are the vector and axial combinations of these chiral zero modes,
\begin{align}
    \frac{R}{2}(j_0+\bar j_0)\ket{e,w}
    &=e\ket{e,w},
    \nonumber\\
    \frac{1}{R}(j_0-\bar j_0)\ket{e,w}
    &=w\ket{e,w}.
    \label{eq:compact_charge_combinations}
\end{align}
Thus $e$ is the compact momentum and $w$ the winding charge. Matching the lattice and continuum symmetry actions below identifies which combination represents the microscopic spin $U(1)$ symmetry of the XXZ chain.

\noindent\emph{Conformal boundaries and current reflection.---}
On the strip $0\leq x\leq\pi$, with open-channel Euclidean time $\tau$, a boundary couples the two chiral sectors by reflecting a current excitation from one chirality into the other. At generic radius, the elementary boundaries considered here preserve the affine $U(1)$ current algebra and belong to the Dirichlet or Neumann family,
\begin{align}
    D_\varphi(\varphi_B):\quad
    &\varphi\big|_B=\varphi_B,
    \quad j+\bar j=0,
    \label{eq:dirichlet_current_gluing}\\
    N_\varphi(\vartheta_B):\quad
    &\partial_x\varphi\big|_B=0
    \quad\Longleftrightarrow\quad
    \vartheta\big|_B=\vartheta_B,
    \quad j-\bar j=0.
    \label{eq:neumann_current_gluing}
\end{align}
These current relations follow from the derivatives of the scalar field,
\begin{equation}
    \partial_\tau\varphi=-\ii(j+\bar j),
    \qquad
    \partial_x\varphi=j-\bar j.
\end{equation}
Dirichlet reflection therefore fixes $\varphi$ along the boundary, while Neumann reflection sets its normal gradient to zero. The two conditions identify the left and right affine currents through the automorphisms $j=\pm\bar j$.

Their conformal invariance follows at the local operator level from the stress-tensor condition
\begin{equation}
    T=\bar T
    \qquad\text{at the boundary},
    \label{eq:compact_conformal_boundary}
\end{equation}
which expresses the absence of energy-momentum flux through the boundary. The Sugawara form $T=\frac12:j^2:$ implies that either reflection condition $j=\pm\bar j$ satisfies Eq.~\eqref{eq:compact_conformal_boundary}. Both boundary families therefore preserve conformal invariance together with the current symmetry.

\noindent\emph{Open-channel sectors.---}
The open-channel spectra follow by imposing the boundary conditions on the scalar-field operator. Consider first a Dirichlet boundary at the left edge,
\begin{equation}
    \varphi(\tau,0)=\varphi_O.
\end{equation}
On the strip, the allowed values of the mode index $n$ are fixed by the second boundary condition. At $x=0$, Eq.~\eqref{eq:compact_chiral_field} and its antiholomorphic counterpart give
\begin{equation}
    \varphi(\tau,0)
    =q_L+q_R
     -\ii(j_0+\bar j_0)\tau
     +\ii\sum_{n\neq0}\frac{j_n+\bar j_n}{n}\e^{-n\tau}.
\end{equation}
Requiring the boundary value to be fixed as an operator identity at every $\tau$ gives
\begin{equation}
    \bar j_n=-j_n,
    \qquad
    \bar j_0=-j_0,
    \qquad
    q_L+q_R=\varphi_O.
    \label{eq:compact_Dleft_current_constraint}
\end{equation}
Writing the surviving open-channel modes as $J_n=j_n$ and substituting Eq.~\eqref{eq:compact_Dleft_current_constraint} into the scalar field yields
\begin{equation}
    \varphi(\tau,x)
    =\varphi_O+2J_0x
     +2\sum_{n\neq0}\frac{J_n}{n}
       \e^{-n\tau}\sin(nx).
    \label{eq:compact_strip_field_Dleft}
\end{equation}
The sine dependence thus follows from the boundary relation between the two chiral currents.

The boundary at $x=\pi$ determines the allowed frequencies and zero mode. A second Dirichlet condition, $\varphi(\tau,\pi)=\varphi_E$ modulo $2\pi R$, gives
\begin{equation}
    n\in\mathbb Z,
    \qquad
    J_0=A_D
    =\frac{\Delta\varphi}{2\pi}+wR,
    \qquad
    \Delta\varphi=\varphi_E-\varphi_O,
    \qquad w\in\mathbb Z.
    \label{eq:AD}
\end{equation}
The zero mode $A_D$ measures the net change of the compact field across the strip. The field connects the two pinned boundary values while winding an integer number of times around the target-space circle. Imposing a Neumann condition at the right edge instead gives
\begin{equation}
    J_0=0,
    \qquad
    n\in\mathbb Z+\frac12.
    \label{eq:compact_DN_zero_half}
\end{equation}
The different reflection conditions in the mixed $DN$ sector therefore remove the current zero mode and leave half-integer-moded current fluctuations.

Applying the same argument to two Neumann boundaries, or using T-duality, gives the $NN$ sector. Its oscillators are integer moded, with zero mode
\begin{equation}
    J_0=A_N
    =\frac{\Delta\vartheta}{2\pi}+\frac{2e}{R},
    \qquad
    \Delta\vartheta=\vartheta_E-\vartheta_O,
    \qquad e\in\mathbb Z.
    \label{eq:AN}
\end{equation}
The surviving zero mode is the compact momentum, which is equivalently the winding of the dual field between its pinned boundary values. Table~\ref{tab:compact_boundaries} summarizes the three types of open-channel sector.

\begin{table}[t]
    \caption{Compact-boson open-channel sectors. $J_0$ denotes the surviving current zero mode. Matching gluing conditions at the two edges retain an integer-moded current algebra and a compact zero-mode lattice. Mixed gluing removes the zero mode and gives a half-integer-moded current module.}
    \label{tab:compact_boundaries}
    \centering
    \begin{tabular}{@{}ccc@{}}
       \toprule
       $B_OB_E$ & current modes & $J_0$\\
       \midrule
       $DD$ & $n\in\mathbb Z$ & $A_D$\\
       $NN$ & $n\in\mathbb Z$ & $A_N$\\
       $DN$ or $ND$ & $n\in\mathbb Z+\tfrac12$ & none\\
       \bottomrule
    \end{tabular}
\end{table}

For matching gluing conditions, the Sugawara construction gives the open-channel Hamiltonians
\begin{align}
    D_\varphi D_\varphi:\quad
    L_0-\frac1{24}
    &=\frac{A_D^2}{2}
      +\sum_{n>0}J_{-n}J_n-\frac1{24},
    \nonumber\\
    N_\varphi N_\varphi:\quad
    L_0-\frac1{24}
    &=\frac{A_N^2}{2}
      +\sum_{n>0}J_{-n}J_n-\frac1{24}.
    \label{eq:compact_same_boundary_hamiltonians}
\end{align}
Each allowed zero mode labels a $\widehat{U(1)}$ primary sector, and the negative integer modes generate its descendants. The zero-mode contribution records the compact-field interpolation and winding in $DD$, or the T-dual momentum in $NN$.

Mixed boundaries give a single half-integer-moded current module,
\begin{equation}
    D_\varphi N_\varphi:\quad
    L_0-\frac1{24}
    =\sum_{\substack{n>0\\n\in\mathbb Z+1/2}}
      J_{-n}J_n+\frac1{48},
    \label{eq:compact_boundary_hamiltonians}
\end{equation}
with the same spectrum for $ND$. Its ground state has
\begin{equation}
    h_{DN}=h_{ND}=\frac1{16},
    \label{eq:DN_twist_weight}
\end{equation}
and corresponds to the lowest $D\leftrightarrow N$ boundary-condition-changing operator. The character of this current module is
\begin{equation}
    Z_{DN}(q)
    =q^{1/48}
      \prod_{\substack{n>0\\n\in\mathbb Z+1/2}}
      \frac{1}{1-q^n}.
    \label{eq:compact_mixed_partition}
\end{equation}
The extended current algebra organizes the entire mixed sector into one twisted module. Under the Virasoro algebra, this module decomposes into infinitely many conformal families,
\begin{equation}
    Z_{DN}(q)
    =\sum_{m=0}^{\infty}
      \chi^{\mathrm{Vir}}_{h_m}(q),
    \qquad
    h_m=\frac{(2m+1)^2}{16}.
    \label{eq:compact_mixed_virasoro}
\end{equation}
Measured from the ground level, the excitation energies are $k/2$ with degeneracies $1,1,1,2,2,3,\ldots$. Here the half-integer moding belongs to the \emph{oscillators}. In the $DD$ BCFT, the half-integer shift acts on the zero-mode lattice, while the oscillators remain integer modes.

%To compare with the numerical entanglement levels, we remove the nonuniversal scale and additive shift relating $-\log\rho_A$ to the BCFT Hamiltonian. Denoting the resulting $L_0$ eigenvalue by $\widetilde E$, the spectrum for two Dirichlet boundaries is
%\begin{equation}
%    \widetilde E_{DD}[w;N]
%    =\frac12\left[
      %\frac{\Delta\varphi}{2\pi}+wR
      %\right]^2+N,
    %\qquad
    %w\in\mathbb Z,
    %\quad N\in\mathbb Z_{\geq0}.
    %\label{eq:DD_rescaled_spectrum}
%\end{equation}

\subsection{Bosonization and microscopic symmetry action}
\label{sec:symmetry_matching}

Bosonization connects the lattice spin operators to the continuum fields and thereby determines how microscopic symmetries act in the CFT. In the present normalization, the leading XXZ operator expansions involve the vertex primaries of Eq.~\eqref{eq:vertex_operator}~\cite{giamarchi_2003_q1d,lukyanov_1998_xxz},
\begin{equation}
\begin{aligned}
    S_j^z
    &\sim c_0\,\partial_x\varphi
      +(-1)^j\!
       \left(c_+V_{1,0}+c_-V_{-1,0}\right)
      +\cdots,
    \nonumber\\
    S_j^+
    &\sim b_0(-1)^jV_{0,1}
      +b_+V_{1,1}
      +b_-V_{-1,1}
      +\cdots .
    \label{eq:bosonization_dictionary}
\end{aligned}
\end{equation}
The coefficients are nonuniversal, with those in $S_j^z$ constrained by Hermiticity. Shifting the origins of $\varphi$ and $\vartheta$ redistributes constant phases among these coefficients. The symmetry quantum numbers of the vertex primaries are independent of these choices once the compactification convention is fixed, and they suffice to determine the symmetry action below.

The onsite $U(1)_z$ action follows from Eq.~\eqref{eq:xxz_u1}: every primary in $S_j^+$ carries one unit of $Q_z$, while those in $S_j^z$ are neutral. Comparing these charges with Eq.~\eqref{eq:compact_charge_combinations} identifies the microscopic generator with the axial combination of the chiral current zero modes,
\begin{equation}
    Q_z=\frac{j_0-\bar j_0}{R},
    \label{eq:xxz_charge_current_zero_modes}
\end{equation}
on the compact-boson Hilbert space. Its action on the vertex primaries is
\begin{equation}
    U_\alpha V_{e,w}U_\alpha^{-1}
    =\e^{\ii w\alpha}V_{e,w},
    \label{eq:u1_vertex_action}
\end{equation}
or, equivalently, on the compact fields,
\begin{equation}
    U_\alpha:\qquad
    \varphi\mapsto\varphi,
    \qquad
    \vartheta\mapsto\vartheta+\frac{2\alpha}{R}.
    \label{eq:u1_field_action}
\end{equation}
The winding label $w$, the eigenvalue of $(j_0-\bar j_0)/R$, is therefore the physical $U(1)_z$ charge in the XXZ realization. The orthogonal combination $R(j_0+\bar j_0)/2$ has eigenvalue $e$ and generates the other compact-boson shift symmetry.

To determine one-site translation, we separate the displacement by one lattice spacing from the internal action on the compact fields. Writing the latter as compact shifts gives
\begin{equation}
    T_1:\quad
    \varphi(x)\mapsto\varphi(x+a)+\delta_\varphi,
    \qquad
    \vartheta(x)\mapsto\vartheta(x+a)+\delta_\vartheta .
    \label{eq:translation_general_shift}
\end{equation}
The induced action on a vertex primary is
\begin{equation}
    T_1V_{e,w}(x)T_1^{-1}
    =\exp\!\left[
       \ii\frac{e\delta_\varphi}{R}
       +\ii\frac{wR\delta_\vartheta}{2}
      \right]
      V_{e,w}(x+a).
    \label{eq:translation_vertex_general}
\end{equation}
The staggered factor in Eq.~\eqref{eq:bosonization_dictionary} changes sign under translation, $(-1)^j\mapsto(-1)^{j+1}$. Consistency of the longitudinal expansion therefore requires $V_{1,0}$ to acquire a minus sign under $T_1$, so
\begin{equation}
    \e^{\ii\delta_\varphi/R}=-1
    \quad\Longrightarrow\quad
    \delta_\varphi=\pi R
    \pmod{2\pi R}.
\end{equation}
The leading staggered transverse primary $V_{0,1}$ must acquire the same sign, independently fixing
\begin{equation}
    \e^{\ii R\delta_\vartheta/2}=-1
    \quad\Longrightarrow\quad
    \delta_\vartheta=\frac{2\pi}{R}
    \pmod{4\pi/R}.
\end{equation}
Together, these conditions fix the physical one-site translation to act as
\begin{align}
    T_1:\quad&
    \varphi(x)\mapsto\varphi(x+a)+\pi R,
    \nonumber\\
    &\vartheta(x)\mapsto\vartheta(x+a)+\frac{2\pi}{R},
    \label{eq:lattice_cft_symmetry}
\end{align}
or, on the vertex primaries,
\begin{equation}
    T_1V_{e,w}(x)T_1^{-1}
    =(-1)^{e+w}V_{e,w}(x+a).
    \label{eq:translation_vertex_action}
\end{equation}
The two signs cancel for $V_{1,1}$ and $V_{-1,1}$, consistent with their appearance in the uniform part of $S_j^+$ at zero crystal momentum. Equation~\eqref{eq:translation_vertex_action} thus fixes the lattice--CFT translation dictionary at half filling.

For a boundary preserving the spin rotation symmetry, this action simplifies upon composing translation with an onsite $\pi$ spin rotation. Defining
\begin{equation}
    \widetilde T_1\equiv U_{-\pi}T_1,
\end{equation}
we obtain
\begin{align}
    \widetilde T_1:\quad&
    \varphi(x)\mapsto\varphi(x+a)+\pi R,
    \qquad
    \vartheta(x)\mapsto\vartheta(x+a),
    \nonumber\\
    &V_{e,w}(x)\mapsto(-1)^eV_{e,w}(x+a).
    \label{eq:residual_translation}
\end{align}
The combined operation differs from $T_1$ by an exact onsite $U(1)_z$ symmetry. Since $U_{-\pi}$ leaves a $U(1)_z$-preserving boundary condition invariant, $T_1$ and $\widetilde T_1$ act identically on its boundary data. The translation remnant on such a boundary is therefore represented by the half-period shift $\varphi\mapsto\varphi+\pi R$.

The remaining symmetry needed for boundary selection is spin flip. We choose the origin of $\varphi$ so that its action takes the form
\begin{align}
    F:\quad&
    \varphi\mapsto\pi R-\varphi,
    \qquad
    \vartheta\mapsto-\vartheta,
    \nonumber\\
    &V_{e,w}\mapsto(-1)^eV_{-e,-w}.
    \label{eq:spinflip_vertex_action}
\end{align}
Changing the boson origin changes the constant shift in the first line while leaving the boundary orbit and all spectra invariant.

\subsection{Symmetry selection of compact-boson boundaries}
\label{sec:xxz_boundary_selection}

These symmetry actions constrain the Dirichlet and Neumann boundaries introduced above. Within the current-preserving families considered here at generic radius, Eq.~\eqref{eq:u1_field_action} gives
\begin{align}
    U_\alpha:\quad
    D_\varphi(\varphi_0)&\longmapsto D_\varphi(\varphi_0),
    \nonumber\\
    N_\varphi(\vartheta_0)&\longmapsto
    N_\varphi\!\left(\vartheta_0+\frac{2\alpha}{R}\right).
    \label{eq:u1_boundary_action}
\end{align}
A boundary preserving the full $U(1)_z$ symmetry must allow arbitrary shifts of $\vartheta$ and therefore belongs to the Dirichlet family for $\varphi$,
\begin{equation}
    U(1)_z\ \text{preserved}
    \quad\Longrightarrow\quad
    B=D_\varphi(\varphi_0)
    \equiv N_\vartheta .
    \label{eq:u1_selects_dirichlet}
\end{equation}
If the microscopic boundary breaks $U(1)_z$, a Neumann boundary $N_\varphi(\vartheta_0)\equiv D_\vartheta$ is also allowed. The preserved internal symmetry thus constrains the \emph{gluing class} accessible in the long-distance boundary description, and the entanglement spectrum resolves the associated tower content.

Spin flip further constrains the Dirichlet modulus. Equation~\eqref{eq:spinflip_vertex_action} gives
\begin{equation}
    F:\quad
    D_\varphi(\varphi_0)
    \longmapsto
    D_\varphi(\pi R-\varphi_0).
\end{equation}
Invariance under $F$ then restricts an elementary Dirichlet boundary to
\begin{equation}
    \varphi_0=\varphi_\star
    \quad\text{or}\quad
    \varphi_\star+\pi R,
    \qquad
    \varphi_\star=\frac{\pi R}{2}.
    \label{eq:spinflip_dirichlet_pair}
\end{equation}
One-site translation exchanges these two values,
\begin{equation}
    T_1:\qquad
    D_\varphi(\varphi_\star)
    \longleftrightarrow
    D_\varphi(\varphi_\star+\pi R).
    \label{eq:translation_dirichlet}
\end{equation}
Their separation $\pi R$ is physical, while the reference value $\varphi_\star$ depends on the choice of boson origin.

Equation~\eqref{eq:translation_dirichlet} expresses the LSM anomaly at the boundary~\cite{cheng_2023_lsm,else_thorngren_2020_lsm,furuya_oshikawa_2017_global_anomaly}. Translation places each $U(1)_z$- and $F$-preserving elementary conformal boundary in a two-element orbit. A translation-invariant boundary can be formed from the reducible direct sum
\begin{equation}
    D_\varphi(\varphi_\star)
    \oplus
    D_\varphi(\varphi_\star+\pi R),
    \label{eq:translation_invariant_direct_sum}
\end{equation}
which retains both elementary boundary sectors. The absence of a single nondegenerate boundary preserving all these symmetries is the BCFT expression of the obstruction to a unique symmetric short-range state.

For the symmetry-preserving preparations in both quantizations, the microscopic initial boundary and the entanglement cut preserve $U(1)_z$ and $F$. Their conformal boundaries therefore belong to the pair in Eq.~\eqref{eq:spinflip_dirichlet_pair}, with the microscopic termination fixing the member at each edge. The two edges can select the same Dirichlet value or values separated by $\pi R$. These relative displacements give the two $DD$ spectra through which the LSM constraint appears in spatial and temporal entanglement.

\subsection{Anomaly-induced sector pumping in spatial and temporal entanglement}
\label{sec:sector_pumping}

The spatial and temporal protocols sample the same pair of $DD$ sectors in different ways. In the spatial state at fixed $\tau$, moving the entanglement cut by one site changes its termination relative to the two-site singlet preparation. Equation~\eqref{eq:translation_dirichlet} then shifts the relative Dirichlet modulus by $\pi R$, so the $A$--$B$ and $B$--$A$ cuts realize the two boundary sectors within a single state. In the temporal protocol, we instead hold the entanglement cut fixed and compare states at successive spatial transfer depths $L$.

The transverse transfer operator preserves $U(1)_z$, and each application of $\mathbb T_x$ advances the original spin system by one spatial lattice site. Thus $L$ in Eq.~\eqref{eq:temporal_state_def} counts one-site transfer steps. Each step shifts the relative Dirichlet modulus by $\pi R$ according to Eq.~\eqref{eq:translation_dirichlet}; after $L$ steps, the accumulated displacement is
\begin{equation}
    \Delta\varphi_L
    =\Delta\varphi_0+L\pi R
    \pmod{2\pi R}.
    \label{eq:delta_phi_pump}
\end{equation}
The factor $L$ counts the repeated half-period shifts. Compactness reduces the accumulated displacement to $\Delta\varphi_0$ for even $L$ and $\Delta\varphi_0+\pi R$ for odd $L$. The orientation convention fixes the sign of the shift, which leaves the spectrum unchanged. Choosing a reference termination with $\Delta\varphi_0=0$, we write
\begin{equation}
    q=w+\frac{\Delta\varphi_L}{2\pi R},
    \qquad
    \gamma(g)=\frac{R^2}{2}
    =\frac{\arccos(-g)}{\pi}.
    \label{eq:gamma_xxz}
\end{equation}
The quantum number $q$ includes the boundary offset in the $U(1)$-resolved open-channel zero mode. The winding label $w$ remains integral, while the offset shifts the allowed values of $q$. The entanglement energies then follow
\begin{equation}
    \begin{aligned}
        \Delta_{\mathrm{even}}[q;N]
        &=\gamma(g)q^2+N,
        & q&\in\mathbb Z,
        \\
        \Delta_{\mathrm{odd}}[q;N]
        &=\gamma(g)q^2+N,
        & q&\in\mathbb Z+\frac12 .
        \label{eq:xxz_even_odd_spectra}
    \end{aligned}
\end{equation}
For the temporal state, the even and odd labels refer to the parity of $L$. For the spatial state, the same two spectra occur at the two inequivalent cuts at fixed $\tau$. Interchanging the two microscopic sublattices exchanges the parity labels while preserving this correspondence. The lowest primary is the unique $q=0$ state in the integer sector and the doublet $q=\pm1/2$ in the half-integer sector when spin flip is preserved. Each zero-mode sector carries integer-moded oscillator descendants with partition-number multiplicities. The correspondence concerns the universal rescaled levels; the nonuniversal scale and additive shift of the entanglement spectrum depend on the state.

\begin{figure}[t]
    \centering
    \includegraphics[width=\linewidth]{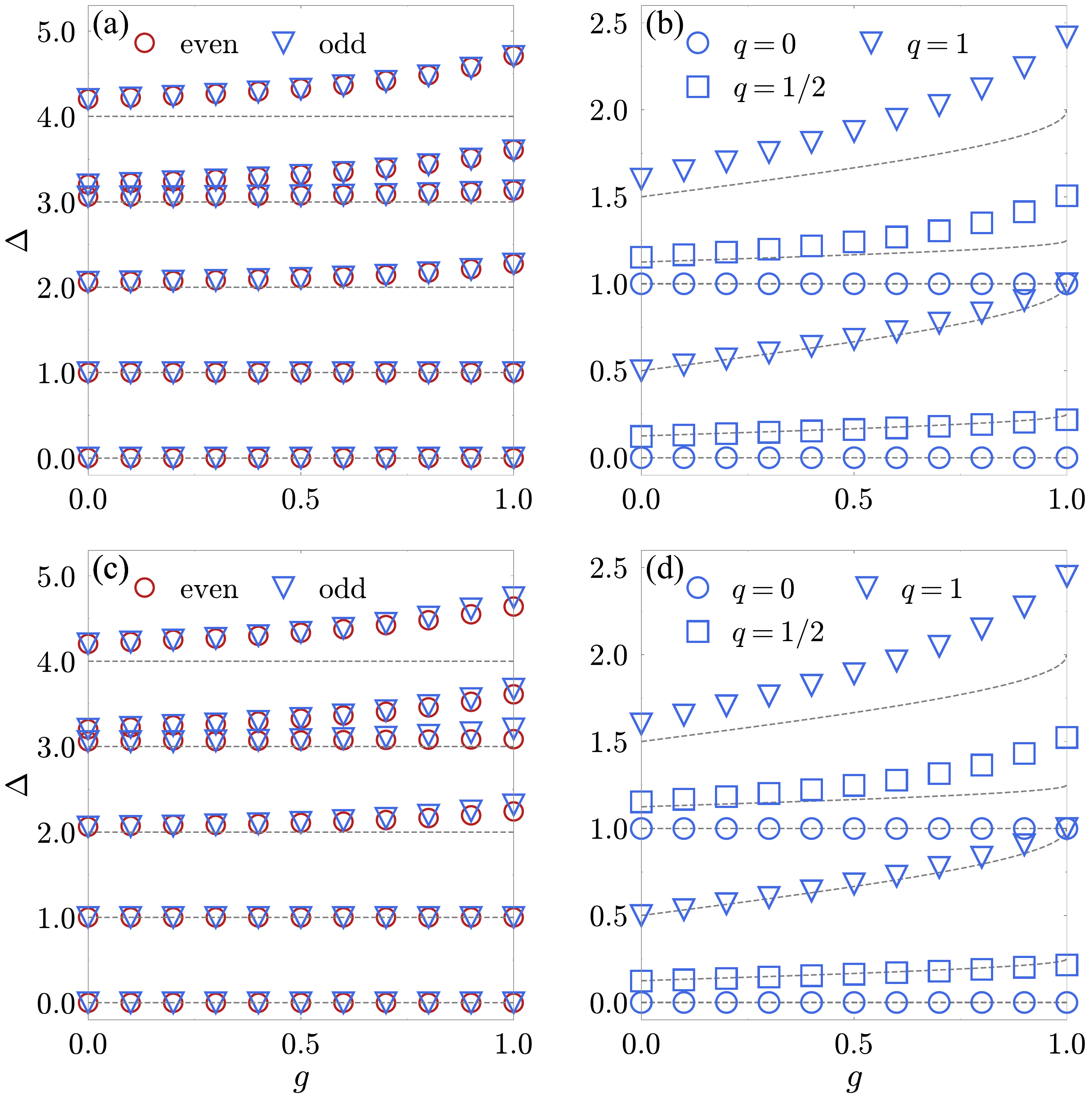}
    \caption{Spatial (upper) and temporal (lower) entanglement spectrum for the XXZ chain, Eq.~\eqref{eq:xxz_hamiltonian}. For the $U(1)_z$-breaking preparation in (a) and (c), the mixed $N_\varphi D_\varphi$ spectrum is radius independent and has no corresponding parity alternation. With $U(1)_z$ and spin flip preserved in the initial state in (b) and (d), the two inequivalent cuts of a single spatial state and a fixed cut of the temporal states at $L=199$ and $L=200$ resolve the same pair of $DD$ spectra in Eq.~\eqref{eq:xxz_even_odd_spectra}. Spatial odd and even labels specify cut positions within one state, whereas $L$ counts the one-site spatial transfer steps generating distinct temporal states. Each temporal state remains translation invariant along Euclidean time. Dashed lines are BCFT predictions. Deviations close to $g=1$ reflect logarithmic corrections near the $\mathrm{SU}(2)_1$ point. For the mixed $DN$ case in (a) and (c), we shifted the vaccum energy to be zero and normalized energy levels with the first energy gap for simplicity. }
    \label{fig:xxz_temporal}
\end{figure}

Figure~\ref{fig:xxz_temporal} compares these two realizations of the LSM constraint throughout the Gaussian regime. For the symmetry-preserving spatial preparation, the two cuts of $\ket{\Psi_{\mathrm{sp}}(\tau)}$ resolve the integer and half-integer $DD$ sectors. The temporal spectra reproduce these two tower structures at a fixed cut in $\ket{\Psi_{\mathrm{tm}}(L)}$ and $\ket{\Psi_{\mathrm{tm}}(L+1)}$. Thus the spatial comparison is between cuts within one state, while the temporal comparison is between states at consecutive transfer steps. %Both temporal parities have the universal entropy slope $c/6=1/6$. %Their difference decays algebraically with a logarithmic prefactor. The zero-mode partition functions give
%\begin{equation}
%    S_{\mathrm{odd}}-S_{\mathrm{even}} \sim(\log L)L^{-1/R^2} =(\log L)L^{-K},
%\label{eq:entropy_parity_asymptotic}
%\end{equation}
%up to a convention-dependent sign, nonuniversal shifts of $L$, and additional logarithmic corrections near $g=1$. Appendix~\ref{app:entropy_parity} derives this result and compares it with the numerical fits.

The temporal uniformity and the parity alternation concern different operations. At each fixed $L$, a two-site temporal \iiMPS{} optimization gives identical Schmidt values on its two temporal bonds within numerical accuracy, consistent with translation invariance along Euclidean time. Advancing the number of \emph{spatial transfer steps} by one exchanges the two $DD$ sectors while preserving this temporal uniformity. The even and odd values of $L$ therefore define two interlaced sequences of evolving states. After two steps the boundary sector returns, while the state and its entanglement scale continue to evolve. The LSM constraint is therefore visible as spatial dimerization or temporal sector pumping, according to the microscopic Hilbert-space factorization.

Breaking $U(1)_z$ in the microscopic initial state removes the restriction in Eq.~\eqref{eq:u1_selects_dirichlet}. The initial boundary may then be described at long distances by $B_O=N_\varphi(\vartheta_0)$, while the symmetry-preserving entanglement boundary remains $B_E=D_\varphi$. For the symmetry-breaking preparation used here, the low-lying spectrum identifies this mixed boundary condition and follows
\begin{equation}
    \widetilde E_{ND}[k]
    =\frac1{16}+\frac{k}{2},
    \qquad
    k=0,1,2,\ldots ,
    \label{eq:xxz_mixed_spectrum}
\end{equation}
with degeneracies $1,1,1,2,2,3,\ldots$. Its $h=1/16$ ground level corresponds to the lowest $N_\varphi\leftrightarrow D_\varphi$ boundary-condition-changing operator described in Sec.~\ref{sec:compact_boson}. The full spectrum forms a half-integer-moded current module containing multiple Virasoro families. The spatial and temporal state then gives the same entanglement spectrum structure as shown in Fig.~\ref{fig:xxz_temporal}. This was verified in With no current zero mode, the spectrum is independent of $R$ and has no odd--even alternation.

A transverse boundary perturbation can realize the same change of conformal boundary in a conventional boundary-Hamiltonian realization. Here, as in the Ising and Potts calculations, we realize the two boundary classes by choosing symmetry-preserving and symmetry-breaking initial states. The enhanced $\mathrm{SU}(2)$ symmetry and marginal logarithmic corrections at $g=1$ require a separate analysis, so the quantitative mixed-boundary identification is restricted to the Gaussian regime.

The mixed-boundary spectrum also appears at the unconstrained \iiMPS{} fixed point for $\ell \gtrsim \xi$ with a finite bond dimension $D$. We interpret this as evidence that the finite-entanglement variational state selects a $U(1)_z$-breaking deformation. When evolving at large steps, there is crossover between different BCFT entanglement spectrum. The role of symmetry in selecting the finite-entanglement boundary is discussed in the next section.

%Table~\ref{tab:spatial_temporal_comparison} summarizes how the same boundary symmetry action constrains the two entanglement spectra. The bulk CFT and the pair of $DD$ sectors are shared by both quantizations. In the spatial state, the two-site preparation places the sectors at inequivalent cuts within one state. In the uniform temporal state, successive spatial transfer steps exchange the sectors at a fixed cut. The LSM constraint is therefore visible as spatial dimerization or temporal sector pumping, according to the microscopic Hilbert-space factorization.

\section{Finite-length to finite-entanglement crossover}
\label{sec:crossover}

Finite bond dimension introduces a second infrared scale and therefore a natural continuation of the finite-length analysis. At fixed $D$, early Euclidean evolution follows the critical finite-length regime. As the evolution length grows, the finite-$D$ variational manifold eventually becomes the active infrared regulator and the trajectory crosses into the finite-entanglement regime. At every step we apply the evolution operator and compress the result back to the finite-$D$ MPS manifold,
\begin{equation}
    \ket{\Psi_D(\ell+\delta\ell)}
    \propto
    \mathcal P_D\mathcal T_{\delta\ell}
    \ket{\Psi_D(\ell)},
    \qquad
    S[\Psi_D]\leq \log D,
    \label{eq:fixed_D_evolution_crossover}
\end{equation}
where $\ell=\tau$ for the spatial state and $\ell=L$ for the temporal state. As long as the critical entanglement requirement $S(\ell)\sim(c/6)\log\ell$ remains well below the finite-$D$ capacity, the projection $\mathcal P_D$ is subleading and the evolution follows the critical trajectory discussed above. The initial-state boundary $B_O$ remains visible, and the low-lying entanglement spectrum is governed by $\mathcal H_{B_OB_E}$.

Once the required critical entanglement approaches the capacity of the fixed-$D$ ansatz, the exact state continues to build entanglement while the MPS remains bounded by Eq.~\eqref{eq:fixed_D_evolution_crossover}. The compression then affects the long-distance state. Repeated evolution and compression approach a fixed point inside the finite-$D$ MPS manifold,
\begin{equation}
    \ket{\Psi_D^\ast}
    \propto
    \mathcal P_D\mathcal T
    \ket{\Psi_D^\ast}.
    \label{eq:finite_D_fixed_point}
\end{equation}
For ordinary imaginary-time evolution this is the same finite-$D$ variational fixed point targeted by an MPS ground-state optimization. In the temporal quantization it is the corresponding finite-$D$ dominant transfer state. Variational optimization within the finite-$D$ MPS manifold reaches this long-evolution fixed point directly rather than by iterating all intermediate steps~\cite{zauner_2018_VUMPS,vanderstraeten_2019_tangnet}. Finite-entanglement scaling thus arises as the large-$\ell$ limit of the same Euclidean evolution restricted to a fixed finite-$D$ variational manifold.

The finite-$D$ fixed point has a finite correlation length $\xi$~\cite{tagliacozzo_2008_fes,frank_2009_ee_scaling,pirvu_2012_finite_size_entanglement,stojevic_2015_fes}. In the finite-entanglement scaling regime, its long-distance effect can be modeled as a weak relevant deformation of the critical theory~\cite{Cho2017,huang_2024_emergent_bc},
\begin{equation}
    H_D
    \simeq
    H_{\mathrm{CFT}}
    +\sum_a u_a\,\xi^{\Delta_a-2}
      \int dy\,\Phi_a(y)+\cdots,
    \label{eq:finite_entanglement_deformation}
\end{equation}
where $u_a$ are dimensionless nonuniversal amplitudes, and $\Delta_a<2$. The leading nonzero relevant deformation becomes strong at the scale $\xi$ and selects the nearby massive phase approached by the finite-$D$ variational state.

An important point is that the allowed amplitudes $u_a$ are determined by the symmetry of the \emph{finite-$D$ MPS manifold}. If the evolution is compressed with generic, nonsymmetric MPS tensors~\footnote{Nonsymmetric MPS here means there is no symmetry constraint, rather than the MPS is a symmetry-breaking state.}, then the projection $\mathcal P_D$ is not restricted to the symmetry-preserving submanifold. Even when the microscopic Hamiltonian and initial state preserve an internal symmetry $G$, symmetry-breaking relevant deformations are therefore not excluded by the variational ansatz and can be selected by the long-evolution fixed point. At a conventional symmetry-breaking critical point the order-parameter field is often the most relevant such deformation, so an unconstrained MPS optimization can converge to an ordered finite-$D$ state and hence to a fixed, symmetry-breaking conformal boundary $B_D$. By contrast, if $G$ is imposed explicitly on the MPS tensors, all $G$-breaking amplitudes vanish and the leading $G$-invariant relevant deformation determines a different $B_D$. Thus the symmetry relevant for the finite-entanglement boundary is the symmetry enforced on the variational manifold, not merely the symmetry of the microscopic evolution.

\begin{figure}[t]
    \centering
    \includegraphics[width=0.75\linewidth]{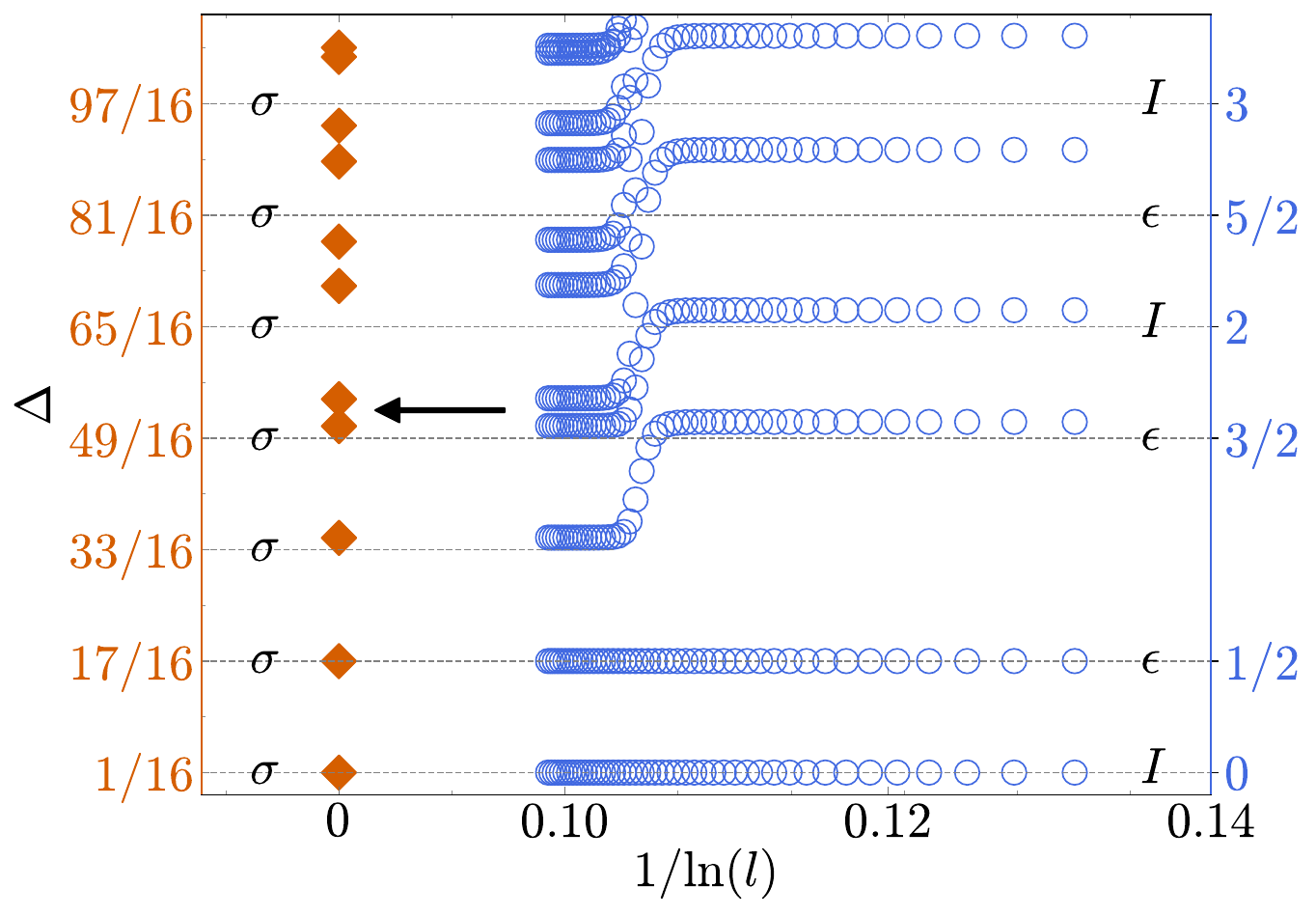}
    \caption{Crossover from free-free BCFT to free-fixed BCFT in the temporal entanglement of the Eulidean-time evolved state for the critical Ising model Eq.~\ref{eq:ising_hamiltonian}. The largest bond dimension for the MPS is $D=20$ and a symmetric initial state is used. The $1/\ln(\ell) = 0$ data is from the leading eigen state of the transfer operator via the variational calculation.  } \label{fig:ising_cross_over}
\end{figure}

The conformal-perturbation-theory form of the entanglement Hamiltonian makes clear why the finite-entanglement regime is not directly a BCFT. Using the same convention as Eq.~\eqref{eq:entanglement_hamiltonian_def}, $\rho=Z^{-1}\exp(-2\pi\mathcal H_E)$, perform the logarithmic map $w=\log(z/\epsilon)=x+\ii\tau$ for a half-space reduced density matrix. The perturbed CFT action becomes
\begin{align}
    S_{\mathrm R}
    =S_\ast
    +\sum_a u_a
      \int_0^\infty dx\int_0^{2\pi}d\tau\,
      \e^{(2-\Delta_a)(x-x_D)}
      \Phi_a(x,\tau)+\cdots,
    \label{eq:fes_rindler_action}
\end{align}
where $x_D=\log\!\left(\frac{\xi}{\epsilon}\right)$. Since the coefficient is independent of the Rindler time $\tau$, the corresponding entanglement Hamiltonian is
\begin{equation}
    \mathcal H_E^{(D)}
    =\mathcal H_{E,\ast}
    +\sum_a u_a\int_0^\infty dx\,
      \e^{(2-\Delta_a)(x-x_D)}
      \Phi_a(x)+\cdots .
    \label{eq:fes_entanglement_hamiltonian}
\end{equation}
The position-dependent coefficient in Eq.~\eqref{eq:fes_entanglement_hamiltonian} is essential. The entanglement Hamiltonian still contains an inhomogeneous relevant \emph{bulk} perturbation, so it is not itself a BCFT Hamiltonian after the conformal map. For $x\ll x_D$ the perturbation is weak and the theory is approximately critical, whereas near $x\sim x_D$ it becomes of order unity and flows rapidly to strong coupling. Only at energies below this crossover can the strong-coupling region be replaced by the conformal boundary $B_D$ selected by the massive phase. The entanglement boundary $B_E$ remains at the ultraviolet end.

Consequently the BCFT description in the finite-entanglement regime is an infrared approximation,
\begin{equation}
    \mathcal H_E^{(D)}
    \overset{\mathrm{IR}}{\simeq}
    \frac{\pi}{W_D}
    \left(
      L_0^{(B_DB_E)}-\frac{c}{24}
    \right)
    +\mathrm{const.}
    +\delta\mathcal H_D,
    \label{eq:fes_bcft_hamiltonian}
\end{equation}
where $W_D\sim\log\xi$ and $\delta\mathcal H_D$ denotes corrections from the smooth strong-coupling crossover and subleading perturbations. This differs conceptually from the finite-length construction in Sec.~\ref{sec:bcft_framework}, in which the bulk path integral remains critical and the annulus BCFT follows directly from a conformal map, whereas here the second boundary $B_D$ emerges only as a low-energy description of a nonconformal finite-entanglement deformation.

The crossover is therefore controlled by which infrared scale is encountered first:
\begin{align}
    1\ll\ell\ll\xi &: \quad
    \mathcal H_E\simeq\mathcal H_{B_OB_E},
    \nonumber\\
    \ell\sim\xi &: \quad
    \text{finite-length/finite-entanglement crossover},
    \nonumber\\
    \ell\gg\xi &: \quad
    \mathcal H_{E,\mathrm{low}}^{(D)}
    \simeq\mathcal H_{B_DB_E}.
    \label{eq:three_regimes}
\end{align}
Around $\ell\sim\xi$ the full entanglement Hamiltonian interpolates between the two limits and need not coincide with a BCFT fixed-point Hamiltonian.

The Ising data in Fig.~\ref{fig:ising_cross_over} make both the crossover and the role of the MPS symmetry explicit. Starting from the symmetric initial state, the finite-length regime realizes the free--free BCFT. If the fixed-$D$ evolution is performed with generic nonsymmetric tensors, the Ising spin field is allowed as a finite-entanglement deformation; the variational fixed point can therefore break $\mathbb Z_2$ and select a fixed boundary $B_D=I$ or $\epsilon$. This gives the free--free to fixed--free crossover. If the MPS tensors are instead constrained to be $\mathbb Z_2$ symmetric, the spin deformation is forbidden; the leading energy deformation preserves the symmetry and selects the free physical boundary. The same microscopic critical evolution can therefore reach different finite-entanglement boundaries depending on the symmetry built into the MPS manifold.

Similar analysis also applies to the Potts and XXZ chain. The long-evolution states is controlled by the finite entanglement scaling regime. It can be summarized as finite-entanglement selection rule controlled by the symmetry of the variational manifold. In the unconstrained finite-$D$ calculation, the physical boundary is determined by the most relevant operator consistent with the lattice theory and there is usually crossover between different BCFT entanglement structures if one starts with a symmetric initital state.

%The XXZ result has the same interpretation. In the unconstrained finite-$D$ calculation, $U(1)_z$-breaking perturbations are allowed and the long-evolution fixed point has the radius-independent mixed-boundary spectrum, consistent with an effective $N_\varphi D_\varphi$ BCFT at low energies. Enforcing $U(1)_z$ on the MPS tensors removes this deformation and can select a different symmetry-allowed $B_D$. This finite-entanglement selection rule is controlled by the symmetry of the variational manifold. It is distinct from the microscopic LSM constraint discussed in Sec.~\ref{sec:xxz}, which ties $U(1)_z$ to one-site translation already at the lattice level.

\section{Conclusions}
\label{sec:conclusions}

We have studied how microscopic lattice symmetry shapes the infrared BCFT seen in the entanglement of quantum-critical lattice models. Our basic probes are spatial and temporal Euclidean-evolution states constructed from the same critical lattice model. Within the finite-length scaling regime, the short-range preparation boundary is described by a conformal boundary at large length scales, and the annulus representation of the reduced density matrix turns the universal low-lying entanglement spectrum into boundary spectroscopy. This gives a common continuum framework in which the microscopic Hilbert-space factorization, symmetry realization, and anomaly data enter through restrictions on accessible conformal boundaries.

A first technical result is a regular lattice construction of the temporal transfer representation. The temporal translation invariance enforce the temporal state to be uniform and the transfer protocol in a split-gate basis. Under a naive $90^\circ$ reshuffling, the identity part of an infinitesimal physical gate becomes a rank-one maximally entangled projector and dominates the local transverse Schmidt structure. We remove this rank deficiency by factorizing each physical gate and regrouping neighboring half-gates before forming the one-dimensional transfer problem. In the resulting local basis, the trivial rank-one structure is absorbed into local basis and normalization factors while the nontrivial transfer dynamics remains well conditioned. The transverse evolution is then represented by a uniform \iiMPO{} acting on a uniform temporal \iiMPS{}, with the internal symmetry inherited through the composite representation $u_g\otimes\bar u_g$.

The Ising and three-state Potts chains provide controlled benchmarks for this construction. Their relevant microscopic symmetries are onsite, so the two lattice quantizations inherit equivalent symmetry constraints and approach the same BCFT data. Standard boundary classifications predict the allowed tower contents, and changing the symmetry of the microscopic initial state can realize different long-distance conformal boundaries. The Potts example also demonstrates why symmetry and spectroscopy play complementary roles. Fixed and mixed boundaries share the same residual $\mathbb Z_2$, while free and new both preserve the full $S_3$, yet their BCFT spectra distinguish them. These models therefore validate the lattice framework and the boundary-engineering protocol against known BCFT structure.

The central physical result is the compact-boson boundary restriction imposed by the LSM anomaly. Bosonization fixes the physical translation action to be $T_1:V_{e,w}\mapsto(-1)^{e+w}V_{e,w}$. On a $U(1)_z$-preserving elementary boundary, this action exchanges two Dirichlet values separated by half a compactification period. The allowed elementary boundary therefore belongs to a two-element translation orbit, providing a direct BCFT encoding of the microscopic LSM constraint.

The two lattice quantizations realize this boundary orbit in complementary ways. A spatial slice consists of the physical spin-$1/2$ sites, where the LSM constraint obstructs a one-site injective, $U(1)_z$-preserving short-range state; our preparation consequently has a doubled unit cell and two inequivalent entanglement cuts. A temporal slice consists of composite transfer degrees of freedom, naturally of the form $\mathcal H\otimes\mathcal H^*$ before the local basis transformation, with $U(1)$ represented by a charge difference. The temporal \iiMPS{} can therefore remain uniform along Euclidean time, while the original spatial translation acts between successive evolved states and pumps the entanglement spectrum through the two allowed boundary sectors,
\begin{equation}
    q\in\mathbb Z
    \quad\longleftrightarrow\quad
    q\in\mathbb Z+\frac12.
\end{equation}
The doubled spatial unit cell and the temporal odd--even sector pumping are thus complementary lattice signatures of the same infrared boundary constraint. The spatial one is encoded in the short-range spatial factorization, while the other appears as evolution between translation-related temporal boundary sectors.

The same Euclidean-evolution framework also connects smoothly to finite-entanglement scaling. Within the finite-length window, the bulk path integral remains critical and the BCFT entanglement Hamiltonian follows directly from the conformal map. Once $\ell\gtrsim\xi$, finite-$D$ compression acts as a relevant deformation. The logarithmic map then produces a position-dependent bulk coupling, and a BCFT with boundary $B_D$ emerges only as the low-energy description after this deformation reaches strong coupling. The symmetry imposed on the finite-$D$ variational manifold determines which deformations, and hence which $B_D$, are available.

Taken together, the two main additions are the regular transverse lattice formulation of the temporal state and the explicit XXZ identification of the LSM-induced restriction on compact-boson boundary sectors, including its spatial and temporal lattice realizations. The analysis combines tensor-network calculations with BCFT, bosonization, and lattice--CFT symmetry matching. Two directions invite a more direct analytical treatment: deriving the transverse Hilbert-space structure and its regularization from the lattice transfer problem, and formulating the anomaly-induced boundary restriction together with its spatial and temporal realizations without using the numerical entanglement spectra as an identifying step. Progress along these lines could clarify which parts of the lattice-to-BCFT correspondence are model dependent and which follow generally from symmetry and anomaly data.

More broadly, the framework connects two complementary perspectives on quantum criticality. From the condensed-matter side, the microscopic symmetry realization and anomaly data are part of the physical specification of a critical lattice model and can restrict symmetry-preserving boundary conditions and variational states beyond the bulk universality class. From the field-theory side, BCFT provides a natural language for expressing these restrictions as constraints on accessible conformal boundaries and boundary operator contents. In the XXZ chain, one-site translation acts nontrivially on the compact-boson boundary orbit, while spatial and temporal entanglement reveal two lattice realizations of that same structure. A more systematic development of this lattice--BCFT viewpoint may therefore help connect microscopic studies of quantum criticality with field-theoretic descriptions of symmetry, anomalies, and conformal boundaries.

\section{Acknowledgments}
We thank Long Zhang for helpful discussions. R.Z.H is supported by the National Natural Science Foundation of China (Grant No.~12504183). X.H.Y is supported by the Innovation Program for Quantum Science and Technology (Grant No.~2021ZD0301900) and the National Natural Science Foundation of China (Grant No.~12504175).

\bibliographystyle{apsrev4-2}
\bibliography{reference}

@misc{tirrito_2018_temporal_mps,
      title={Characterizing the quantum field theory vacuum using temporal Matrix Product states}, 
      author={Emanuele Tirrito and Neil J. Robinson and Maciej Lewenstein and Shi-Ju Ran and Luca Tagliacozzo},
      year={2022},
      eprint={1810.08050},
      archivePrefix={arXiv},
      primaryClass={cond-mat.stat-mech},
      url={https://arxiv.org/abs/1810.08050}, 
}

@article{affleck_1998_bc,
  title = {Boundary Critical Phenomena in the Three-State {{Potts}} Model},
  author = {Affleck, Ian and Oshikawa, Masaki and Saleur, Hubert},
  year = 1998,
  month = jul,
  journal = {Journal of Physics A: Mathematical and General},
  volume = {31},
  number = {28},
  pages = {5827--5842},
  issn = {0305-4470, 1361-6447},
  doi = {10.1088/0305-4470/31/28/003},
  urldate = {2026-08-28}
}

@article{affleck_gepner_1989_spin_chains,
  title = {Critical Behaviour of Spin-s {{Heisenberg}} Antiferromagnetic Chains: Analytic and Numerical Results},
  shorttitle = {Critical Behaviour of Spin-s {{Heisenberg}} Antiferromagnetic Chains},
  author = {Affleck, I and Gepner, D and Schulz, H J and Ziman, T},
  year = 1989,
  month = mar,
  journal = {Journal of Physics A: Mathematical and General},
  volume = {22},
  number = {5},
  pages = {511--529},
  issn = {0305-4470, 1361-6447},
  doi = {10.1088/0305-4470/22/5/015},
  urldate = {2026-08-28}
}

@article{affleck_ludwig_1991_boundary_entropy,
  title = {Universal Noninteger ``Ground-State Degeneracy'' in Critical Quantum Systems},
  author = {Affleck, Ian and Ludwig, Andreas W. W.},
  year = 1991,
  month = jul,
  journal = {Physical Review Letters},
  volume = {67},
  number = {2},
  pages = {161--164},
  issn = {0031-9007},
  doi = {10.1103/PhysRevLett.67.161},
  urldate = {2026-08-28},
  copyright = {http://link.aps.org/licenses/aps-default-license},
  langid = {english}
}

@article{banuls_2009_transverse,
  title = {Matrix {{Product States}} for {{Dynamical Simulation}} of {{Infinite Chains}}},
  author = {Ba{\~n}uls, M. C. and Hastings, M. B. and Verstraete, F. and Cirac, J. I.},
  year = 2009,
  month = jun,
  journal = {Physical Review Letters},
  volume = {102},
  number = {24},
  pages = {240603},
  issn = {0031-9007, 1079-7114},
  doi = {10.1103/PhysRevLett.102.240603},
  urldate = {2026-08-28},
  copyright = {http://link.aps.org/licenses/aps-default-license},
  langid = {english}
}

@article{behrend_2000_rcft_boundaries,
  title = {Boundary Conditions in Rational Conformal Field Theories},
  author = {Behrend, Roger E. and Pearce, Paul A. and Petkova, Valentina B. and Zuber, Jean-Bernard},
  year = 2000,
  month = mar,
  journal = {Nuclear Physics B},
  volume = {570},
  number = {3},
  pages = {525--589},
  issn = {05503213},
  doi = {10.1016/S0550-3213(99)00592-1},
  urldate = {2026-08-28},
  copyright = {https://www.elsevier.com/tdm/userlicense/1.0/},
  langid = {english}
}

@article{betsuyaku_1985_transfer_matrix_2,
  title = {Study of {{One-Dimensional Quantum Spin Systems}} by the {{Transfer-Matrix Method}}},
  author = {Betsuyaku, H.},
  year = 1985,
  month = feb,
  journal = {Progress of Theoretical Physics},
  volume = {73},
  number = {2},
  pages = {319--331},
  issn = {0033-068X, 1347-4081},
  doi = {10.1143/PTP.73.319},
  urldate = {2026-08-28},
  langid = {english}
}

@article{calabrese_lefevre_2008_entanglement_spectrum,
  title = {Entanglement Spectrum in One-Dimensional Systems},
  author = {Calabrese, Pasquale and Lefevre, Alexandre},
  year = 2008,
  month = sep,
  journal = {Physical Review A},
  volume = {78},
  number = {3},
  pages = {032329},
  issn = {1050-2947, 1094-1622},
  doi = {10.1103/PhysRevA.78.032329},
  urldate = {2026-08-28},
  copyright = {http://link.aps.org/licenses/aps-default-license},
  langid = {english}
}

@article{calabrese_cardy_2016_quench_review,
  title = {Quantum Quenches in 1 + 1 Dimensional Conformal Field Theories},
  author = {Calabrese, Pasquale and Cardy, John},
  year = 2016,
  month = jun,
  journal = {Journal of Statistical Mechanics: Theory and Experiment},
  volume = {2016},
  number = {6},
  pages = {064003},
  issn = {1742-5468},
  doi = {10.1088/1742-5468/2016/06/064003},
  urldate = {2026-08-28}
}

@article{calabrese_cardy_2007_quenches,
  title = {Quantum Quenches in Extended Systems},
  author = {Calabrese, Pasquale and Cardy, John},
  year = 2007,
  month = jun,
  journal = {Journal of Statistical Mechanics: Theory and Experiment},
  volume = {2007},
  number = {06},
  pages = {P06008-P06008},
  issn = {1742-5468},
  doi = {10.1088/1742-5468/2007/06/P06008},
  urldate = {2026-08-28}
}

@article{calabrese_cardy_2006_quench,
  title = {Time {{Dependence}} of {{Correlation Functions Following}} a {{Quantum Quench}}},
  author = {Calabrese, Pasquale and Cardy, John},
  year = 2006,
  month = apr,
  journal = {Physical Review Letters},
  volume = {96},
  number = {13},
  pages = {136801},
  issn = {0031-9007, 1079-7114},
  doi = {10.1103/PhysRevLett.96.136801},
  urldate = {2026-08-28},
  copyright = {http://link.aps.org/licenses/aps-default-license},
  langid = {english}
}

@article{cardy_1989_bc_fusion,
  title = {Boundary Conditions, Fusion Rules and the {{Verlinde}} Formula},
  author = {Cardy, John L.},
  year = 1989,
  month = oct,
  journal = {Nuclear Physics B},
  volume = {324},
  number = {3},
  pages = {581--596},
  issn = {05503213},
  doi = {10.1016/0550-3213(89)90521-X},
  urldate = {2026-08-28},
  copyright = {https://www.elsevier.com/tdm/userlicense/1.0/},
  langid = {english}
}

@incollection{cardy_2008_bcft,
  title = {Boundary {{Conformal Field Theory}}},
  booktitle = {Encyclopedia of {{Mathematical Physics}}},
  author = {Cardy, J.},
  year = 2006,
  pages = {333--340},
  publisher = {Elsevier},
  doi = {10.1016/B0-12-512666-2/00398-9},
  urldate = {2026-08-28},
  copyright = {https://www.elsevier.com/tdm/userlicense/1.0/},
  isbn = {978-0-12-512666-3},
  langid = {english}
}

@article{cardy_1984_surface,
  title = {Conformal Invariance and Surface Critical Behavior},
  author = {Cardy, John L.},
  year = 1984,
  month = nov,
  journal = {Nuclear Physics B},
  volume = {240},
  number = {4},
  pages = {514--532},
  issn = {05503213},
  doi = {10.1016/0550-3213(84)90241-4},
  urldate = {2026-08-28},
  copyright = {https://www.elsevier.com/tdm/userlicense/1.0/},
  langid = {english}
}

@article{cardy_1986_boundary_operator,
  title = {Effect of Boundary Conditions on the Operator Content of Two-Dimensional Conformally Invariant Theories},
  author = {Cardy, John L.},
  year = 1986,
  month = oct,
  journal = {Nuclear Physics B},
  volume = {275},
  number = {2},
  pages = {200--218},
  issn = {05503213},
  doi = {10.1016/0550-3213(86)90596-1},
  urldate = {2026-08-28},
  copyright = {https://www.elsevier.com/tdm/userlicense/1.0/},
  langid = {english}
}

@article{Cardy_2016_EntanglementHamiltonian,
  title = {Entanglement {{Hamiltonians}} in Two-Dimensional Conformal Field Theory},
  author = {Cardy, John and Tonni, Erik},
  year = 2016,
  month = dec,
  journal = {Journal of Statistical Mechanics: Theory and Experiment},
  volume = {2016},
  number = {12},
  pages = {123103},
  issn = {1742-5468},
  doi = {10.1088/1742-5468/2016/12/123103},
  urldate = {2026-08-28}
}

@article{carignano_tagliacozzo_2025_temporal_cft,
  title = {Loschmidt Echo, Emerging Dual Unitarity and Scaling of Generalized Temporal Entropies after Quenches to the Critical Point},
  author = {Carignano, Stefano and Tagliacozzo, Luca},
  year = 2025,
  month = sep,
  journal = {Quantum},
  volume = {9},
  pages = {1859},
  issn = {2521-327X},
  doi = {10.22331/q-2025-09-16-1859},
  urldate = {2026-08-28},
  langid = {english}
}

@article{carignano_2024_temporal_entropy,
  title = {Temporal Entropy and the Complexity of Computing the Expectation Value of Local Operators after a Quench},
  author = {Carignano, Stefano and Marim{\'o}n, Carlos Ramos and Tagliacozzo, Luca},
  year = 2024,
  month = jul,
  journal = {Physical Review Research},
  volume = {6},
  number = {3},
  pages = {033021},
  issn = {2643-1564},
  doi = {10.1103/PhysRevResearch.6.033021},
  urldate = {2026-08-28},
  langid = {english}
}

@article{cheng_2023_lsm,
  title = {Lieb-{{Schultz-Mattis}}, {{Luttinger}}, and 't {{Hooft}} - Anomaly Matching in Lattice Systems},
  author = {Cheng, Meng and Seiberg, Nathan},
  year = 2023,
  month = aug,
  journal = {SciPost Physics},
  volume = {15},
  number = {2},
  pages = {051},
  issn = {2542-4653},
  doi = {10.21468/SciPostPhys.15.2.051},
  urldate = {2026-08-28}
}

@article{cho_2017_universal_ES,
  title = {Universal Entanglement Spectra of Gapped One-Dimensional Field Theories},
  author = {Cho, Gil Young and Ludwig, Andreas W. W. and Ryu, Shinsei},
  year = 2017,
  month = mar,
  journal = {Physical Review B},
  volume = {95},
  number = {11},
  pages = {115122},
  issn = {2469-9950, 2469-9969},
  doi = {10.1103/PhysRevB.95.115122},
  urldate = {2026-08-28},
  copyright = {http://link.aps.org/licenses/aps-default-license},
  langid = {english}
}

@article{cirac_2021_mps_review,
  title = {Matrix Product States and Projected Entangled Pair States: {{Concepts}}, Symmetries, Theorems},
  shorttitle = {Matrix Product States and Projected Entangled Pair States},
  author = {Cirac, J. Ignacio and {P{\'e}rez-Garc{\'i}a}, David and Schuch, Norbert and Verstraete, Frank},
  year = 2021,
  month = dec,
  journal = {Reviews of Modern Physics},
  volume = {93},
  number = {4},
  pages = {045003},
  issn = {0034-6861, 1539-0756},
  doi = {10.1103/RevModPhys.93.045003},
  urldate = {2026-08-28},
  langid = {english}
}

@book{difrancesco_1997_CFT_book,
  title = {Conformal {{Field Theory}}},
  author = {Di Francesco, Philippe and Mathieu, Pierre and S{\'e}n{\'e}chal, David},
  year = 1997,
  series = {Graduate {{Texts}} in {{Contemporary Physics}}},
  publisher = {Springer New York},
  address = {New York, NY},
  doi = {10.1007/978-1-4612-2256-9},
  urldate = {2026-08-28},
  copyright = {https://www.springer.com/tdm},
  isbn = {978-1-4612-7475-9 978-1-4612-2256-9},
  langid = {english}
}

@article{else_thorngren_2020_lsm,
  title = {Topological Theory of {{Lieb-Schultz-Mattis}} Theorems in Quantum Spin Systems},
  author = {Else, Dominic V. and Thorngren, Ryan},
  year = 2020,
  month = jun,
  journal = {Physical Review B},
  volume = {101},
  number = {22},
  pages = {224437},
  issn = {2469-9950, 2469-9969},
  doi = {10.1103/PhysRevB.101.224437},
  urldate = {2026-08-28},
  langid = {english}
}

@article{fateev_zamolodchikov_1987_z3,
  title = {Conformal Quantum Field Theory Models in Two Dimensions Having {{Z3}} Symmetry},
  author = {Fateev, V.A. and Zamolodchikov, A.B.},
  year = 1987,
  journal = {Nuclear Physics B},
  volume = {280},
  pages = {644--660},
  issn = {05503213},
  doi = {10.1016/0550-3213(87)90166-0},
  urldate = {2026-08-28},
  langid = {english}
}

@article{foligno_2023_temporal_entanglement,
  title = {Temporal {{Entanglement}} in {{Chaotic Quantum Circuits}}},
  author = {Foligno, Alessandro and Zhou, Tianci and Bertini, Bruno},
  year = 2023,
  month = oct,
  journal = {Physical Review X},
  volume = {13},
  number = {4},
  pages = {041008},
  issn = {2160-3308},
  doi = {10.1103/PhysRevX.13.041008},
  urldate = {2026-08-28},
  langid = {english}
}

@article{frohlich_2004_kw_defects,
  title = {Kramers-{{Wannier Duality}} from {{Conformal Defects}}},
  author = {Fr{\"o}hlich, J{\"u}rg and Fuchs, J{\"u}rgen and Runkel, Ingo and Schweigert, Christoph},
  year = 2004,
  month = aug,
  journal = {Physical Review Letters},
  volume = {93},
  number = {7},
  pages = {070601},
  issn = {0031-9007, 1079-7114},
  doi = {10.1103/PhysRevLett.93.070601},
  urldate = {2026-08-28},
  langid = {english}
}

@article{fuchs_1998_potts_boundaries,
  title = {Completeness of Boundary Conditions for the Critical Three-State {{Potts}} Model},
  author = {Fuchs, J{\"u}rgen and Schweigert, Christoph},
  year = 1998,
  month = nov,
  journal = {Physics Letters B},
  volume = {441},
  number = {1-4},
  pages = {141--146},
  issn = {03702693},
  doi = {10.1016/S0370-2693(98)01185-X},
  urldate = {2026-08-28},
  copyright = {https://www.elsevier.com/tdm/userlicense/1.0/},
  langid = {english}
}

@article{furuya_oshikawa_2017_global_anomaly,
  title = {Symmetry {{Protection}} of {{Critical Phases}} and a {{Global Anomaly}} in 1 + 1 {{Dimensions}}},
  author = {Furuya, Shunsuke C. and Oshikawa, Masaki},
  year = 2017,
  month = jan,
  journal = {Physical Review Letters},
  volume = {118},
  number = {2},
  pages = {021601},
  issn = {0031-9007, 1079-7114},
  doi = {10.1103/PhysRevLett.118.021601},
  urldate = {2026-08-28},
  copyright = {http://link.aps.org/licenses/aps-default-license},
  langid = {english}
}

@book{giamarchi_2003_q1d,
  title = {Quantum {{Physics}} in {{One Dimension}}},
  author = {Giamarchi, Thierry},
  year = 2003,
  month = dec,
  publisher = {Oxford University Press},
  doi = {10.1093/acprof:oso/9780198525004.001.0001},
  urldate = {2026-08-28},
  isbn = {978-0-19-852500-4}
}

@article{giudice_2022_temporal_entanglement,
  title = {Temporal {{Entanglement}}, {{Quasiparticles}}, and the {{Role}} of {{Interactions}}},
  author = {Giudice, Giacomo and Giudici, Giuliano and Sonner, Michael and Thoenniss, Julian and Lerose, Alessio and Abanin, Dmitry A. and Piroli, Lorenzo},
  year = 2022,
  month = jun,
  journal = {Physical Review Letters},
  volume = {128},
  number = {22},
  pages = {220401},
  issn = {0031-9007, 1079-7114},
  doi = {10.1103/PhysRevLett.128.220401},
  urldate = {2026-08-28},
  langid = {english}
}

@article{haegeman_2017_transfer_mpo,
  title = {Diagonalizing {{Transfer Matrices}} and {{Matrix Product Operators}}: {{A Medley}} of {{Exact}} and {{Computational Methods}}},
  shorttitle = {Diagonalizing {{Transfer Matrices}} and {{Matrix Product Operators}}},
  author = {Haegeman, Jutho and Verstraete, Frank},
  year = 2017,
  month = mar,
  journal = {Annual Review of Condensed Matter Physics},
  volume = {8},
  number = {1},
  pages = {355--406},
  issn = {1947-5454, 1947-5462},
  doi = {10.1146/annurev-conmatphys-031016-025507},
  urldate = {2026-08-28},
  langid = {english}
}

@article{haegeman_2016_tdvp,
  title = {Unifying Time Evolution and Optimization with Matrix Product States},
  author = {Haegeman, Jutho and Lubich, Christian and Oseledets, Ivan and Vandereycken, Bart and Verstraete, Frank},
  year = 2016,
  month = oct,
  journal = {Physical Review B},
  volume = {94},
  number = {16},
  pages = {165116},
  issn = {2469-9950, 2469-9969},
  doi = {10.1103/PhysRevB.94.165116},
  urldate = {2026-08-28},
  copyright = {http://link.aps.org/licenses/aps-default-license},
  langid = {english}
}

@article{hastings_2015_time_space,
  title = {Connecting Entanglement in Time and Space: {{Improving}} the Folding Algorithm},
  shorttitle = {Connecting Entanglement in Time and Space},
  author = {Hastings, M. B. and Mahajan, R.},
  year = 2015,
  month = mar,
  journal = {Physical Review A},
  volume = {91},
  number = {3},
  pages = {032306},
  issn = {1050-2947, 1094-1622},
  doi = {10.1103/PhysRevA.91.032306},
  urldate = {2026-08-28},
  copyright = {http://link.aps.org/licenses/aps-default-license},
  langid = {english}
}

@article{huang_2024_emergent_bc,
  title = {Emergent {{Conformal Boundaries}} from {{Finite-Entanglement Scaling}} in {{Matrix Product States}}},
  author = {Huang, Rui-Zhen and Zhang, Long and L{\"a}uchli, Andreas M. and Haegeman, Jutho and Verstraete, Frank and Vanderstraeten, Laurens},
  year = 2024,
  month = feb,
  journal = {Physical Review Letters},
  volume = {132},
  number = {8},
  pages = {086503},
  issn = {0031-9007, 1079-7114},
  doi = {10.1103/PhysRevLett.132.086503},
  urldate = {2026-08-28},
  langid = {english}
}

@article{ishibashi_1989_boundary_crosscap,
  title = {{{THE BOUNDARY AND CROSSCAP STATES IN CONFORMAL FIELD THEORIES}}},
  author = {Ishibashi, Nobuyuki},
  year = 1989,
  month = feb,
  journal = {Modern Physics Letters A},
  volume = {04},
  number = {03},
  pages = {251--264},
  issn = {0217-7323, 1793-6632},
  doi = {10.1142/S0217732389000320},
  urldate = {2026-08-28},
  langid = {english}
}

@misc{lauchli_2013_opContent,
  title = {Operator Content of Real-Space Entanglement Spectra at Conformal Critical Points},
  author = {L{\"a}uchli, Andreas M.},
  year = 2013,
  month = mar,
  number = {arXiv:1303.0741},
  eprint = {1303.0741},
  primaryclass = {cond-mat.stat-mech},
  publisher = {arXiv},
  doi = {10.48550/arXiv.1303.0741},
  urldate = {2026-08-28},
  archiveprefix = {arXiv}
}

@article{lerose_2023_spacetime_duality,
  title = {Overcoming the Entanglement Barrier in Quantum Many-Body Dynamics via Space-Time Duality},
  author = {Lerose, Alessio and Sonner, Michael and Abanin, Dmitry A.},
  year = 2023,
  month = feb,
  journal = {Physical Review B},
  volume = {107},
  number = {6},
  pages = {L060305},
  issn = {2469-9950, 2469-9969},
  doi = {10.1103/PhysRevB.107.L060305},
  urldate = {2026-08-28},
  langid = {english}
}

@article{lerose_2021_temporal_scaling,
  title = {Scaling of Temporal Entanglement in Proximity to Integrability},
  author = {Lerose, Alessio and Sonner, Michael and Abanin, Dmitry A.},
  year = 2021,
  month = jul,
  journal = {Physical Review B},
  volume = {104},
  number = {3},
  pages = {035137},
  issn = {2469-9950, 2469-9969},
  doi = {10.1103/PhysRevB.104.035137},
  urldate = {2026-08-28},
  langid = {english}
}

@article{Lieb_1961_LSM,
  title = {Two Soluble Models of an Antiferromagnetic Chain},
  author = {Lieb, Elliott and Schultz, Theodore and Mattis, Daniel},
  year = 1961,
  month = dec,
  journal = {Annals of Physics},
  volume = {16},
  number = {3},
  pages = {407--466},
  issn = {00034916},
  doi = {10.1016/0003-4916(61)90115-4},
  urldate = {2026-08-28},
  copyright = {https://www.elsevier.com/tdm/userlicense/1.0/},
  langid = {english}
}

@article{lukyanov_1998_xxz,
  title = {Low Energy Effective {{Hamiltonian}} for the {{XXZ}} Spin Chain},
  author = {Lukyanov, Sergei},
  year = 1998,
  month = jul,
  journal = {Nuclear Physics B},
  volume = {522},
  number = {3},
  pages = {533--549},
  issn = {05503213},
  doi = {10.1016/S0550-3213(98)00249-1},
  urldate = {2026-08-28},
  copyright = {https://www.elsevier.com/tdm/userlicense/1.0/},
  langid = {english}
}

@article{mullerhermes_2012_folding,
  title = {Tensor Network Techniques for the Computation of Dynamical Observables in One-Dimensional Quantum Spin Systems},
  author = {{M{\"u}ller-Hermes}, Alexander and Ignacio Cirac, J and Ba{\~n}uls, Mari Carmen},
  year = 2012,
  month = jul,
  journal = {New Journal of Physics},
  volume = {14},
  number = {7},
  pages = {075003},
  issn = {1367-2630},
  doi = {10.1088/1367-2630/14/7/075003},
  urldate = {2026-08-28}
}

@article{orus_2008_itebd,
  title = {Infinite Time-Evolving Block Decimation Algorithm beyond Unitary Evolution},
  author = {Or{\'u}s, R. and Vidal, G.},
  year = 2008,
  month = oct,
  journal = {Physical Review B},
  volume = {78},
  number = {15},
  pages = {155117},
  issn = {1098-0121, 1550-235X},
  doi = {10.1103/PhysRevB.78.155117},
  urldate = {2026-08-28},
  copyright = {http://link.aps.org/licenses/aps-default-license},
  langid = {english}
}

@article{oshikawa_2000_commensurability,
  title = {Commensurability, {{Excitation Gap}}, and {{Topology}} in {{Quantum Many-Particle Systems}} on a {{Periodic Lattice}}},
  author = {Oshikawa, Masaki},
  year = 2000,
  month = feb,
  journal = {Physical Review Letters},
  volume = {84},
  number = {7},
  pages = {1535--1538},
  issn = {0031-9007, 1079-7114},
  doi = {10.1103/PhysRevLett.84.1535},
  urldate = {2026-08-28},
  copyright = {http://link.aps.org/licenses/aps-default-license},
  langid = {english}
}

@article{calabrese_cardy_2004_entanglement,
  title = {Entanglement Entropy and Quantum Field Theory},
  author = {{Pasquale Calabrese} and {John Cardy}},
  year = 2004,
  month = jun,
  journal = {Journal of Statistical Mechanics: Theory and Experiment},
  volume = {2004},
  number = {06},
  pages = {P06002},
  issn = {1742-5468},
  doi = {10.1088/1742-5468/2004/06/P06002},
  urldate = {2026-08-28},
}

@article{pirvu_2010_mpo,
  title = {Matrix Product Operator Representations},
  author = {Pirvu, B and Murg, V and Cirac, J I and Verstraete, F},
  year = 2010,
  month = feb,
  journal = {New Journal of Physics},
  volume = {12},
  number = {2},
  pages = {025012},
  issn = {1367-2630},
  doi = {10.1088/1367-2630/12/2/025012},
  urldate = {2026-08-28}
}

@article{pirvu_2012_finite_size_entanglement,
  title = {Matrix Product States for Critical Spin Chains: {{Finite-size}} versus Finite-Entanglement Scaling},
  shorttitle = {Matrix Product States for Critical Spin Chains},
  author = {Pirvu, B. and Vidal, G. and Verstraete, F. and Tagliacozzo, L.},
  year = 2012,
  month = aug,
  journal = {Physical Review B},
  volume = {86},
  number = {7},
  pages = {075117},
  issn = {1098-0121, 1550-235X},
  doi = {10.1103/PhysRevB.86.075117},
  urldate = {2026-08-28},
  copyright = {http://link.aps.org/licenses/aps-default-license},
  langid = {english}
}

@article{frank_2009_ee_scaling,
  title = {Theory of {{Finite-Entanglement Scaling}} at {{One-Dimensional Quantum Critical Points}}},
  author = {Pollmann, Frank and Mukerjee, Subroto and Turner, Ari M. and Moore, Joel E.},
  year = 2009,
  month = jun,
  journal = {Physical Review Letters},
  volume = {102},
  number = {25},
  pages = {255701},
  issn = {0031-9007, 1079-7114},
  doi = {10.1103/PhysRevLett.102.255701},
  urldate = {2026-08-28},
  copyright = {http://link.aps.org/licenses/aps-default-license},
  langid = {english}
}

@article{sanz_2009_mps_lsm,
  title = {Matrix Product States: {{Symmetries}} and Two-Body {{Hamiltonians}}},
  shorttitle = {Matrix Product States},
  author = {Sanz, M. and Wolf, M. M. and {P{\'e}rez-Garc{\'i}a}, D. and Cirac, J. I.},
  year = 2009,
  month = apr,
  journal = {Physical Review A},
  volume = {79},
  number = {4},
  pages = {042308},
  issn = {1050-2947, 1094-1622},
  doi = {10.1103/PhysRevA.79.042308},
  urldate = {2026-08-28},
  copyright = {http://link.aps.org/licenses/aps-default-license},
  langid = {english}
}

@article{schollwock_2011_dmrg_mps,
  title = {The Density-Matrix Renormalization Group in the Age of Matrix Product States},
  author = {Schollw{\"o}ck, Ulrich},
  year = 2011,
  month = jan,
  journal = {Annals of Physics},
  volume = {326},
  number = {1},
  pages = {96--192},
  issn = {00034916},
  doi = {10.1016/j.aop.2010.09.012},
  urldate = {2026-08-28},
  copyright = {https://www.elsevier.com/tdm/userlicense/1.0/},
  langid = {english}
}

@article{sonner_2021_temporal_entanglement,
  title = {Influence Functional of Many-Body Systems: {{Temporal}} Entanglement and Matrix-Product State Representation},
  shorttitle = {Influence Functional of Many-Body Systems},
  author = {Sonner, Michael and Lerose, Alessio and Abanin, Dmitry A.},
  year = 2021,
  month = dec,
  journal = {Annals of Physics},
  volume = {435},
  pages = {168677},
  issn = {00034916},
  doi = {10.1016/j.aop.2021.168677},
  urldate = {2026-08-28},
  langid = {english}
}

@article{stojevic_2015_fes,
  title = {Conformal Data from Finite Entanglement Scaling},
  author = {Stojevic, Vid and Haegeman, Jutho and McCulloch, I. P. and Tagliacozzo, Luca and Verstraete, Frank},
  year = 2015,
  month = jan,
  journal = {Physical Review B},
  volume = {91},
  number = {3},
  pages = {035120},
  issn = {1098-0121, 1550-235X},
  doi = {10.1103/PhysRevB.91.035120},
  urldate = {2026-08-28},
  copyright = {http://link.aps.org/licenses/aps-default-license},
  langid = {english}
}

@article{suzuki_1976_transfer_matrix_1,
  title = {Relationship between D-{{Dimensional Quantal Spin Systems}} and (D+1)-{{Dimensional Ising Systems}}: {{Equivalence}}, {{Critical Exponents}} and {{Systematic Approximants}} of the {{Partition Function}} and {{Spin Correlations}}},
  shorttitle = {Relationship between D-{{Dimensional Quantal Spin Systems}} and (D+1)-{{Dimensional Ising Systems}}},
  author = {Suzuki, M.},
  year = 1976,
  month = nov,
  journal = {Progress of Theoretical Physics},
  volume = {56},
  number = {5},
  pages = {1454--1469},
  issn = {0033-068X, 1347-4081},
  doi = {10.1143/PTP.56.1454},
  urldate = {2026-08-28},
  langid = {english}
}

@article{tagliacozzo_2008_fes,
  title = {Scaling of Entanglement Support for Matrix Product States},
  author = {Tagliacozzo, L. and De Oliveira, {\relax Thiago}. R. and Iblisdir, S. and Latorre, J. I.},
  year = 2008,
  month = jul,
  journal = {Physical Review B},
  volume = {78},
  number = {2},
  pages = {024410},
  issn = {1098-0121, 1550-235X},
  doi = {10.1103/PhysRevB.78.024410},
  urldate = {2026-08-28},
  copyright = {http://link.aps.org/licenses/aps-default-license},
  langid = {english}
}

@misc{tasaki_2022_LSM_topo,
      title={The Lieb-Schultz-Mattis Theorem: A Topological Point of View}, 
      author={Hal Tasaki},
      year={2022},
      eprint={2202.06243},
      archivePrefix={arXiv},
      primaryClass={cond-mat.stat-mech},
      url={https://arxiv.org/abs/2202.06243}, 
}

@article{vanderstraeten_2019_tangnet,
  title = {Tangent-Space Methods for Uniform Matrix Product States},
  author = {Vanderstraeten, Laurens and Haegeman, Jutho and Verstraete, Frank},
  year = 2019,
  month = jan,
  journal = {SciPost Physics Lecture Notes},
  pages = {7},
  issn = {2590-1990},
  doi = {10.21468/SciPostPhysLectNotes.7},
  urldate = {2026-08-28}
}

@article{vidal_2007_tebd,
  title = {Classical {{Simulation}} of {{Infinite-Size Quantum Lattice Systems}} in {{One Spatial Dimension}}},
  author = {Vidal, G.},
  year = 2007,
  month = feb,
  journal = {Physical Review Letters},
  volume = {98},
  number = {7},
  pages = {070201},
  issn = {0031-9007, 1079-7114},
  doi = {10.1103/PhysRevLett.98.070201},
  urldate = {2026-08-28},
  copyright = {http://link.aps.org/licenses/aps-default-license},
  langid = {english}
}

@article{wang_1997_tmrg,
  title = {Transfer-Matrix Density-Matrix Renormalization-Group Theory for Thermodynamics of One-Dimensional Quantum Systems},
  author = {Wang, Xiaoqun and Xiang, Tao},
  year = 1997,
  month = sep,
  journal = {Physical Review B},
  volume = {56},
  number = {9},
  pages = {5061--5064},
  issn = {0163-1829, 1095-3795},
  doi = {10.1103/PhysRevB.56.5061},
  urldate = {2026-08-28},
  copyright = {http://link.aps.org/licenses/aps-default-license},
  langid = {english}
}

@article{yu_2022_conformal_bc,
  title = {Conformal {{Boundary Conditions}} of {{Symmetry-Enriched Quantum Critical Spin Chains}}},
  author = {Yu, Xue-Jia and Huang, Rui-Zhen and Song, Hong-Hao and Xu, Limei and Ding, Chengxiang and Zhang, Long},
  year = 2022,
  month = nov,
  journal = {Physical Review Letters},
  volume = {129},
  number = {21},
  pages = {210601},
  issn = {0031-9007, 1079-7114},
  doi = {10.1103/PhysRevLett.129.210601},
  urldate = {2026-08-28},
  langid = {english}
}

@article{zauner_2018_VUMPS,
  title = {Variational Optimization Algorithms for Uniform Matrix Product States},
  author = {{Zauner-Stauber}, V. and Vanderstraeten, L. and Fishman, M. T. and Verstraete, F. and Haegeman, J.},
  year = 2018,
  month = jan,
  journal = {Physical Review B},
  volume = {97},
  number = {4},
  pages = {045145},
  issn = {2469-9950, 2469-9969},
  doi = {10.1103/PhysRevB.97.045145},
  urldate = {2026-08-28},
  langid = {english}
}

@book{blumenhagen_2009_CFT_book,
  title = {Introduction to {{Conformal Field Theory}}},
  author = {Blumenhagen, Ralph and Plauschinn, Erik},
  year = 2009,
  series = {Lecture {{Notes}} in {{Physics}}},
  volume = {779},
  publisher = {Springer Berlin Heidelberg},
  address = {Berlin, Heidelberg},
  doi = {10.1007/978-3-642-00450-6},
  urldate = {2022-10-18},
  isbn = {978-3-642-00449-0 978-3-642-00450-6},
  langid = {english}
}

@article{obrien_2020_potts_new_bdy,
  title = {``Not-$A$'', representation symmetry-protected topological, and Potts phases in an ${S}_{3}$-invariant chain},
  author = {O'Brien, Edward and Vernier, Eric and Fendley, Paul},
  journal = {Phys. Rev. B},
  volume = {101},
  issue = {23},
  pages = {235108},
  numpages = {13},
  year = {2020},
  month = {Jun},
  publisher = {American Physical Society},
  doi = {10.1103/PhysRevB.101.235108},
  url = {https://link.aps.org/doi/10.1103/PhysRevB.101.235108}
}

@article{vanhecke_2021_vomps,
	title = {Tangent-space methods for truncating uniform MPS},
	pages = {004},
	author = {Vanhecke, Bram and Van Damme, Maarten and Haegeman, Jutho and Vanderstraeten, Laurens and Verstraete, Frank},
	journal = {SciPost Phys. Core},
	volume = {4},
	year = {2021},
	publisher = {SciPost},
	doi = {10.21468/SciPostPhysCore.4.1.004},
	url = {https://scipost.org/10.21468/SciPostPhysCore.4.1.004}
}

@article{ref_appendix,
	journal = {See appendix for more technical details and explanations.}
}

@article{Cho2017,
  title = {Universal entanglement spectra of gapped one-dimensional field theories},
  author = {Cho, Gil Young and Ludwig, Andreas W. W. and Ryu, Shinsei},
  journal = {Physical Review B},
  volume = {95},
  issue = {11},
  pages = {115122},
  numpages = {10},
  year = {2017},
  month = {Mar},
  publisher = {American Physical Society},
  doi = {10.1103/PhysRevB.95.115122},
  url = {https://link.aps.org/doi/10.1103/PhysRevB.95.115122}
}

\appendix
\onecolumngrid

\renewcommand{\thefigure}{\thesection\arabic{figure}}
\makeatletter
\@addtoreset{figure}{section}
\makeatother

\section{The tensor network implementation}
\label{app:transverse-impo}

Here we give a detailed presentation on the way we implement the temporal and spatial quenches and explain why the spatial transfer
evolution of the temporal iMPS should be represented by a regrouped
infinite matrix-product operator (iMPO). The need for this representation
in the present work arises from evolving an infinite, translation-invariant
temporal iMPS. Although a physical Trotter gate approaches the full-rank
two-site identity as $\delta\tau\to0$, the same four-leg tensor, when
regarded as a $d^2\times d^2$ transverse matrix, approaches the rank-one
operator $|I\rangle\langle I|$. The singularity discussed below, however, is the
algorithmic problem that we encounter when this reshuffled tensor is
repeatedly applied in the evolution of an infinite temporal iMPS. It has one $O(1)$ singular
direction, whereas the nontrivial transfer dynamics enters through the
$O(\delta\tau)$ corrections. Direct application therefore imposes an
unsuitable local Schmidt structure and obscures these corrections.
Factoring each gate and regrouping neighboring half-gates avoids this
singular infinite-iMPS evolution and produces a regular transverse iMPO
without changing the two-dimensional Euclidean tensor network.

\subsection{iMPS evolution and variational compression}
\label{app:evolve-compress}

In practice, both quantizations are implemented by an iterative
evolve--compress procedure. Let $|\Psi_D^{(k)}\rangle$ denote the
normalized MPS of bond dimension $D$ after $k$ steps, and let
$\mathcal T_{\delta\ell}$ be one uncompressed evolution step. The
numerical update can be summarized as
\begin{equation}
  |\Psi_D^{(k+1)}\rangle
  =\mathcal{N}_k\,\mathcal{P}_D
   \!\left[\mathcal{T}_{\delta\ell}
   |\Psi_D^{(k)}\rangle\right],
  \qquad \ell_k=k\,\delta\ell ,
  \label{eq:app-evolve-compress}
\end{equation}
where $\mathcal P_D$ denotes compression onto the MPS manifold of bond
dimension $D$ (implemented variationally for the iMPO--iMPS product),
and $\mathcal N_k$ restores the canonical normalization. Here $\mathcal{T}_{\delta\ell}$ can either be the checkerbord Trotter layer as in iTEBD or the temporal/spatial transfer matrix written as an iMPO. In this work we choose the iMPO evolution scheme since the Trotter gates evolution has some singularity in the spatial quench of temporal states, which we will explain in detail below.

If the iMPO transfer matrix $T^{\mathrm{iMPO}}$ has bond dimension $\chi$, its exact action on
an iMPS of bond dimension $D$ produces an intermediate state with bond
dimension up to $\chi D$. Rather than truncating this enlarged state by
a sequence of local SVDs, we use variational compression~\cite{vanderstraeten_2019_tangnet, vanhecke_2021_vomps} to find the uniform bond-$D$ iMPS with maximal normalized
overlap per site with the intermediate target. The state from the
preceding transfer step is used as the initial guess. Once the
variational environments and local tensors have converged, the result
is returned to mixed canonical form, normalized, and used as the input
to the next step. Writing
$|\Phi(A)\rangle=T^{\mathrm{iMPO}}|\Psi(A)\rangle$ for the
uncompressed target state, the corresponding variational problem is explicitly:
\begin{align}
  A_{\mathrm{comp}}
  &=\underset{\operatorname{bond}(\widetilde A)= D}{\arg\max}
    f\!\left(\widetilde A,\Phi(A)\right),
  \notag\\
  f\!\left(\widetilde A,\Phi(A)\right)
  &=
      \frac{|\langle\Psi(\widetilde A)|\Phi(A)\rangle|^2}{\langle\Psi(\widetilde A)|\Psi(\widetilde A)\rangle
        \langle\Phi(A)|\Phi(A)\rangle} .
  \label{eq:app-impo-compression}
\end{align}

During the finite-length scaling calculations, this cycle is stopped at
the desired $\tau$ or $L$. The cutoff in bond dimension will set the correlation length $\xi_D$, under which the finite length scaling is performed, as discussed in the main text.

\subsection{Singularity in spatial quench with reshuffled Trotter gates}

In this section we explain why we choose the iMPO representation and how is the representation constructed from Trotter gates. For a nearest-neighbor Hamiltonian $H_x=\sum_n h_{n,n+1}$, let $n$
label the spatial bond $(n,n+1)$ and let $m$ label a Trotter time slice.
Each slice contains two checkerboard sublayers, denoted by
$\sigma=1,2$. The local two-site gate is
\begin{equation}
  (G_{m,\sigma}^{(n)})_{a'b';ab}
  =\langle a'b'|e^{-\delta\tau h_{n,n+1}}|ab\rangle
  =\delta_{a'a}\delta_{b'b}
   -\delta\tau\,(h_{n,n+1})_{a'b';ab}
   +O(\delta\tau^2).
  \label{eq:app-local-gate}
\end{equation}
In the ordinary temporal transfer orientation, the leading term is simply the identity
on the two sites. The spatial transfer problem instead groups the
indices into the transverse pairs $(b',b)$ and $(a',a)$. Defining
\begin{equation}
  (\widetilde G_{m,\sigma}^{(n)})_{(b',b);(a',a)}
  \equiv (G_{m,\sigma}^{(n)})_{a'b';ab},
  \label{eq:app-transverse-reshuffling}
\end{equation}
the identity contribution becomes
\begin{equation}
  (\widetilde G_{m,\sigma}^{(n)})_{(b',b);(a',a)}
  =\delta_{b'b}\delta_{a'a}+O(\delta\tau)
  =|I\rangle\langle I|+O(\delta\tau),
  \qquad
  |I\rangle=\sum_{s=1}^{d}|s,s\rangle .
  \label{eq:app-transverse-projector}
\end{equation}
Thus the leading transverse map has rank one, even though the same
tensor is the identity operator in the physical orientation. For a
spin-$1/2$ site and the composite-index ordering
$(00,01,10,11)$, this statement reads
\begin{equation}
  \widetilde G_{m,\sigma}^{(n)}
  =
  \begin{pmatrix}
    1&0&0&1\\
    0&0&0&0\\
    0&0&0&0\\
    1&0&0&1
  \end{pmatrix}
  +O(\delta\tau)
  =2|\Phi^+\rangle\langle\Phi^+|+O(\delta\tau),
  \qquad
  |\Phi^+\rangle=\frac{|00\rangle+|11\rangle}{\sqrt2}.
  \label{eq:app-spinhalf-transverse-gate}
\end{equation}

If the reshuffled gates of, for example, the lower sublayer are treated
as ordinary two-site evolution operators and applied on disjoint
temporal bonds, their leading action on a generic temporal iMPS
$|\Psi(A)\rangle$ is
\begin{equation}
  \prod_{m\in\mathbb Z}\widetilde G_{m,\sigma}^{(n)}
  |\Psi(A)\rangle
  =\mathcal N_{\Psi}
   \bigotimes_{j\in\mathbb Z}|I\rangle_{2j,2j+1}
   +O(\delta\tau),
  \label{eq:app-rank-one-action}
\end{equation}
provided the corresponding overlap is nonzero. The leading map of one layer of reshuffled gates therefore
overwrites the local Schmidt structure with the product of
maximally entangled pairs. Entanglement associated with these pairs is
ultraviolet and must not be
identified with the universal conformal entanglement spectrum. The exact
two-dimensional partition function remains regular. In particular, the
rank-one limit of a local reshuffled gate does not by itself imply a
singularity of a finite tensor-network contraction. The problem found here is specific to the infinite construction, since the directly reshuffled gate would be applied uniformly and repeatedly to an
infinite temporal iMPS, followed by a variational compression. In this case, the
rank-one $O(1)$-contribution projects every updated bond onto the same
maximally entangled direction and renders the physically relevant
corrections poorly conditioned. Since the direct rotated-gate evolution contains the singularity, we decompose and regroup them into iMPO for both the temporal and the spatial evolution.

\subsection{iMPO construction by SVD factorization and regrouping}

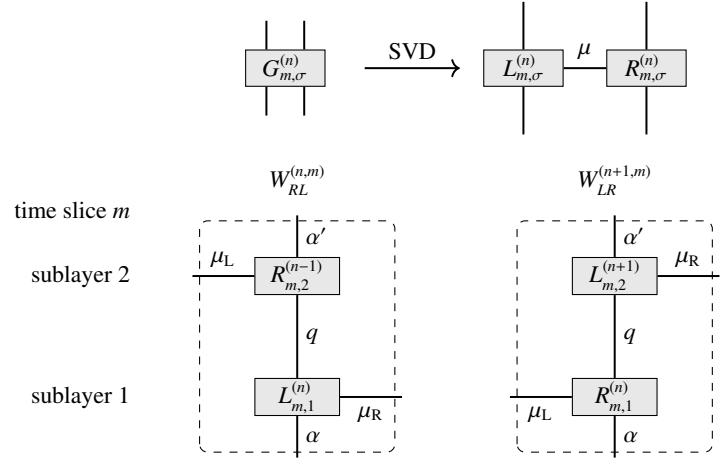
\begin{figure}[h]
  \centering
  \begin{tikzpicture}[
      x=1.05cm,y=1.00cm,
      box/.style={draw,fill=gray!18,minimum width=0.62cm,
                  minimum height=0.48cm,inner sep=1pt},
      line/.style={line width=0.75pt}
    ]
    % Upper panel: SVD of one physical gate.
    \begin{scope}[yshift=0.45cm]
    \node[box,minimum width=1.05cm] (G) at (2.00,2.55)
      {$G_{m,\sigma}^{(n)}$};
    \draw[line] ([xshift=-0.25cm]G.north)--++(0,0.38);
    \draw[line] ([xshift= 0.25cm]G.north)--++(0,0.38);
    \draw[line] ([xshift=-0.25cm]G.south)--++(0,-0.38);
    \draw[line] ([xshift= 0.25cm]G.south)--++(0,-0.38);

    \draw[->,line] (3.00,2.55)--(4.20,2.55)
      node[midway,above] {SVD};

    \node[box,minimum width=1.05cm] (L) at (5.00,2.55)
      {$L_{m,\sigma}^{(n)}$};
    \node[box,minimum width=1.05cm] (R) at (6.55,2.55)
      {$R_{m,\sigma}^{(n)}$};
    \draw[line] (L.east)--(R.west)
      node[midway,above] {$\mu$};
    \draw[line] (L.north)--++(0,0.63);
    \draw[line] (L.south)--++(0,-0.63);
    \draw[line] (R.north)--++(0,0.63);
    \draw[line] (R.south)--++(0,-0.63);
    \end{scope}

    % Lower panel: the two staggered vertical regroupings, drawn apart.
    \node[anchor=east] at (0.15,0.25) {sublayer $2$};
    \node[anchor=east] at (0.15,-1.35) {sublayer $1$};
    \node[anchor=east] at (0.15,1.12) {time slice $m$};

    \node[box,minimum width=1.12cm] (Rup) at (2.15,0.25)
      {$R_{m,2}^{(n-1)}$};
    \node[box,minimum width=1.12cm] (Llow) at (2.15,-1.35)
      {$L_{m,1}^{(n)}$};
    \draw[line] (Rup.south)--(Llow.north)
      node[midway,right] {$q$};
    \draw[line] (Rup.west)--++(-0.78,0)
      node[midway,above] {$\mu_{\mathrm L}$};
    \draw[line] (Llow.east)--++(0.78,0)
      node[midway,below] {$\mu_{\mathrm R}$};
    \draw[line] (Rup.north)--++(0,0.55)
      node[midway,right] {$\alpha'$};
    \draw[line] (Llow.south)--++(0,-0.55)
      node[midway,right] {$\alpha$};
    \draw[dashed,rounded corners=3pt]
      (0.92,-2.08) rectangle (3.38,0.98);
    \node at (2.15,1.52) {$W_{RL}^{(n,m)}$};

    \node[box,minimum width=1.12cm] (Lup) at (6.15,0.25)
      {$L_{m,2}^{(n+1)}$};
    \node[box,minimum width=1.12cm] (Rlow) at (6.15,-1.35)
      {$R_{m,1}^{(n)}$};
    \draw[line] (Lup.south)--(Rlow.north)
      node[midway,right] {$q$};
    \draw[line] (Lup.east)--++(0.78,0)
      node[midway,above] {$\mu_{\mathrm R}$};
    \draw[line] (Rlow.west)--++(-0.78,0)
      node[midway,below] {$\mu_{\mathrm L}$};
    \draw[line] (Lup.north)--++(0,0.55)
      node[midway,right] {$\alpha'$};
    \draw[line] (Rlow.south)--++(0,-0.55)
      node[midway,right] {$\alpha$};
    \draw[dashed,rounded corners=3pt]
      (4.92,-2.08) rectangle (7.38,0.98);
    \node at (6.15,1.52) {$W_{LR}^{(n+1,m)}$};
  \end{tikzpicture}
  \caption{SVD factorization and regrouping of the checkerboard
  gates. The upper panel splits a generic gate $G_{m,\sigma}^{(n)}$ into
  $L_{m,\sigma}^{(n)}$ and $R_{m,\sigma}^{(n)}$. Here $n$ labels a
  spatial bond, $m$ labels a Trotter time slice, and $\sigma=1,2$
  labels its lower and upper sublayers; Euclidean time
  increases upward. In the lower panel, the left cell has
  $R_{m,2}^{(n-1)}$ above $L_{m,1}^{(n)}$, while the right cell has
  $L_{m,2}^{(n+1)}$ above $R_{m,1}^{(n)}$. The dashed boundaries mark the resulting
  rank-four iMPO tensors.}
  \label{fig:app-transverse-impo}
\end{figure}

The regular construction factorizes each physical gate before changing
the contraction direction. Retaining the spatial-bond label $n$, the
time slice label $m$, and the sublayer label $\sigma$, an SVD across
the two spatial sites gives
\begin{equation}
  (G_{m,\sigma}^{(n)})_{a'b';ab}
  =\sum_{\mu}
    (L_{m,\sigma}^{(n)})_{a'a;\mu}
    (R_{m,\sigma}^{(n)})_{\mu;b'b},
  \label{eq:app-transverse-svd}
\end{equation}
where the singular values may be absorbed symmetrically into $L$ and
$R$. The two sublayers belonging to the same Trotter time slice
$m$ are displaced by one spatial site. Consequently, the right half
$R_{m,1}^{(n)}$ of a lower-sublayer gate meets
$L_{m,2}^{(n+1)}$ in the upper sublayer, whereas its left half
$L_{m,1}^{(n)}$ meets $R_{m,2}^{(n-1)}$. The resulting rank-four tensors are therefore~(\ref{fig:app-transverse-impo})
\begin{align}
  (W_{RL}^{(n,m)})_{\mu_{\mathrm L}\mu_{\mathrm R};\,\alpha'\alpha}
  &=\sum_q
    (R_{m,2}^{(n-1)})_{\mu_{\mathrm L};\,\alpha' q}
    (L_{m,1}^{(n)})_{q\alpha;\,\mu_{\mathrm R}},
  \notag\\
  (W_{LR}^{(n+1,m)})_{\mu_{\mathrm L}\mu_{\mathrm R};\,\alpha'\alpha}
  &=\sum_q
    (L_{m,2}^{(n+1)})_{\alpha' q;\,\mu_{\mathrm R}}
    (R_{m,1}^{(n)})_{\mu_{\mathrm L};\,q\alpha}.
  \label{eq:app-half-gate-regrouping}
\end{align}
where $q$ is the contracted leg. In the transverse orientation relevant
to the spatial quench, $\mu_{\mathrm L}$ and $\mu_{\mathrm R}$ are the
left and right physical legs associated with the temporal iMPS on which
the iMPO acts, whereas $\alpha$ and $\alpha'$ are respectively the lower
and upper virtual indices of the iMPO. Thus the
neighboring pair $(W_{RL}^{(n,m)},W_{LR}^{(n+1,m)})$ is the natural
two-site iMPO unit cell for the ordinary temporal evolution of the spatial iMPS. Rotated to the spatial quench direction, the tensors naturally give two layers of iMPOs acting on temporal states, $\mathcal W_1$ and $\mathcal W_2$, with the length of the unit cells being one. With our ordering convention, the two operators are related by
\begin{equation}
  \mathcal W_1=\mathcal W_2\mathcal T,
  \label{eq:app-impo-translation-relation}
\end{equation}
where $\mathcal T$ translates the temporal state by one site. 
Every temporal
iMPS used in this work is one-site translation invariant. Hence
\begin{equation}
  \mathcal T|\Psi(A)\rangle=|\Psi(A)\rangle,
  \qquad
  \mathcal W_1|\Psi(A)\rangle
  =\mathcal W_2\mathcal T|\Psi(A)\rangle
  =\mathcal W_2|\Psi(A)\rangle.
  \label{eq:app-impo-translation-equivalence}
\end{equation}
The one-site shift is therefore trivial on the variational states of
interest. Although $\mathcal W_1$ and $\mathcal W_2$ are distinct as
operators locally, they generate the same updated state from
the one-site translation-invariant temporal iMPS. The regrouping is
exact, and the bond-dimension truncation enters only in the subsequent
variational compression.

\section{Entanglement Entropy Scaling}
\label{app:e4ntanglement_scaling}

Identifying the infrared conformal field theory is the necessary first
step before assigning its conformal boundary conditions. Among the
universal bulk data, the central charge $c$ is particularly important, which is the coefficient of the central extension of the Virasoro
algebra. It measures the effective number of gapless degrees of freedom,
and controls universal finite-size and entanglement scaling. An
independent determination of $c$ therefore tests whether a lattice
calculation has reached the proposed conformal fixed point rather than
merely displaying scale-dependent behavior over a limited range. In 
the present work, the distinct values $c=1/2$, $4/5$, and $1$ characterize the bulk CFTs underlying the Ising, three-state Potts, and critical XXZ chains, respectively.

The main text identifies the infrared boundary conditions primarily
from the universal low-lying entanglement spectrum. Here we complement
that boundary-sensitive analysis by extracting the bulk central charge
from the von Neumann entanglement entropy. As discussed around Eq.~(25) of the main
text, in the finite-length scaling window
$1\ll\ell\ll\xi_D$ the entropy for the quenched state bipartitioned as $A\cup\bar{A}$ obeys
\begin{equation}
  S_A(\ell)
  = \frac{c}{6}\log\!\left(\frac{2\ell}{\pi\epsilon}\right)
    + \log g_O + \log g_E + s_{\mathrm{uv}} + \cdots ,
  \label{app:EE_scaling}
\end{equation}
where $\ell=\tau$ for the spatial state and $\ell=L$ for the temporal
state. The factors $g_O$ and $g_E$ are the Affleck--Ludwig boundary
degeneracies of the initial-state boundary $B_O$ and the entanglement
boundary $B_E$, respectively. Consequently, changing $B_O$ modifies
the constant part of the entropy but leaves the universal logarithmic
coefficient fixed by the bulk central charge. Thus, the slope provides a boundary-independent characterization of
the bulk CFT. Obtaining the same $c$ in both directions is a nontrivial bulk
consistency check, which confirms that the two lattice constructions flow
to the same continuum CFT even when their microscopic boundary
realizations or entanglement spectra differ. The central charge supplies the fundamental
bulk classification that the boundary-spectrum analysis subsequently
refines, thus we fit the central charges for the three cases in the finite-length scaling window.

\begin{figure}
    \centering
    \includegraphics[width=1.0\linewidth]{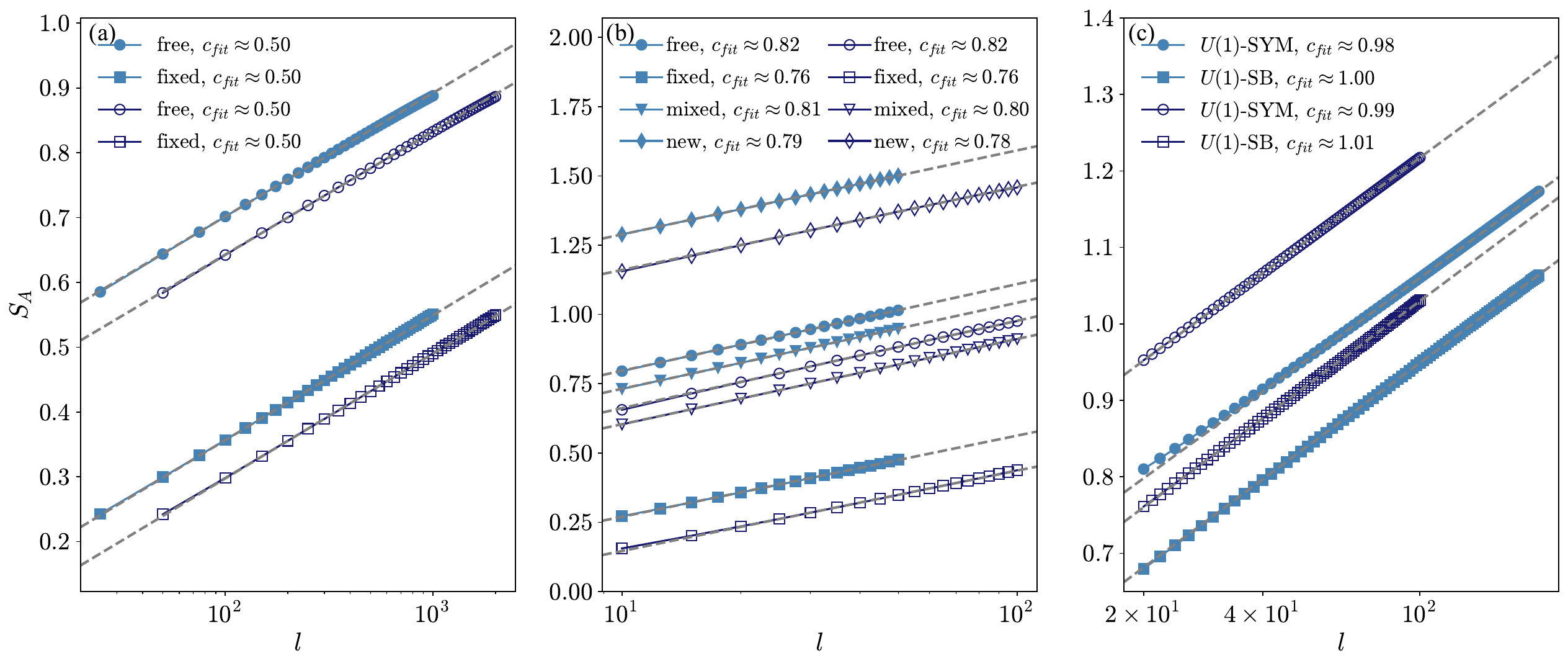}
    \caption{The spatial and temporal Von Neumann entanglement entropy for the critical (a)~Ising~($c=1/2$), (b)~three-state Potts~($c=4/5$) and (c)~XXZ spin chains ($c=1$) with $g=0.5$, with grey dashed lines showing fitting results. 
    The filled markers represent the entropy scaling for the temporal quench of spatial states, while the hollow markers represent
    the entropy scaling for the spatial quench of temporal states. The fittings are implemented for the length scale window $1\ll \ell \ll \xi_{D}$.}
    \label{fig:app_entropy_scaling}
\end{figure}

For the critical Ising chain, the $Z_2$-symmetric initial state flows
to the free boundary, whereas a $Z_2$-breaking initial state selects a
fixed boundary. Fig.~\ref{fig:app_entropy_scaling}~(a) shows that both
boundary classes yield $c_{\mathrm{fit}}\simeq 0.5$ in the spatial and
temporal calculations. The vertical separation of the free and fixed
data is boundary dependent, while their common logarithmic slope
reproduces $c=1/2$ and thereby identifies the bulk fixed point with the
Ising CFT.

For the three-state Potts model, we consider the four $S_3$-inequivalent boundary classes---free, fixed, mixed, and new---that were distinguished spectroscopically in the
main text. As shown in Fig.~\ref{fig:app_entropy_scaling}~(b), their entropy
offsets differ substantially, but the fitted central charges remain
close to $c=4/5$, providing a direct bulk-level characterization of the
three-state Potts CFT. Across the two scaling procedures,
the estimates lie between approximately $0.76$ and $0.82$. The spread comes from the residual finite-length and lattice corrections to
the leading scaling form.

Finally, Fig.~\ref{fig:app_entropy_scaling}~(c) presents the XXZ results at
$g=0.5$. For a $U(1)_z$-preserving initial state, the spatial
construction realizes the LSM constraint through a doubled unit cell,
whereas successive temporal transfer steps pump the boundary sector
between translation-related Dirichlet sectors. Breaking $U(1)_z$
removes this constraint and allows the mixed Neumann--Dirichlet
boundary condition described in the main text. Despite these distinct
boundary structures, the entropy fits give
$c_{\mathrm{fit}}\simeq0.98$ and $1.00$ from the $\tau$ scaling and
$c_{\mathrm{fit}}\simeq0.99$ and $1.01$ from the $L$ scaling for the
$U(1)_z$-preserving and $U(1)_z$-breaking states, respectively. Since the preserving of $U(1)_z$ will enforce the entropy differs spatially for odd and even site, here we choose the average entropy $\overline{S_A}$ of the odd and even lattice sites to perform the fitting. All
four estimates agree well with $c=1$, confirming that both symmetry
classes are governed by the same compact-boson CFT despite their
different boundary conditions.

\section{Control of initial states--Spatial v.s. Temporal}
\label{app:boundary-engineering}

Once the bulk CFT has been identified through its central charge, the
next task is to control the conformal boundary $B_O$ represented by the
microscopic termination of the Euclidean tensor network. In this work
we choose representative microscopic initial states or boundary fields
and identify them with the associated conformal boundary. The critical bulk Hamiltonian, the bulk transfer tensors, and the entanglement cut are kept fixed. The
entanglement boundary $B_E$ is therefore unchanged, so differences in
the universal low-lying spectrum diagnose different realizations of
$B_O$ rather than changes of the bulk CFT or of the bipartition.

The two Euclidean quantizations provide complementary controls. For the
ordinary imaginary-time evolution of a spatial iMPS, we specify the
initial state directly. For the spatial transfer evolution of a
temporal iMPS, a boundary field terminates the spatial end of the
two-dimensional Euclidean network, and the resulting boundary tensor
determines the initial state for the spatial transfer evolution.

\subsection{Control of the spatial-iMPS initial state}

For the ordinary temporal evolution, each calculation begins from a
chosen representative short-range spatial iMPS
$|\Psi_{\mathrm{sp},0}\rangle$, which is evolved with the fixed critical
Hamiltonian,
\begin{equation}
  |\Psi_{\mathrm{sp}}(\tau)\rangle
  =\mathcal N_{\mathrm{sp}}(\tau)
   e^{-\tau H_x}|\Psi_{\mathrm{sp},0}\rangle .
  \label{eq:app-spatial-initial-state}
\end{equation}
Here $|\Psi_{\mathrm{sp},0}\rangle$ denotes one representative selected
for the intended symmetry class and boundary realization. Symmetry provides the most robust distinction among these preparations.
If the
critical Hamiltonian has symmetry $G$, and the initial iMPS preserve a subgroup
$H\leq G$, the entire evolution remains in the $H$-invariant
sector. Symmetry alone, however, does not always identify a unique
conformal boundary. For example, the fixed and mixed Potts boundaries
preserve the same residual subgroup, while the free and new boundaries
are both fully $S_3$ symmetric. They must therefore be distinguished by
using microscopically different representative tensors and by comparing
their low-lying spectra. In particular, the new boundary is represented
by a correlated MPS rather than a simple product state.

We now state the representative states used for each model explicitly.
For the Ising chain, whose order parameter is $\sigma^z$ and whose
onsite $Z_2$ symmetry is generated by $\prod_j\sigma_j^x$, we choose
\begin{equation}
 \begin{aligned}
  \ket{\Psi_{\mathrm{Ising}}^{\mathrm{free}}}
    &=\bigotimes_j|\rightarrow\rangle_j,
    &\sigma_j^x|\rightarrow\rangle_j&=|\rightarrow\rangle_j,\\
  \ket{\Psi_{\mathrm{Ising}}^{\mathrm{fixed},+}}
    &=\bigotimes_j|\uparrow\rangle_j,
    &\sigma_j^z|\uparrow\rangle_j&=|\uparrow\rangle_j,\\
  \ket{\Psi_{\mathrm{Ising}}^{\mathrm{fixed},-}}
    &=\bigotimes_j|\downarrow\rangle_j,
    &\sigma_j^z|\downarrow\rangle_j&=-|\downarrow\rangle_j .
 \end{aligned}
 \label{eq:app-ising-spatial-states}
\end{equation}
The first state is $Z_2$ invariant and represents the free boundary,
whereas the two $z$-polarized states explicitly break $Z_2$ and
represent the two symmetry-related fixed boundaries.

For the Potts chain, with the clock basis $\{\ket{0}, \ket{1}, \ket{2}\}$, the product of fully symmetric local states
\begin{equation}
  \ket{\Psi_{\mathrm{Potts}}^{\mathrm{free}}}=\bigotimes_j\ket{+}_j,\qquad|+\rangle=\frac{|0\rangle+|1\rangle+|2\rangle}{\sqrt{3}}
  \label{eq:app-potts-symmetric-state}
\end{equation}
gives the free-boundary representative, while
$\bigotimes_j|0\rangle_j$ gives one member of the fixed triplet (and
cyclic permutations give the $B$ and $C$ representatives), 
\begin{equation}
    \ket{\Psi_{\mathrm{Potts}}^{A}}=\bigotimes_j\ket{0}_j.
\end{equation}
The complete
Potts boundary classification also contains the mixed triplet
$AB,BC,CA$ and the $S_3$-invariant new boundary
\cite{affleck_1998_bc}. For example, the mixed boundary $BC$ is
represented by the ``not-$A$'' product state
\begin{equation}
  \ket{\Psi_\mathrm{Potts}^{BC}}=\bigotimes_j|\overline 0\rangle_j ,
  \qquad
  |\overline 0\rangle
  =\frac{|1\rangle+|2\rangle}{\sqrt{2}}  ,
  \label{eq:app-potts-mixed-state}
\end{equation}
which excludes $A$ while retaining fluctuations between $B$ and $C$.
It preserves the reflection that exchanges $B$ and $C$, the same
residual $Z_2$ preserved by the fixed-$A$ boundary, although the two
boundary spectra are inequivalent. Its cyclic images represent $CA$
and $AB$.

The $S_3$-invariant new boundary cannot be represented by an uncorrelated
$S_3$-symmetric product state. We use instead the uniform
bond-dimension-two MPS \cite{obrien_2020_potts_new_bdy}
\begin{equation}
  \ket{\Psi_{\mathrm{Potts}}^{\mathrm{new}}}
  =\operatorname{Tr}\!\left[\prod_j R_j\right],
  \qquad
  R_j=
  \begin{pmatrix}
    |q_3=0\rangle_j & |q_3=+1\rangle_j\\
    |q_3=-1\rangle_j & |q_3=0\rangle_j
  \end{pmatrix}.
  \label{eq:app-potts-new-mps}
\end{equation}
Here $q_3=0,\pm1$ labels the $Z_3$ charge basis,
$X|q_3\rangle=\omega^{q_3}|q_3\rangle$, with
$\omega=e^{2\pi i/3}$ and $|q_3=0\rangle=|+\rangle$. This MPS is
$S_3$ invariant but has short-range correlations that distinguish it
from the free-boundary product state. It is the Kramers--Wannier dual
of the not-$A$ state in Eq.~\eqref{eq:app-potts-mixed-state}. Duality relates the mixed and new representatives, while an ordinary
$S_3$ rotation only permutes members within the fixed or mixed
triplets.

Finally, for the XXZ chain we use two preparations that differ by
their $U(1)_z$ symmetry. The symmetry-preserving representative is the
nearest-neighbor singlet product
\begin{equation}
  \ket{\Psi_{\mathrm{XXZ}}^{U(1)}}
   =\bigotimes_j |s\rangle_{2j-1,2j},
  \qquad
  |s\rangle_{i,i+1}
   =\frac{|\uparrow_i\downarrow_{i+1}\rangle
          -|\downarrow_i\uparrow_{i+1}\rangle}{\sqrt2},
  \label{eq:app-xxz-singlet-state}
\end{equation}
The doubled unit cell originates from the LSM theorem contraint for a critical antiferromagnetic spin-$1/2$ chain. For the symmetry-breaking
representative we use the two-site product state
\begin{equation}
  \ket{\Psi_{\mathrm{XXZ}}^{\mathrm{br}}}
  =\bigotimes_j|\rightarrow\rangle_{2j-1}|\leftarrow\rangle_{2j},
  \qquad S_j^x|\rightarrow\rangle_j=\tfrac12|\rightarrow\rangle_j,\qquad S_j^x|\leftarrow\rangle_j=-\tfrac12|\leftarrow\rangle_j 
  \label{eq:app-xxz-breaking-state}
\end{equation}
which breaks $U(1)_z$ and pins down the corresponding compact-boson boundary field, selecting a single Dirichlet boundary condition.

\subsection{Control of the temporal-iMPS initial state}

The temporal iMPS is evolved in the orthogonal, spatial direction. Its
microscopic degrees of freedom are Euclidean-time transfer sites rather
than the original spins, so its initial state should not be chosen directly the same way as the ordinary spatial iMPS. Instead, we terminate the spatial edge
of the same Trotterized Euclidean network by a boundary Hamiltonian
\begin{equation}
  h_{\partial}(\boldsymbol\lambda)
  =\delta H_{\partial}(\boldsymbol\lambda)
   +\sum_a\mathcal O_{\partial,a},
  \label{eq:app-boundary-field}
\end{equation}
where $\mathcal O_{\partial,a}$ are the original local boundary
operators of the lattice model and
$\delta H_{\partial}(\boldsymbol\lambda)$ is the tunable boundary
deformation. We restrict attention to the single-site boundary terms
used in the main text. After Trotterization, the ordinary boundary gates are
factorized by an SVD into three-leg tensors
$\mathsf B^{s_n}_{\alpha_{n-1}\alpha_n}$, with one temporal-site leg
$s_n$ and two virtual legs. By contrast, the strong-field weight defines
a two-leg map
$$
  \mathsf V_{\partial}(\boldsymbol\lambda)
  =
  e^{-\delta\tau\,\delta H_{\partial}(\boldsymbol\lambda)}
$$
on the virtual space. Here the operator $\mathsf{V}_{\partial}$ acts on the virtual Hilbert spaces, which are the original local-site Hilbert spaces of the lattice model on different time slices. The initial temporal iMPS is
therefore obtained by inserting $\mathsf V_{\partial}$ between
neighboring three-leg boundary tensors,
\begin{equation}
  |\Psi_{\mathrm{tm},0}(\boldsymbol\lambda)\rangle
  =\sum_{\{s_n\}}
   \operatorname{Tr}\!\left[
    \cdots \mathsf B^{s_n}
    \mathsf V_{\partial}(\boldsymbol\lambda)
    \mathsf B^{s_{n+1}}
    \mathsf V_{\partial}(\boldsymbol\lambda)\cdots
   \right]
   |\ldots s_n s_{n+1}\ldots\rangle .
  \label{eq:app-temporal-boundary-state}
\end{equation}
Thus the boundary field dresses the virtual bonds of the original
iMPS rather than generating a separate MPS by contracting a product of
boundary Boltzmann weights. An equivalent convention can absorb
$\mathsf V_{\partial}^{1/2}$ into each adjacent $\mathsf B$ tensor, but
the bond-insertion form makes the two-leg character of the field
operator explicit.

\begin{figure}[tbp]
  \centering
  \begin{tikzpicture}[
      x=0.65cm,y=0.65cm,
      tensor/.style={draw,fill=gray!15,minimum width=0.55cm,
                     minimum height=0.42cm,inner sep=1pt},
      insertion/.style={draw,rounded corners=1pt,fill=gray!30,
                        minimum width=0.72cm,minimum height=0.36cm,
                        inner sep=1pt},
      delta/.style={circle,fill=black,minimum size=0.12cm,inner sep=0pt},
      line/.style={line width=0.7pt}
    ]
    % Finite boundary-field insertion.
    \node[tensor] (A1) at (0,0) {$\mathsf B$};
    \node[insertion] (V) at (1.55,0) {$\mathsf V_{\partial}$};
    \node[tensor] (A2) at (3.10,0) {$\mathsf B$};
    \draw[line] (-0.75,0)--(A1.west);
    \draw[line] (A1.east)--(V.west);
    \draw[line] (V.east)--(A2.west);
    \draw[line] (A2.east)--(3.85,0);
    \draw[line] (A1.north)--++(0,0.68);
    \draw[line] (A2.north)--++(0,0.68);
    \node at (1.55,-0.70) {finite $\boldsymbol\lambda$};

    \draw[->,line] (4.45,0)--(6.20,0)
      node[midway,above] {$|\boldsymbol\lambda|\to\infty$};

    % Strong-field form: delta vertices and a two-leg insertion.
    \node[delta] (D1) at (8.10,0) {};
    \node[insertion] (VS) at (9.65,0) {$\mathsf V_{\partial}$};
    \node[delta] (D2) at (11.20,0) {};
    \draw[line] (7.35,0)--(D1.west);
    \draw[line] (D1.east)--(VS.west);
    \draw[line] (VS.east)--(D2.west);
    \draw[line] (D2.east)--(11.95,0);
    \draw[line] (D1.north)--++(0,0.68);
    \draw[line] (D2.north)--++(0,0.68);
  \end{tikzpicture}
  \caption{Boundary-field construction of the initial temporal iMPS.
  At finite $\boldsymbol\lambda$, the two-leg map
  $\mathsf V_{\partial}$ is inserted on the virtual bond between
  neighboring three-leg SVD tensors $\mathsf B$ (left). When the boundary
  field dominates, the ordinary boundary terms can be neglected and
  $\mathsf B$ reduces to the three-leg delta tensor $\Delta$, while
  $\mathsf V_{\partial}$ remains a two-leg bond operator (right). The
  pattern is repeated along the temporal iMPS.}
  \label{fig:app-boundary-imps-insertion}
\end{figure}
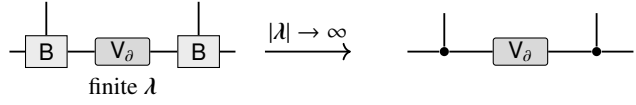

When $|\boldsymbol\lambda|$ is the dominant boundary energy scale, the
remaining terms $\sum_a\mathcal O_{\partial,a}$ are negligible in the
normalized boundary transfer matrix. The corresponding ordinary
boundary gate tends to the identity, and the tensors on the vertices become three-leg copy-tensors~(Fig.~\ref{fig:app-boundary-imps-insertion}),
\begin{equation}
  \mathsf B^{s}_{\alpha\beta}
  \xrightarrow{\ |\boldsymbol\lambda|\to\infty\ }
  \Delta_{\alpha\beta s}
  \equiv\delta_{\alpha s}\delta_{\beta s},
  \qquad
  (\mathsf V_{\partial})_{\alpha\beta}
  :\ \mathcal V\longrightarrow\mathcal V .
  \label{eq:app-boundary-delta-tensor}
\end{equation}
The tensor $\Delta_{\alpha\beta s}$ has two virtual legs and one
temporal-site leg. This identification with the delta function tensor can be clarified by Trotter-decomposing original local boundary gates into the diagonal part and the non-diagonal part, and put them on the temporal-site legs and spatial-site legs, respectively. The strong-field tensor
$\mathsf V_{\partial}$ remains a two-index operator inserted on the bond, whose matrix elements determine which
boundary states are chosen or projected out.

The temporal state is then generated by repeated application of the
fixed spatial transfer operator,
\begin{equation}
  |\Psi_{\mathrm{tm}}(L;\boldsymbol\lambda)\rangle
  =\mathcal N_{\mathrm{tm}}(L)
   T_x^L|\Psi_{\mathrm{tm},0}(\boldsymbol\lambda)\rangle .
  \label{eq:app-temporal-evolution}
\end{equation}
Equations~\eqref{eq:app-temporal-boundary-state} and
\eqref{eq:app-temporal-evolution} make explicit that changing the
virtual-bond insertion changes the initial state of the spatial
transfer evolution without altering $T_x$ or the bulk CFT.

The symmetry of $\delta H_{\partial}$ directly controls that of the resulting
temporal iMPS. With the symmetry $G$, a $H$-invariant field~($H\leq G$) gives an $H$-symmetric boundary
tensor, whereas a noninvariant component explicitly breaks that
symmetry. The direction or sign of the field distinguishes
symmetry-related microscopic terminations, and its magnitude fixes their
local weights. In the calculations reported here, we choose
representative field values and assign the resulting $B_O$ from the
low-lying entanglement spectrum.

We finish by specifying the boundary fields used for the three models.
For the Ising chain, the relevant distinction is whether the boundary
term preserves the onsite $Z_2$ symmetry. We use
\begin{equation}
  \delta H_{\partial}^{\mathrm{Ising}}(h_z)
  =-h_z\sigma_{\partial}^{z}.
  \label{eq:app-ising-boundary-field}
\end{equation}
The choice $h_z=0$ gives the $Z_2$-symmetric, free-boundary
representative. A nonzero longitudinal field pins the boundary spin, with $h_z>0$ and $h_z<0$ giving the two symmetry-related fixed-boundary
representatives. Only representative values are used and no continuous
field-strength dependence is inferred from the calculation. For the $|h_z|\rightarrow\infty$ limit, the gate $\mathsf{V}_{\partial}(h_z)$ will approach the projector $\ket{\uparrow}\bra{\uparrow}$ or $\ket{\downarrow}\bra{\downarrow}$, depending on the sign of $h_z$, giving a temporal state:
\begin{equation}
    \ket{\Psi^{\mathrm{fixed}}_{\mathrm{Ising},\mathrm{tm}}(h_z)}\xrightarrow[]{|h_z|\rightarrow\infty}
    \left\{\begin{aligned}
         &\bigotimes_i\ket{\uparrow}_i,\quad h_z>0,\\
         &\bigotimes_i\ket{\downarrow}_i,\quad h_z<0.
    \end{aligned}\right.
\end{equation}

For the Potts chain, fixed and mixed terminations are produced by the
complex longitudinal field
\begin{equation}
  \delta H_{\partial}^{\mathrm{Potts}}(h)
  =-h Z_{\partial}-h^*Z_{\partial}^{\dagger}.
  \label{eq:app-potts-longitudinal-field}
\end{equation}
For a generic phase of $h$, one color has the lowest boundary energy
and the insertion represents a color-fixed boundary. Along
$\arg h=\pi,\pm\pi/3$, two colors remain degenerate and the third is
suppressed. These three choices give the mixed $BC$, $CA$, and $AB$
representatives, respectively. For example, $\arg h=\pi$ suppresses
$A$ and retains the reflection symmetry that exchanges $B$ and $C$,
in direct correspondence with the not-$A$ state in
Eq.~\eqref{eq:app-potts-mixed-state}, with $\mathsf{V}_{\partial}(h)\rightarrow\ket{1}\bra{1}+\ket{2}\bra{2}$ (as $|h|\rightarrow\infty$). The corresponding temporal state is:
\begin{equation}
    \ket{\Psi^{\mathrm{mixed}}_{\mathrm{Potts},\mathrm{tm}}(h)}\xrightarrow[\arg h=\pi]{|h|\rightarrow\infty}\bigotimes_i\frac{|1\rangle+|2\rangle}{\sqrt{2}},
    \label{eq:app-potts-not-A-temporal-state}
\end{equation}
which is exactly a not-A mixed state as~\ref{eq:app-potts-mixed-state}. For $\arg h\neq \pi, \pm\pi/3$, two colors will be suppressed and the initial temporal will flow to the product state of $\ket{0}$, $\ket{1}$ or $\ket{2}$, as representative states related to fixed boundaries A, B or C.

The free and new Potts terminations are instead distinguished by the
transverse boundary field
\begin{equation}
  \delta H_{\partial}^{\mathrm{Potts}}(h_T)
  =-h_T X_{\partial}-h_T^*X_{\partial}^{\dagger}.
  \label{eq:app-potts-transverse-field}
\end{equation}
This boundary fields can be obtained via the duality transformation of the fixed and mixed boundary fields~\ref{eq:app-potts-longitudinal-field}~\cite{affleck_1998_bc}. For $\arg h_T=\pi,\pm\pi/3$, the $\ket{q_3=0}$ or $\ket{q_3=\pm1}$ states will be excluded, leading to the new boundary. Otherwise the boundary will flow to the free one. We use the real $h_T$ as an example. For real $h_T>0$ and $|h_T|\rightarrow\infty$, the boundary state is the symmetric state
$\otimes_i|q_3=0\rangle_i$, giving the free-boundary representative. For real
$h_T<0$ and $|h_T|\rightarrow\infty$, $|q_3=0\rangle$ is excluded in the strong-field limit and the
two-dimensional subspace
$P_{\perp}=1-\ket{q_3=0}\bra{q_3=0}$ is retained. Inserting the projector $P_{\perp}$ to the virtual bonds, one obtrain the new-boundary representative temporal state,
\begin{equation}
    \ket{\Psi^{\mathrm{new}}_{\mathrm{Potts},\mathrm{tm}}}=\operatorname{Tr}\!\left[
    \cdots \mathsf{\Delta}_{s_n}
    \mathsf P_{\perp}
    \mathsf{\Delta}_{s_{n+1}}
    \mathsf P_{\perp}\cdots
   \right]
   |\ldots s_n s_{n+1}\ldots\rangle,
    \label{eq:app-potts-temporal-new}
\end{equation}
Under the Fourier transformation to the charge basis, the $\mathrm{\Delta}$ tensor becomes:
\begin{equation}
    \mathrm{\Delta}=\frac{1}{\sqrt{3}}\left(\begin{array}{ccc}
        \ket{q_3=0} & \ket{q_3=+1} & \ket{q_3=-1} \\
        \ket{q_3=-1} & \ket{q_3=0} & \ket{q_3=+1} \\
        \ket{q_3=+1} & \ket{q_3=-1} & \ket{q_3=0}
    \end{array}\right)
    \label{eq:app-delta-charge-basis}
\end{equation}
Inserting $P_\perp$ to the virtual legs select the lower right $2\times 2$ block, which is exactly the ordinary new boundary state of the three-states Potts model.
Both $h_T>0$ and $h_T<0$ preserve $S_3$, just as the fixed-$A$ and
mixed-$BC$ choices preserve the same residual $Z_2$. Their differences therefore rely on the distinct entanglement spectra rather than on symmetry alone.

For the XXZ chain, we use three concrete prescriptions rather than the general field parametrization. For the $U(1)_z$-symmetric calculation, we add no boundary field. The resulting boundary insertion therefore
preserves the $U(1)_z$ symmetry of the bulk transfer operator. In long distance the state from the field-free boundary gate will fall into the same boundary class as the polarized boundary with $\delta H_\partial^{\mathrm{XXZ}}=-h_z S^z$, which as $h_z\rightarrow\infty$ will pin the state to the product state of $\ket{\uparrow}$ or $\ket{\downarrow}$, depending on the sign of $h_z$. In the field-free temporal construction, one spatial transfer step implements the
microscopic one-site translation and exchanges the two LSM-related
Dirichlet sectors, producing the observed alternation between integer
and half-integer sectors.
The $S_{\partial}^{x}$ insertion gives the
$U(1)_z$-breaking mixed Neumann--Dirichlet spectrum for $g<1$, $\delta H_{\partial}=-h_xS_{\partial}^{x}$. At the $g=1$ point, where
the bulk spin symmetry is enlarged to $\mathrm{SU}(2)$, we instead use a
deliberately non-Hermitian boundary term
$\delta H_{\partial}=-h_+S_{\partial}^{+}$, where
$S_{\partial}^{+}=S_{\partial}^{x}+iS_{\partial}^{y}$. This choice
removes the residual continuous symmetry that would remain for a
Hermitian field along a single spin axis and thus breaks the full
$\mathrm{SU}(2)$ symmetry. The $S_{\partial}^{+}$ insertion is used as a
boundary-state preparation device and is not interpreted as a physical
Hermitian boundary Hamiltonian. For the $S^x$ insertion case with $|h_x|\rightarrow\infty$, the boundary temporal state is pinned to a product state of $\ket{\rightarrow}$ or $\ket{\leftarrow}$, depending on the sign of $h_x$, which explicitly breaks the $U(1)_z$ symmetry. For $g=1$, the bond tensor becomes
$$
\mathrm{V}_\partial(h_+)=\exp(h_+\delta\tau S^+)=\ket{\uparrow}\bra{\uparrow}+\ket{\downarrow}\bra{\downarrow}+h_+\delta\tau\ket{\uparrow}\bra{\downarrow}.
$$
Inserting the operators into the virtual bonds, and taking $|h_+|\rightarrow\infty$ gives:
\begin{equation}
    \ket{\Psi_{\mathrm{XXX},\mathrm{tm}}^{\mathrm{br}}(h_+)}\xrightarrow[]{|h_+|\rightarrow\infty}\sum_i\cdots\ket{\uparrow}_{i-1}\ket{\uparrow}_{i}\ket{\downarrow}_{i+1}\ket{\downarrow}_{i+2}\cdots,
\end{equation}
which is the equal-weight superposition of the domain-wall configurations, and explicitly breaks the whole global spin rotation symmetry.

\end{document}